\documentclass[11pt]{article}
\usepackage{graphicx}
\usepackage{subcaption}
\usepackage{mathtools}
\usepackage{soul}
\usepackage{amsthm}
\usepackage{amsmath,amssymb}
\usepackage{systeme}
\usepackage{gensymb}
\usepackage[ruled,vlined,noend]{algorithm2e}
\SetKwComment{Comment}{/* }{ */}
\SetKwInOut{Input}{input}\SetKwInOut{Output}{output}
\LinesNumbered

\graphicspath{{./Figures/}}
\usepackage[absolute]{textpos}

\usepackage{lineno}

\usepackage{cite}
\usepackage[top=0.8in, bottom=0.9in, left=0.9in, right=0.9in]{geometry}

\usepackage[plainpages = false, pdfpagelabels,
                 bookmarks=false,
                 bookmarksopen = true,
                 bookmarksnumbered = true,
                 breaklinks = true,
                 linktocpage,
                 pagebackref,
                 colorlinks = true,  
                 linkcolor = blue,
                 urlcolor  = cyan,
                 citecolor = red,
                 anchorcolor = green,
                 hyperindex = true,
                 hyperfigures
                 ]{hyperref}

\usepackage{todonotes,cleveref}

\usepackage{tikz}
\usepackage{pgfplots}
\pgfplotsset{compat=1.15}
\usetikzlibrary{arrows}

\usepackage{thmtools}
\usepackage{thm-restate}

\DeclareMathOperator*{\argmin}{arg\,min}

\newtheorem{lemma}{Lemma}
\newtheorem{corollary}{Corollary}
\newtheorem{observation}{Observation}
\newtheorem{definition}{Definition}
\newtheorem{theorem}{Theorem}

\newcommand{\os}[1]{\overset{(#1)}}

\newcommand{\G}{G(S)} 
\newcommand{\R}{\text{Vor}} 
\newcommand{\E}{\mathcal{E}} 
\newcommand{\DT}{\mathcal{DT}} 
\newcommand{\vd}{\mathcal{VD}}
\newcommand{\N}{N_{\DT}}

\newcommand{\ca}{^\circ}
\newcommand{\sa}{^\diamond}

\newcommand{\seg}[1]{\overline{#1}}

\newcommand{\la}{<_{v^*}}

\newcommand{\ea}{=_{v^*}}

\newcommand{\W}{\mathcal{W}}
\newcommand{\lw}{<_{\W_t}}
\newcommand{\lp}{<_{\partial P}}

\newcommand{\sector}{%
  \mathord{%
    \begin{tikzpicture}[baseline=0]
      \draw[line width=0.08em, line join=round]
        (0,0) -- (60:1.6ex) arc (60:120:1.6ex) -- cycle;
    \end{tikzpicture}%
  }%
}

\title{Implicit Additively Weighted Geodesic Voronoi Diagrams and Applications\thanks{This paper will appear in {\em Proceedings of the 38th ACM-SIAM Symposium on Discrete Algorithms (SODA 2027).}}}
\author{
Bruce W. Brewer\thanks{Kahlert School of Computing,
University of Utah, Salt Lake City, UT 84112, USA. {\tt bruce.brewer@utah.edu}}
\and
Haitao Wang\thanks{Kahlert School of Computing,
University of Utah, Salt Lake City, UT 84112, USA. {\tt haitao.wang@utah.edu}}
}
\date{}

\begin{document}

\maketitle

\vspace{-0.2in}
\begin{abstract}
Let $P$ be a simple polygon with $m$ vertices. After $O(m)$-time preprocessing of $P$, given any set $S$ of $n$ additively weighted point sites in $P$, we present an algorithm that computes the additively weighted geodesic Voronoi diagram $\vd(S)$ implicitly in $O(n \log n + n \log^2 m)$ time. The resulting representation supports nearest-neighbor queries: for any query point $q \in P$, a nearest site of $q$ in $S$ can be found in $O(\log(n + m))$ time.
In contrast, explicitly constructing $\vd(S)$ requires $\Omega(m)$ time in the worst case. Thus, our implicit approach avoids the inherent dependence on the complexity of $P$, and is particularly advantageous when many diagrams need to be constructed in the same polygon and when $n$ is much smaller than $m$.

Our algorithm introduces a new sweep-line paradigm for additively weighted geodesic Voronoi diagrams that operates without explicitly constructing the diagram. A key technical ingredient is a novel method for point location in implicitly represented subdivisions. These techniques may be of independent interest.
We demonstrate that our results lead to several applications.

\begin{itemize}
    \item We obtain dynamic data structures for additively weighted geodesic nearest neighbor searching in simple polygons. After $O(m)$-time preprocessing of $P$, we can maintain an insertion-only data structure for a set $S$ of $n$ weighted sites supporting insertions in amortized $O(\log^2 n + \log n \log^2 m)$ time and queries in $O(\log n \log (n + m))$ time. We also obtain a fully dynamic data structure supporting insertions in amortized $O(\log n + \log^2 m)$ time, deletions in amortized $O(\sqrt{n} (\log n + \log^2 m))$ time, and queries in $O(\sqrt{n} \log (n + m))$ time.

    \item We obtain the first efficient algorithms for shortest paths in geodesic disk graphs in simple polygons. Given a set $S$ of $n$ points in $P$, where each point $p \in S$ is associated with a radius $r_p$, the geodesic disk graph $G(S)$ has an edge between two vertices $u,v \in S$ if $d(u,v) \le r_u + r_v$, where $d(u,v)$ is the geodesic distance between $u$ and $v$ in $P$. For a source $s \in S$, we consider the single-source shortest path problem in $G(S)$. In the unweighted case, the length of a path is the number of edges in the path. In the weighted case, each edge $(u,v)$ in $G(S)$ has weight equal to $d(u,v)$, and the length of a path is the sum of its edge weights.

    After $O(m)$-time preprocessing of $P$, given any $S$ and a source $s \in S$, we solve this problem in $O(n \log n + n \log^2 m)$ time in the unweighted case, and in $O(n^{3/2}(\log n + \log^2 m))$ time in the weighted case. Our approach provides a new and conceptually simple framework that may be useful for other shortest path problems in geometric intersection graphs.
\end{itemize}

Finally, we extend our techniques to computing implicit farthest geodesic Voronoi diagrams for unweighted sites, achieving the same asymptotic bounds as above. As a consequence, we also obtain dynamic data structures for farthest neighbor searching.
\end{abstract}


\newpage
\tableofcontents
\newpage

\section{Introduction}
\label{sec:intro}

\begin{figure}[t]
    \centering
    \includegraphics[width=.35\linewidth]{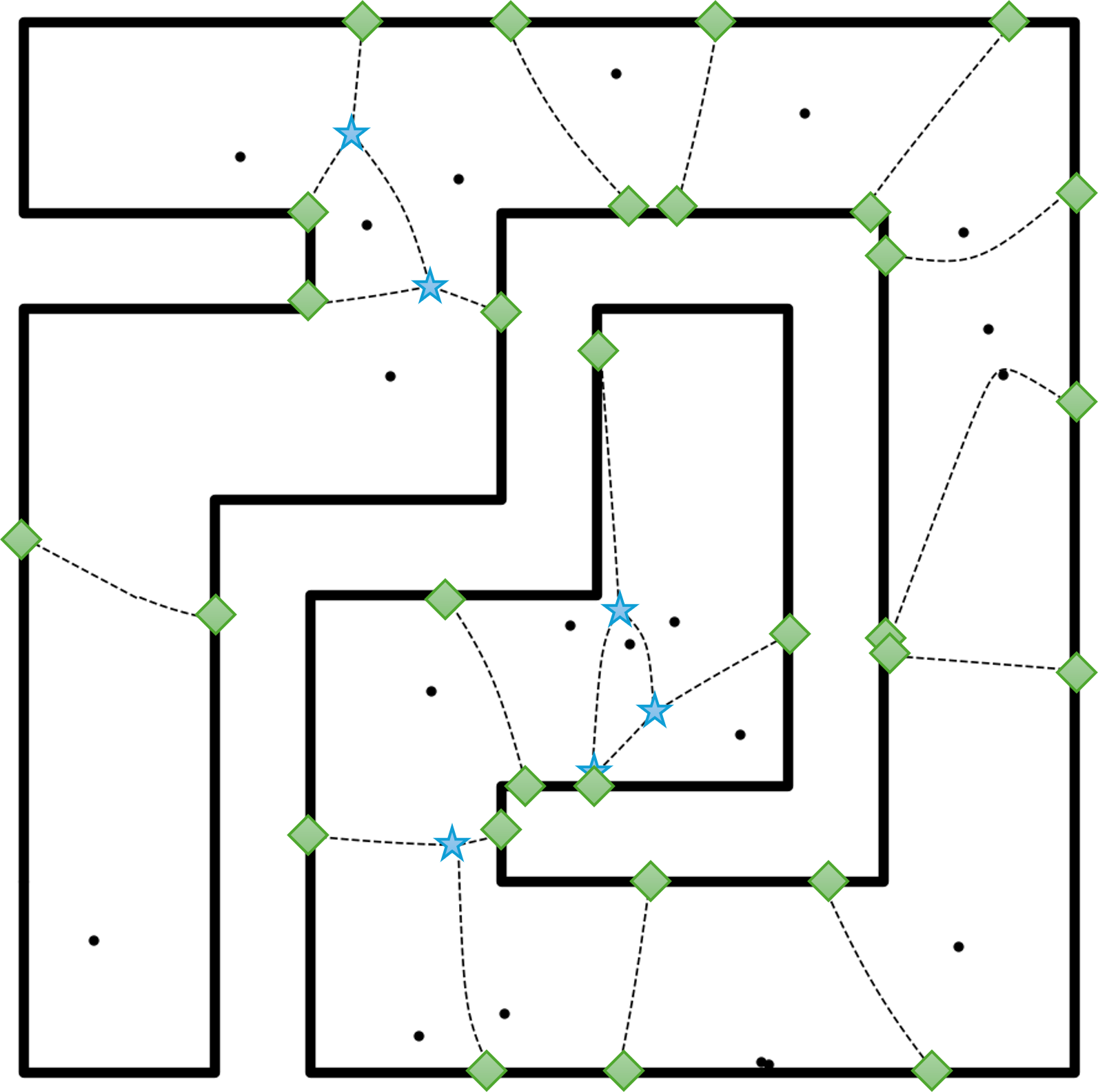}
    \caption{Illustration of an additively weighted geodesic Voronoi diagram in a simple polygon.
        The sites are black points,
        the polygon boundary is thick solid black,
        the Voronoi edges are dashed black,
        the Voronoi vertices are light blue stars,
        and the Voronoi terminals are light green diamonds.} \label{fig:big_voronoi}
\end{figure}

Let $P$ be a simple polygon with $m$ vertices.
The \emph{geodesic distance} $d(p, q)$ between any two points $p, q \in P$
is the length of the shortest path from $p$ to $q$ contained in $P$.
Let $S$ be a set of $n$ point sites in $P$, where each site $s \in S$ has an associated weight $w(s)$.
The \emph{additively weighted geodesic distance} (or simply \emph{weighted distance}) from a site $s \in S$ to a point $p \in P$ is defined as $d(s, p) + w(s)$.
We consider the problem of constructing the Voronoi diagram of $S$ under the weighted distance, which is called the \emph{additively weighted geodesic Voronoi diagram} (see Figure~\ref{fig:big_voronoi}); we denote it by $\vd(S)$.

A fundamental application of $\vd(S)$ is nearest neighbor searching: given a query point $q \in P$, find a nearest neighbor of $q$ in $S$ with respect to the weighted distance. This can be achieved by first constructing $\vd(S)$ and then building a point location data structure on $\vd(S)$~\cite{ref:KirkpatrickOp83,ref:EdelsbrunnerOp86}.
Since the combinatorial complexity of $\vd(S)$ is $\Theta(n+m)$ in the worst case~\cite{ref:AronovOn89,ref:HershbergerAn99}, explicitly constructing $\vd(S)$ requires $\Omega(m)$ time. The algorithm of Hershberger and Suri~\cite{ref:HershbergerAn99} can construct $\vd(S)$ explicitly in $O((n+m)\log(n+m))$ time.

In many applications, however, $m$ can be much larger than $n$, and we need to construct $\vd(S)$ repeatedly for different sets $S$ to support nearest neighbor queries. A typical example arises in dynamic data structures, where the set $S$ is updated over time. For instance, the logarithmic method of Bentley and Saxe~\cite{ref:BentleyDe80} yields an insertion-only structure, but requires rebuilding $\vd(S)$ for sets of sizes $n = 1, 2, 4, 8, \ldots$. In such settings, the $\Omega(m)$ construction cost per instance is prohibitive.

This motivates preprocessing the polygon $P$ so that, for any given set $S$, $\vd(S)$ can be constructed \emph{implicitly} in time depending primarily on $n$ and only logarithmically on $m$. This is the main problem studied in this paper, which we call the \emph{implicit additively weighted geodesic Voronoi diagram problem} (or the \emph{implicit GVD problem}).

\subsection{Our results}

We summarize our main result in the following theorem.

\begin{restatable}{theorem}{mainthm} \label{thm:mainthm}
Let $P$ be a simple polygon with $m$ vertices. After $O(m)$-time preprocessing of $P$, given any set $S$ of $n$ additively weighted point sites in $P$, we can compute the additively weighted geodesic Voronoi diagram $\vd(S)$ implicitly in $O(n \log n + n \log^2 m)$ time. Subsequently, a point location data structure can be built on the implicit $\vd(S)$ in an additional $O(n \log m)$ time, supporting nearest neighbor queries for $S$ in $O(\log(n + m))$ time per query.
\end{restatable}

Since $\vd(S)$ is represented implicitly, standard point location algorithms~\cite{ref:KirkpatrickOp83,ref:EdelsbrunnerOp86} do not apply. We therefore develop a new point location method for the implicit diagram, which may be of independent interest.
Note that once the implicit diagram is computed as in the theorem, the explicit diagram can be constructed in an additional $O(m + n \log m)$ time.

Brewer and Wang~\cite{ref:BrewerSh26} recently studied the implicit GVD problem as a subroutine in their algorithm for computing shortest paths in geodesic unit-disk graphs. After $O(m)$-time preprocessing of $P$, given any set $S$, they constructed a data structure for $S$ in $O(n \log n \log^2 m + n \log^3 m)$ time that supports nearest-neighbor queries in $O(\log n \log^2 m)$ time. In comparison, our result improves the construction time by a factor of $\log m$ and improves the query time by a factor of $\log^2 m$.

In order to design dynamic nearest neighbor data structures, Agarwal, Arge, and Staals~\cite{ref:AgarwalIm18} studied the {\em unweighted} version of the implicit GVD problem. After $O(m)$-time preprocessing of $P$, given any set $S$ of unweighted sites, they constructed a data structure in $O(n \log n \log m + n \log^3 m)$ time that supports queries in $O(\log n \log^2 m)$ time. Our result, which applies to the more general weighted setting, improves both the preprocessing and query bounds.

Extending the techniques of Agarwal, Arge, and Staals~\cite{ref:AgarwalIm18} to the weighted setting presents several challenges. For example, their algorithm relies crucially on the following two properties. First, for any two unweighted sites on the same side of a diagonal $d$ of $P$, their bisector intersects $d$ at most once. Second, the bisectors of any three unweighted sites have at most one common intersection (equivalently, there is at most one point in $P$ equidistant from the three sites). Neither property holds in the weighted setting.

More generally, extending algorithms from the unweighted setting to the weighted one is by no means routine. While some algorithms admit relatively simple extensions, e.g., Fortune's sweep-line algorithm for Euclidean Voronoi diagrams~\cite{ref:FortuneA87}, others do not. Consider the dynamic Euclidean nearest neighbor problem~\cite{ref:ChanA10,ref:ChanDy20,ref:KaplanDy20,ref:LiuNe22}. Deterministic data structures are known for the unweighted case~\cite{ref:ChanDy20}, while for the weighted case only randomized data structures with asymptotically matching bounds are currently known~\cite{ref:KaplanDy20,ref:LiuNe22}. Whether the deterministic techniques of~\cite{ref:ChanDy20} can be extended to the weighted setting to achieve the same bounds remains an important open problem.

Oh and Ahn~\cite{ref:OhVo20} also considered the unweighted implicit GVD problem. After $O(m)$-time preprocessing of $P$, they constructed a data structure for $S$ in $O(n \log n \log^2 m)$ time.\footnote{They claimed $O(\log(n+m))$ query time. However, we found an issue in their point location algorithm~\cite[Lemma~8.3]{ref:OhVo20}. Specifically, they claimed to support arbitrary ray-shooting queries among $n$ disjoint segments in the plane in $O(\log n)$ time after $O(n \log n)$ preprocessing and using $O(n)$ space. While this is possible when all query rays are vertical~\cite{ref:SarnakPl86}, such performance is unlikely for arbitrary query directions~\cite{ref:ChanSi23}. We were unable to fix this issue after several attempts. Instead, we develop a new method that achieves asymptotically the same query time. Our technique may have broader impact because it reduces the problem of locating a point in an implicit GVD to that of locating a point in geodesic triangles and geodesic sectors, which is a problem that might find other applications.}

While Theorem~\ref{thm:mainthm} is the main result of this paper, we also study the farthest version of the problem but on unweighted point sites only, i.e., construct the geodesic farthest Voronoi diagram implicitly to answer farthest neighbor queries. We obtain the same bounds as in Theorem~\ref{thm:mainthm}, summarized in the following theorem.

\begin{theorem}\label{thm:farthest}
Let $P$ be a simple polygon with $m$ vertices. After $O(m)$-time preprocessing of $P$, given any set $S$ of $n$ point sites in $P$, we can compute the farthest geodesic Voronoi diagram $\vd_f(S)$ implicitly in $O(n \log n + n \log^2 m)$ time.
Subsequently, a point location data structure can be built on the implicit $\vd_f(S)$ in an additional $O(n \log m)$ time, supporting farthest neighbor queries for $S$ in $O(\log(n + m))$ time per query.
\end{theorem}

The problem of constructing $\vd_f(S)$ implicitly was also studied by Oh and Ahn~\cite{ref:OhVo20}, whose algorithm constructs the diagram implicitly in $O(n \log n \log^2 m)$ time after $O(m)$-time preprocessing of $P$. Thus, our algorithm improves their result by a logarithmic factor.
Extending our result to the weighted setting appears to be significantly more challenging, and we leave it for future work.

\subsubsection{Our approach}
\label{sec:ourapproach}

Our algorithm for constructing the implicit $\vd(S)$ in Theorem~\ref{thm:mainthm} is inspired by Fortune's sweep-line algorithm for Euclidean Voronoi diagrams~\cite{ref:FortuneA87}.
In Fortune's algorithm (and its variants~\cite{ref:GuibasTh88,ref:deBergCo08}), a line $\ell$ sweeps the plane from top to bottom while maintaining the portion of the Voronoi diagram above $\ell$, treating $\ell$ as a special site.
The boundary of the Voronoi region of $\ell$, called the \emph{beach line}, is $x$-monotone and is maintained in a data structure that supports efficient updates as $\ell$ moves.

To extend this idea to our setting, we proceed as follows.
Pick any convex vertex of $P$, denoted by $v^*$, and assign it a weight $w(v^*) = t$, where $t$ is a parameter.
Let $\vd_t(S)$ be the Voronoi diagram of $S \cup \{v^*\}$ with respect to these weights.
We maintain $\vd_t(S)$ as $t$ increases from $-\infty$ to $\infty$.
In the unweighted case (i.e., when all sites have equal weight), this process is equivalent to sweeping a geodesic circle centered at $v^*$ with radius decreasing from $\infty$ to $0$, while maintaining the Voronoi diagram outside the circle by treating the circle as a special site.
In the weighted case, however, there is no direct geometric interpretation; instead, we conceptually sweep the parameter $t$.
The boundary of the Voronoi cell of $v^*$ in $\vd_t(S)$ serves as the beach line of our algorithm.
As in classical sweep-line algorithms, the efficiency of our approach hinges on handling events efficiently.


As mentioned above, Brewer and Wang~\cite{ref:BrewerSh26} gave an algorithm for supporting
nearest-neighbor queries. Their algorithm takes a different approach and does
not follow the direction of generalizing Fortune's sweep-line algorithm.
Instead, it uses a balanced polygon decomposition~\cite{ref:GuibasLi87,ref:GuibasOp89,ref:ChazelleVi89}
to reduce the problem to a ``diagonal-separated'' subproblem, which is
relatively easier to handle. The drawback, however, is that this approach
leads to a slower algorithm. In contrast, our approach is based on a new
generalization of Fortune's sweep-line method and avoids the polygon
decomposition altogether, resulting in a faster algorithm. This constitutes
one of the main novelties of our approach.

\paragraph{\bf The farthest diagram.}
Our algorithm for computing the farthest implicit diagram $\vd_f(S)$ does
not follow the same approach as its nearest counterpart. Instead, we build
on the algorithm of Oh and Ahn~\cite{ref:OhVo20}, which uses a ``reverse
geodesic sweep'' to construct $\vd_f(S)$ from the boundary of $P$ towards
the geodesic center $c^*$ of $S$. The bottleneck of their algorithm is
computing $c^*$, which takes $O(n \log n \log^2 m)$ time after $O(m)$-time
preprocessing. The remainder of their algorithm runs in
$O(n \log n + n \log^2 m)$ time. We present a new algorithm that computes
$c^*$ in $O(n \log n + n \log^2 m)$ time. Consequently, $\vd_f(S)$ can be
computed in $O(n \log n + n \log^2 m)$ total time after $O(m)$-time
preprocessing.

\paragraph{\bf Implicit point locations.}
To answer point location queries in $\vd(S)$, the traditional approach is
to construct $\vd(S)$ explicitly and then build a point location data
structure on it~\cite{ref:KirkpatrickOp83,ref:EdelsbrunnerOp86}.
This approach, however, requires $\Omega(m)$ preprocessing time, since
explicitly constructing $\vd(S)$ already requires $\Omega(m+n)$ space.
In our setting, we must instead support point location queries directly
on the implicit representation of $\vd(S)$.

We show that, after $O(n \log m)$ preprocessing of the implicit $\vd(S)$,
a point location query can be answered in
$O(\log(n+m))$ time, which is optimal. To achieve this result, we develop
a more general technique that applies to a decomposition of $P$ into so-called
\emph{geodesic triangles} and \emph{geodesic sectors}; see
Section~\ref{sec:point_location} for details. We believe that this technique
may be useful in other settings where point location must be performed
on an implicit representation. Our point location data structure for the
farthest diagram $\vd_f(S)$ also relies on this technique.

\subsubsection{Applications}

Nearest/farthest neighbor searching is a fundamental problem, and thus Theorems~\ref{thm:mainthm} and~\ref{thm:farthest} have the potential to find many applications.
In general, our results are useful in settings where Voronoi diagrams need to be constructed repeatedly.
For example, in robotics applications, $S$ may represent a set of moving robots in a workspace $P$, and $\vd(S)$ or $\vd_f(S)$ needs to be recomputed at each time step.
Below we present several concrete applications; for these problems, our algorithms are either the first known or improve upon the previous best results.

\paragraph{Dynamic nearest neighbor searching.}

An immediate application is the design of data structures for dynamic nearest neighbor queries.
Let $S$ be a set of weighted point sites in $P$. We seek a \emph{fully dynamic} data structure supporting the following operations:
(1) \textbf{Insertions}: insert a weighted site into $S$;
(2) \textbf{Deletions}: delete a site from $S$;
(3) \textbf{Queries}: given a query point $q$, find a nearest neighbor of $q$ in $S$ with respect to the weighted distance.

Applying standard static-to-dynamic transformations~\cite{ref:BentleyDe80,ref:OvermarsTh83} to Theorem~\ref{thm:mainthm}, we obtain $O(\log n + \log^2 m)$ amortized insertion time, $O(\sqrt{n} (\log n + \log^2 m))$ amortized deletion time, and $O(\sqrt{n} \log (n + m))$ query time.
For the insertion-only case, we achieve $O(\log^2 n + \log n \log^2 m)$ amortized insertion time and $O(\log n \log (n + m))$ query time.
These results also serve as building blocks for further applications discussed below.

\paragraph{Shortest paths in unweighted geodesic disk graphs.}

For a point $p \in P$ and a radius $r_p$, a \emph{geodesic disk} centered at $p$ with radius $r_p$ consists of all points in $P$ whose geodesic distance from $p$ is at most $r_p$.

Let $S$ be a set of $n$ points in $P$, where each point $p \in S$ is associated with a radius $r_p$.
The \emph{geodesic disk graph} $G(S)$ has vertex set $S$ and contains an edge between two vertices if their corresponding disks intersect, i.e., if $d(u,v) \le r_u + r_v$ for $u, v \in S$.

Given a source point $s \in S$, we consider the single-source shortest path (SSSP) problem of computing shortest paths from $s$ to all other vertices in $G(S)$, where the length of a path is the number of edges.

To the best of our knowledge, this problem has not been studied before.
In the Euclidean setting (i.e., when $P$ is the entire plane), it can be solved in $O(n \log n)$ time~\cite{ref:BrewerAn25,ref:deBergAn25}, which matches an $\Omega(n \log n)$ lower bound~\cite{ref:CabelloSh15}.
For the geodesic unit-disk case (where all radii are equal), Brewer and Wang~\cite{ref:BrewerSh26} gave an $O(m + n \log n \log^2 m)$-time algorithm.

We present an $O(m + n \log n + n \log^2 m)$-time algorithm for this SSSP problem, with Theorem~\ref{thm:mainthm} as a key ingredient.
Our running time matches the Euclidean bound when $m = O(1)$ and improves upon the result of~\cite{ref:BrewerSh26}.
More generally, after $O(m)$-time preprocessing of $P$, given any set $S$ and a source $s \in S$, we can solve the SSSP problem in $O(n \log n + n \log^2 m)$ time. This result is particularly useful when solving the problem repeatedly for different sets $S$, where each instance can be viewed as a query with input $(S, s)$.

\paragraph{Shortest paths in weighted geodesic disk graphs.}

The above corresponds to the unweighted case. In the weighted version, each edge $(u,v)$ of $G(S)$ has weight equal to the geodesic distance $d(u,v)$ between $u$ and $v$, and the length of a path is the sum of its edge weights.

For the Euclidean case, the current best algorithm runs in $O(n \log^4 n)$ time~\cite{ref:AnSi25}.
For the geodesic unit-disk case, Brewer and Wang~\cite{ref:BrewerSh26} gave an $O(m + n \log^2 n \log^2 m)$-time algorithm.

We present the first algorithm for this weighted SSSP problem in simple polygons, with a running time of $O(m + n^{3/2} (\log n + \log^2 m))$.
Our approach builds on our dynamic nearest-neighbor data structures.
More generally, after $O(m)$-time preprocessing of $P$, given any set $S$ and a source $s \in S$, we can solve the weighted SSSP problem in $O(n^{3/2} (\log n + \log^2 m))$ time.
Our technique introduces a new and conceptually simple framework that may be useful for other variants of shortest path problems in disk graphs and, more broadly, geometric intersection graphs.

\paragraph{Dynamic farthest neighbor searching.}

As in the nearest neighbor searching problem, by applying standard static-to-dynamic transformations~\cite{ref:BentleyDe80,ref:OvermarsTh83} to Theorem~\ref{thm:farthest}, we immediately obtain dynamic data structures for geodesic farthest neighbor searching in simple polygons. The time bounds are exactly the same as the above for the nearest-neighbor version.

\subsection{Other related work}

Voronoi diagrams are among the most studied topics in computational geometry~\cite{ref:AurenhammerVo13}.
One of their primary applications is nearest neighbor searching, which typically requires building a point location data structure on the diagram~\cite{ref:KirkpatrickOp83,ref:EdelsbrunnerOp86}.

\paragraph{Static (nearest) Voronoi diagrams.}
Computing the Voronoi diagram of $n$ points in the Euclidean plane can be done in optimal $O(n \log n)$ time, either by the divide-and-conquer algorithm of Shamos and Hoey~\cite{ref:ShamosCl75} or by Fortune's sweep-line algorithm~\cite{ref:FortuneA87}.
Fortune's algorithm extends to the additively weighted case with the same time bound.

Constructing explicit unweighted geodesic Voronoi diagrams in simple polygons has been studied extensively.
The first algorithm, due to Takao Asano and Tetsuo Asano~\cite{ref:AsanoVo87}, runs in $O(mn + n \log n)$ time.
Aronov~\cite{ref:AronovOn89} improved this to $O((n + m) \log m \log (n + m))$.
Papadopoulou and Lee~\cite{ref:PapadopoulouA98} subsequently obtained an $O((n + m) \log (n + m))$-time algorithm.
More recently, Oh and Ahn~\cite{ref:OhVo20} achieved $O(m + n \log n \log^2 m)$ time, and Liu~\cite{ref:LiuA20} gave an $O(m + n \log n + n \log^2 m)$-time algorithm.
Finally, Oh~\cite{ref:OhOp19} obtained an optimal $O(m + n \log n)$-time algorithm, matching the $\Omega(m + n \log n)$ lower bound.

To the best of our knowledge, the additively weighted geodesic Voronoi diagram in simple polygons has not been studied directly.
However, the continuous Dijkstra framework of Hershberger and Suri~\cite{ref:HershbergerAn99} can be extended to construct it in $O((n + m) \log (n + m))$ time, even for polygons with holes.

\paragraph{Dynamic nearest neighbor searching.}
In the dynamic setting, the set $S$ is updated over time.
The insertion-only case can be handled using the logarithmic method of Bentley and Saxe~\cite{ref:BentleyDe80}, yielding amortized $O(\log^2 n)$ insertion time and $O(\log^2 n)$ query time in the Euclidean setting.
For the unweighted case, a recent breakthrough by Iacono and Nekrich~\cite{ref:IaconoIn25} achieves amortized $O(\log^{1+\epsilon} n)$ insertion time for any $\epsilon > 0$ and $O(\log n)$ query time, although their approach does not extend to weighted settings.

Using the equal-block method of Overmars~\cite{ref:OvermarsTh83}, the fully dynamic problem can be solved with $O(\sqrt{n} \log n)$ update time and $O(\sqrt{n} \log n)$ query time.
This can be improved significantly using shallow cuttings.
For the unweighted case, the best result is a data structure by Chan~\cite{ref:ChanDy20} that supports amortized $O(\log^2 n)$ insertion time, amortized $O(\log^4 n)$ deletion time, and $O(\log^2 n)$ query time.
For the additively weighted case, similar bounds can be achieved with randomized expected update time~\cite{ref:LiuNe22,ref:KaplanDy20}.

For dynamic nearest neighbor searching in simple polygons (with unweighted sites), Agarwal, Arge, and Staals~\cite{ref:AgarwalIm18} combined their implicit GVD structure with shallow cuttings~\cite{ref:ChanDy20,ref:LiuNe22,ref:KaplanDy20} to obtain a data structure with $O(\log^5 n \log m + \log^4 n \log^3 m)$ amortized expected insertion time, $O(\log^7 n \log m + \log^6 n \log^3 m)$ amortized expected deletion time, and $O(\log^2 n \log^2 m)$ query time. By extending the techniques from \cite{ref:AgarwalIm18,ref:ChanDy20,ref:LiuNe22,ref:KaplanDy20}, de Berg and Staals~\cite{ref:BergDy23} studied dynamic $k$-nearest neighbor queries in simple polygons.

\paragraph{Shortest paths in disk graphs.}

The problem of computing shortest paths in Euclidean disk graphs has been studied extensively, with variants depending on whether the disks have equal radii (unit-disk vs.\ general-disk) and whether edges are weighted.

For the unweighted unit-disk case, the problem can be solved in $O(n \log n)$ time~\cite{ref:CabelloSh15,ref:ChanAl16,ref:EfratGe01}, matching an $\Omega(n \log n)$ lower bound~\cite{ref:CabelloSh15}.
For the weighted unit-disk case, Roditty and Segal~\cite{ref:RodittyOn11} gave an $O(n^{4/3+\epsilon})$-time algorithm, which was improved to $O(n^{1+\epsilon})$ by Cabello and Jej\v{c}i\v{c}~\cite{ref:CabelloSh15}.
Further improvements were obtained by Kaplan, Mulzer, Roditty, Seiferth, and Sharir~\cite{ref:KaplanDy20} and Liu~\cite{ref:LiuNe22}, whose algorithms follow the algorithmic framework of~\cite{ref:CabelloSh15} using dynamic bichromatic closest-pair (BCP) data structures.
More recently, Wang and Xue~\cite{ref:WangNe20} proposed an $O(n \log^2 n)$-time algorithm that avoids such BCP structures, and Brewer and Wang~\cite{ref:BrewerAn26} improved this to $O(n \log^2 n / \log \log n)$  by refining a key bottleneck subproblem.

For the unweighted general-disk case, the problem can be solved via breadth-first search combined with dynamic nearest neighbor structures~\cite{ref:KaplanDy20,ref:KaplanTh23}.
Using the randomized structure of~\cite{ref:LiuNe22}, this yields $O(n \log^4 n)$ expected time.
Klost~\cite{ref:KlostAn23} gave a deterministic $O(n \log^2 n)$-time algorithm, and the problem was later solved optimally in $O(n \log n)$ time~\cite{ref:BrewerAn25,ref:deBergAn25}.

For the weighted general-disk case, Kaplan, Katz, Saban, and Sharir~\cite{ref:KaplanTh23} gave an $O(n \log^6 n)$ expected-time algorithm, which was improved to a deterministic $O(n \log^4 n)$-time algorithm by An, Oh, and Xue~\cite{ref:AnSi25}.

The shortest path problem in geodesic disk graphs within simple polygons was recently introduced by Brewer and Wang~\cite{ref:BrewerSh26}. Only the unit-disk case was studied in~\cite{ref:BrewerSh26}, and the results were mentioned above.

Several other problems on geodesic disks in polygons have also been considered, including clustering~\cite{ref:BorgeltGe07}, separators~\cite{ref:deBergCl22}, and piercing~\cite{ref:BosePi21}.

\paragraph{Farthest Voronoi diagrams.}
The problem of computing the farthest Voronoi diagram $\vd_f(S)$ explicitly has been extensively studied.
In the Euclidean plane, the problem can be solved in $O(n \log n)$ time~\cite{ref:PreparataCo85}, matching the $\Omega(n \log n)$ lower bound~\cite{ref:LeeFa80}. For the geodesic case in simple polygons, Aronov, Fortune, and Wilfong~\cite{ref:AronovTh93} showed that the complexity of the diagram is $\Theta(n + m)$ and gave an $O(m \log m + n \log n)$-time algorithm, which is near-optimal since $\Omega(m + n \log n)$ is a lower bound. More recently, Oh, Barba, and Ahn~\cite{ref:OhTh20} presented an $O(m \log\log m + n \log n)$-time algorithm, and subsequently, Oh and Ahn~\cite{ref:OhVo20} gave an $O(m + n \log n + n \log^2 m)$-time algorithm.
Barba~\cite{ref:BarbaOp19} proposed a randomized algorithm with $O(m + n \log n)$ expected time, and Wang~\cite{ref:WangAn23DCG} later obtained a deterministic $O(m + n \log n)$-time algorithm.

\paragraph{Outline.} The rest of the paper is organized as follows. After introducing some notation and concepts in Section~\ref{sec:pre}, Section~\ref{sec:high_level} gives an overview of our sweeping algorithm for Theorem~\ref{thm:mainthm}, with the details presented in the subsequent sections.
Section~\ref{sec:basics} introduces the geometric primitives (e.g., computing Voronoi vertices) used in our algorithm, along with the data structures needed during the algorithm (e.g., to maintain the sweeping status).
The algorithms for processing the events are described in Section~\ref{sec:site_events} and Section~\ref{sec:nonsite_events}.
In Section~\ref{sec:wrap_up}, we analyze the time complexity of the algorithm and prove its correctness.
Our point location algorithm is presented in Section~\ref{sec:point_location}.
Section~\ref{sec:applications} discusses the applications of Theorem~\ref{thm:mainthm}, including the dynamic data structures and the shortest path algorithms for geodesic disk graphs. Theorem~\ref{thm:farthest} is proved in Section~\ref{sec:farthest}, which is the only section dealing with farthest Voronoi diagrams.
Section~\ref{sec:conclude} concludes, with some directions for future work.

\section{Preliminaries}
\label{sec:pre}

We follow the notation introduced earlier, including $P$, $m$, $S$, and $n$.
For any region $R$ in the plane, let $\partial R$ denote its boundary; in particular, $\partial P$ is the boundary of $P$.
Let $V_P$ denote the set of vertices of $P$.
A vertex of $P$ is \emph{reflex} if its interior angle is greater than $\pi$, and \emph{convex} otherwise.

We use $\seg{pq}$ to denote the line segment connecting two points $p$ and $q$, and $|\seg{pq}|$ to denote its length.

For any two points $p, q \in P$, let $\pi(p, q)$ denote the shortest path in $P$ from $p$ to $q$, oriented from $p$ to $q$.
It is well known that every interior vertex of $\pi(p, q)$ (i.e., excluding $p$ and $q$) is a reflex vertex of $P$.
The vertex of $\pi(p, q)$ adjacent to $p$ (resp., $q$) is called the \emph{anchor} of $p$ (resp., $q$) in $\pi(p, q)$.
Let $d(p, q)$ denote the length of $\pi(p, q)$, i.e., the \emph{geodesic distance} between $p$ and $q$, and refer to $\pi(p, q)$ as a \emph{geodesic path}.

For a site $s \in S$ with weight $w(s)$ and a point $p \in P$, define the \emph{weighted distance} $d_s(p) = d(s, p) + w(s)$.
We further define $d_S(p) = \min_{s \in S} d_s(p)$. Except in Section~\ref{sec:farthest}, sites in $S$ always have weights.

\begin{figure}[t]
\begin{minipage}[t]{0.49\linewidth}
\begin{center}
\includegraphics[totalheight=2.6in]{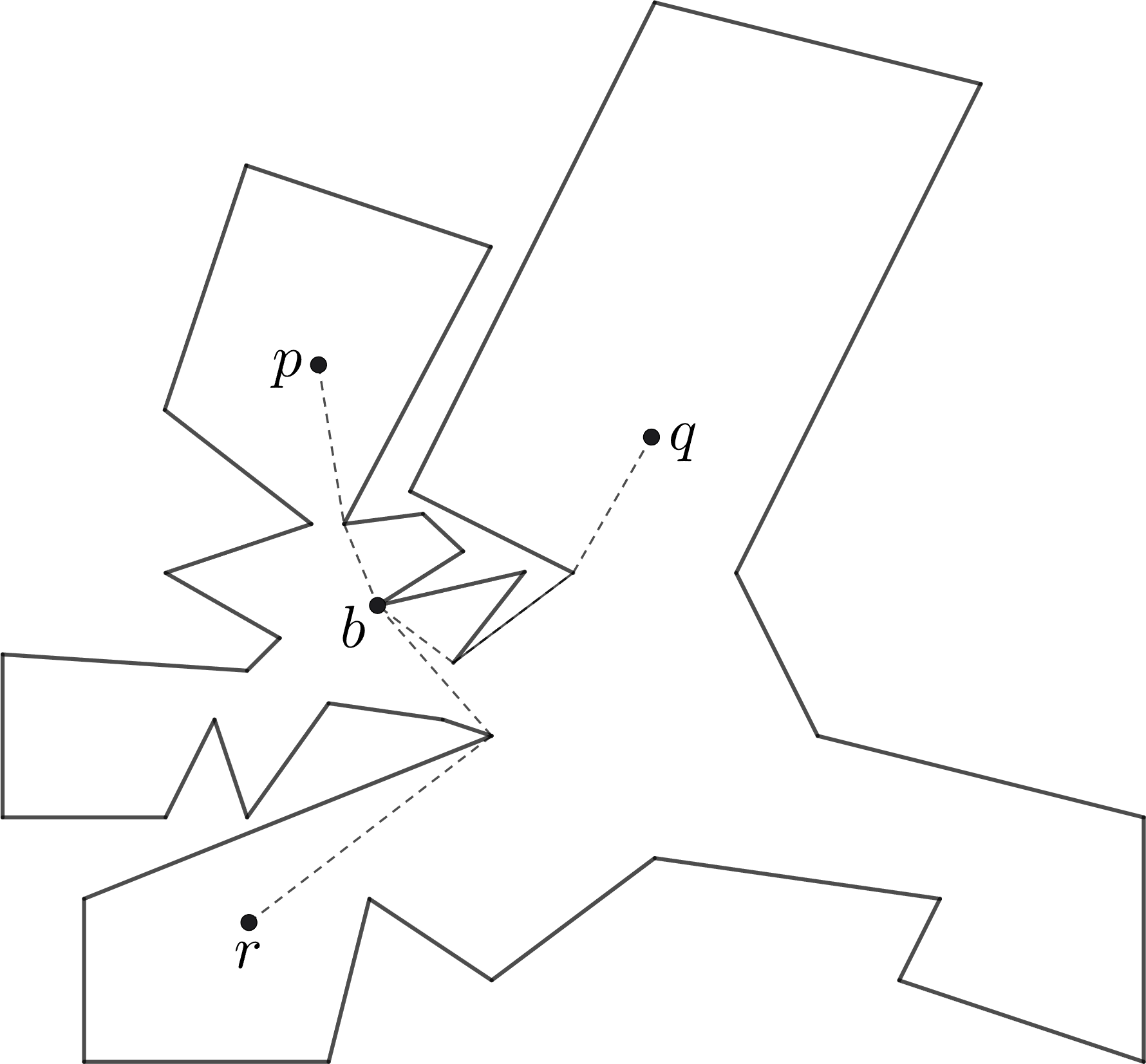}
\caption{$b$ is the junction vertex of $\pi(p, r)$ and $\pi(p, q)$.}
\label{fig:junction_vertex}
\end{center}
\end{minipage}
\hspace{0.05in}
\begin{minipage}[t]{0.49\linewidth}
\begin{center}
\includegraphics[totalheight=2.6in]{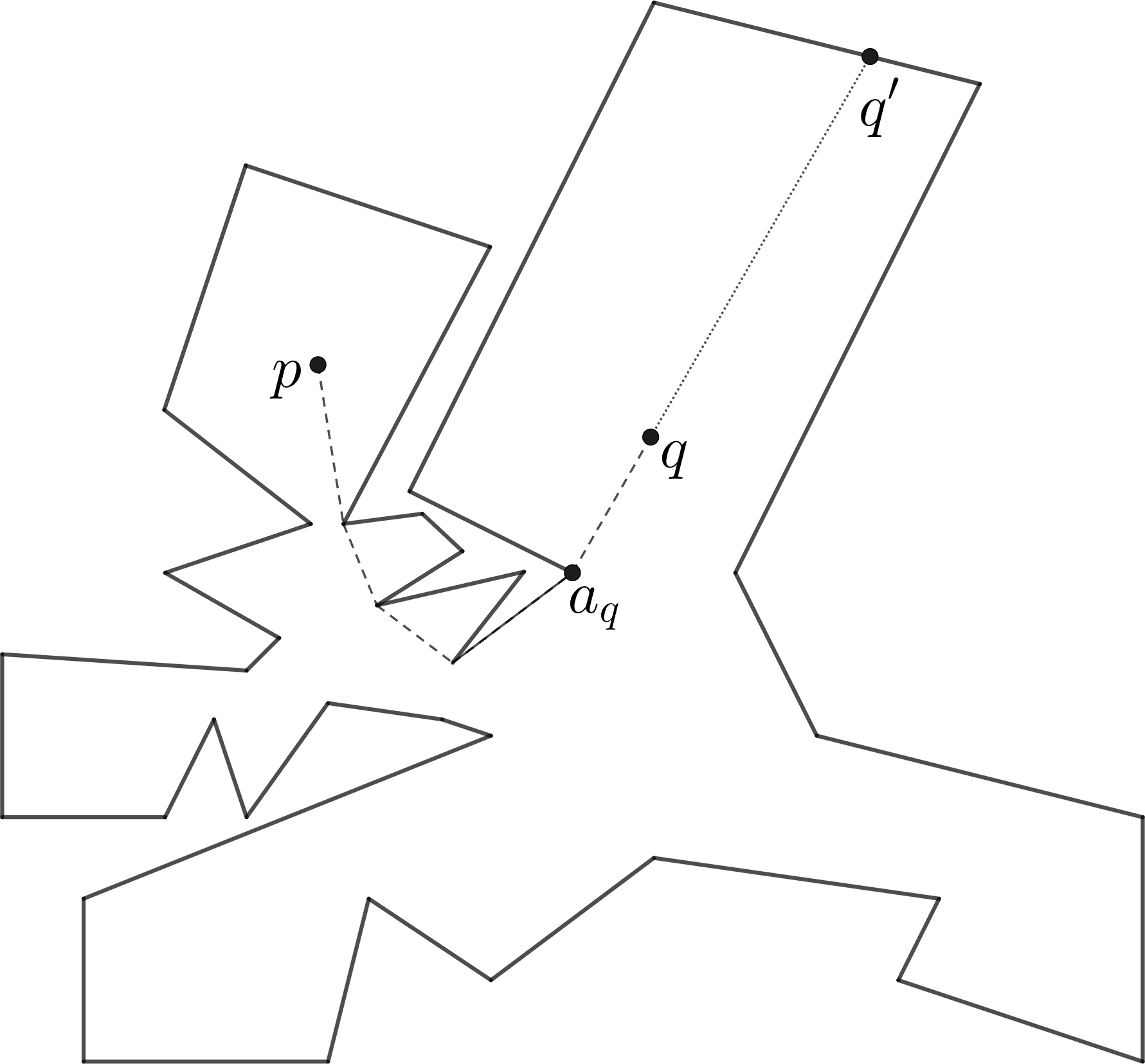}
\caption{$\seg{q q'}$ is the last-edge extension of $\pi(p, q)$.}
\label{fig:last_edge_extension}
\end{center}
\end{minipage}
\end{figure}

For any three points $p, q, r \in P$, the vertex of the common subpath $\pi(p,r) \cap \pi(p,q)$ that is farthest from $p$ is called the \emph{junction vertex} of $\pi(p,r)$ and $\pi(p,q)$; see Figure~\ref{fig:junction_vertex}.

\paragraph{The GH data structure and ray shooting.}
We frequently use the data structure of Guibas and Hershberger~\cite{ref:GuibasOp89,ref:HershbergerA91} for simple polygons, referred to as the \emph{GH data structure}.
It can be constructed in $O(m)$ time and supports computing $d(p, q)$ in $O(\log m)$ time for any two points $p, q \in P$.
It can also report the anchors of $p$ and $q$ on $\pi(p, q)$, as well as a compact representation of $\pi(p, q)$ that supports binary search along the path.
In addition, it can compute the junction vertex of $\pi(p, r)$ and $\pi(p, q)$ for any three points $p, q, r \in P$.
All these operations take $O(\log m)$ time.

Given a point $p \in P$ and a direction, a \emph{ray-shooting query} asks for the first point on $\partial P$ hit by the ray from $p$ in that direction.
Such queries can be supported in $O(\log m)$ time after $O(m)$ preprocessing~\cite{ref:ChazelleVi89,ref:ChazelleRa94,ref:HershbergerA95}.

\paragraph{Last-edge extensions.}

Given a path $\pi(p, q)$, we often extend its last edge beyond $q$ in the direction from $a_q$ to $q$, until reaching a point $q' \in \partial P$, where $a_q$ is the anchor of $q$ in $\pi(p, q)$.
We refer to the segment $\overline{qq'}$ as the \emph{last-edge extension} of $\pi(p, q)$ (see Figure~\ref{fig:last_edge_extension}).
Given $p$ and $q$, the last-edge extension of $\pi(p, q)$ can be computed in $O(\log m)$ time using the GH data structure together with the ray-shooting data structure.

\paragraph{Bisectors and geodesic Voronoi diagrams.}
For two sites $s_1, s_2 \in S$, let $B(s_1, s_2)$ denote their \emph{bisector}, i.e., the set of points equidistant from $s_1$ and $s_2$:
\[
B(s_1, s_2) = \{p \in P : d_{s_1}(p) = d_{s_2}(p)\}.
\]
Note that $B(s_1, s_2) = \emptyset$ if $d_{s_1}(s_2) < d_{s_2}(s_2)$ or $d_{s_2}(s_1) < d_{s_1}(s_1)$.
The following properties of $B(s_1, s_2)$ were proved by Aronov~\cite{ref:AronovOn89} for the unweighted case and extend to the weighted setting~\cite{ref:BrewerSh26}.

\begin{lemma}[Aronov~\cite{ref:AronovOn89}]\label{lem:bisector}
If $B(s_1, s_2)$ is nonempty and does not contain any vertex of $P$, then the following hold:
\begin{enumerate}
\item $B(s_1, s_2)$ is a smooth curve connecting two points on $\partial P$ and does not intersect $\partial P$ in its interior.
\item $B(s_1, s_2)$ consists of $O(m)$ hyperbolic arcs (where line segments are considered a special case), and each arc endpoint is the intersection of $B(s_1, s_2)$ with the last-edge extension of $\pi(s_1, p)$ or $\pi(s_2, p)$ for some vertex $p$ of $P$.
\end{enumerate}
\end{lemma}

For a site $s \in S$, let $\R(s)$ denote its \emph{Voronoi region}, i.e., the set of points at least as close to $s$ as to any other site in $S$:
\[
\R(s) = \{p \in P : d_s(p) = d_S(p)\}.
\]
Note that $\R(s)$ may be empty.
The collection of all Voronoi regions forms the \emph{additively weighted geodesic Voronoi diagram} of $S$, denoted by $\vd(S)$.

For two sites $s_1, s_2 \in S$, the common boundary $\E(s_1, s_2)$ of their Voronoi regions is the set of points equidistant from both and no farther from them than from any other site:
\[
\E(s_1, s_2) = \{p \in P : d_{s_1}(p) = d_{s_2}(p) = d_S(p)\}.
\]
Note that $\E(s_1, s_2) \subseteq B(s_1, s_2)$, and it may have $\Omega(n)$ connected components even in the Euclidean case~\cite{ref:SharirIn85}; each such component is called a \emph{Voronoi edge}.
The intersection of three Voronoi edges is a \emph{Voronoi vertex}, which is equidistant from at least three sites.
The intersection of a Voronoi edge with $\partial P$ is called a \emph{Voronoi terminal}.
By definition, $\R(s)$ is closed and includes its incident Voronoi edges.
A maximal connected portion of $\partial P$ on the boundary of $\R(s)$ is called a \emph{$\partial P$-portion} of $\R(s)$.
See Figure~\ref{fig:big_voronoi} for an illustration.

If $v$ is a Voronoi vertex or lies on a Voronoi edge (including the case where $v$ is a Voronoi terminal) and $s$ is a nearest site of $v$, we say that $s$ is \emph{associated with} $v$.
The following observation follows directly from the triangle inequality.

\begin{observation}\label{obser:nonempty}
For any set $S'$ of sites in $P$, the Voronoi region of a site $s$ is nonempty if and only if $s$ itself is in the Voronoi region.
\end{observation}

\section{Algorithm overview for constructing $\boldsymbol{\vd(S)}$}
\label{sec:high_level}

In this section, we give an overview of our algorithm for computing $\vd(S)$ implicitly.
For convenience, we assume that all site weights are positive; otherwise, we can add the same constant to all weights without affecting $\vd(S)$.

Our algorithm is inspired by Fortune's~\cite{ref:FortuneA87} sweep-line algorithm for Euclidean Voronoi diagrams.
In that algorithm (and its variants~\cite{ref:GuibasTh88,ref:deBergCo08}), a line $\ell$ sweeps the plane from top to bottom while maintaining the portion of the Voronoi diagram above $\ell$, treating $\ell$ as a special site.
The boundary of the Voronoi region of $\ell$, called the \emph{beach line}, is $x$-monotone and is maintained in a data structure that supports efficient updates as $\ell$ moves.

To extend this idea to our setting, we proceed as follows.
Pick any convex vertex of $P$, denoted by $v^*$, and assign it a weight $w(v^*) = t$, where $t$ is a parameter.
Let $\vd_t(S)$ be the Voronoi diagram of $S \cup \{v^*\}$ with respect to these weights.
We maintain $\vd_t(S)$ as $t$ increases from $-\infty$ to $\infty$.
In the unweighted case (i.e., when all sites have equal weight), this process is equivalent to sweeping a geodesic circle $C_t$ centered at $v^*$ with radius decreasing from $\infty$ to $0$, while maintaining the Voronoi diagram outside $C_t$ by treating $C_t$ as a special site.
In the weighted case, however, there is no direct geometric interpretation; instead, we conceptually sweep the parameter $t$.

For any site $s \in S$, let $\R_t(s)$ denote the Voronoi region of $s$ in $\vd_t(S)$ (while $\R(s)$ denotes its region in $\vd(S)$).
In particular, let $\R_t(v^*)$ denote the Voronoi region of $v^*$.
Define $\W_t$ as the boundary of $\R_t(v^*)$, which serves as the \emph{beach line} of our algorithm.

In general, $\W_t$ is a cycle consisting of Voronoi edges and $\partial P$-portions, and $v^* \in \W_t$.
We store $\W_t$ in a binary search tree ordered by the counterclockwise traversal of $\W_t$ starting and ending at $v^*$.
We represent $\vd_t(S)$ using a doubly-connected edge list (DCEL), where all vertex coordinates and edges are maintained implicitly.
Details of $\W_t$ and $\vd_t(S)$ are given in Section~\ref{sec:basics}.

\subsection{Defining events}
As $t$ increases from $-\infty$ to $\infty$, both $\W_t$ and $\vd_t(S)$ are updated at discrete events.
We distinguish five types of events, each of which corresponds to a value for $t$.

\begin{figure}
    \centering
    \includegraphics[width=.35\linewidth]{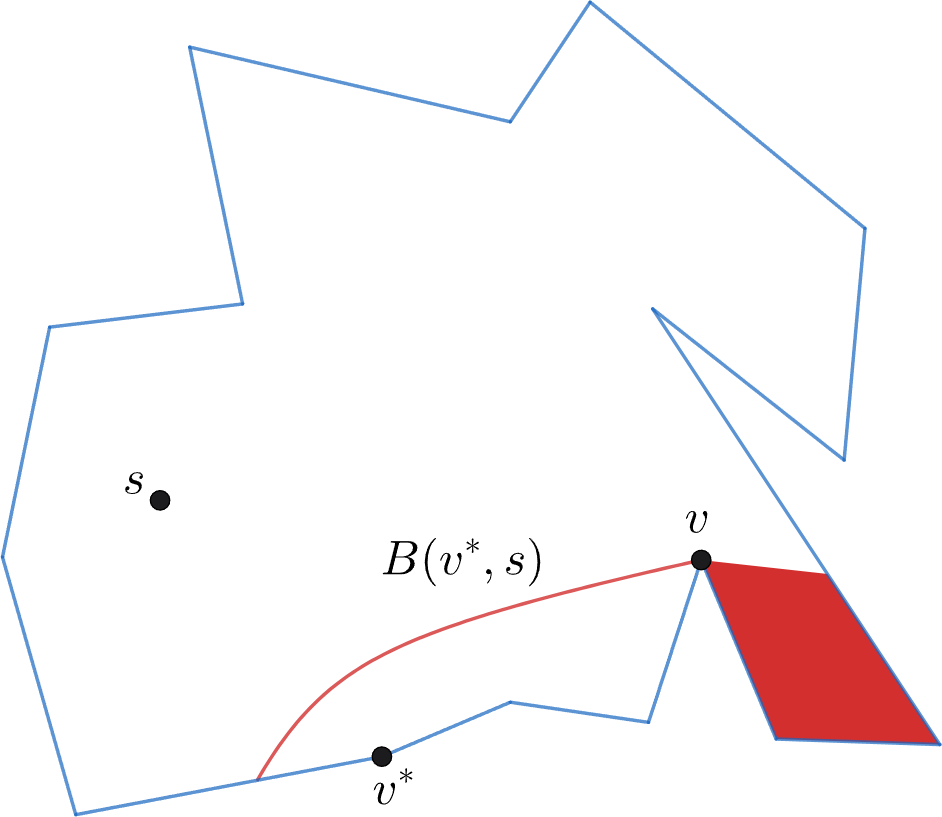}
    \caption{Pocket event with $v \in B(v^*, s)$. The red region is $pocket(B(v^*, s))$} \label{fig:bis_deg2}
\end{figure}

\begin{enumerate}
\item \emph{Site events.}
A site event occurs when $\R_t(s)$ changes from empty to nonempty for a site $s \in S$.
This happens when $d_s(s) = d(v^*, s) + t$, i.e., $t = d_s(s) - d(v^*, s) = w(s) - d(v^*, s)$.
We define $t_s = w(s) - d(v^*, s)$ as the site event of $s$.

Lemma~\ref{lem:events} shows that if $t < t_s$, then $\R_t(s) = \emptyset$; otherwise, $\R_t(s) \neq \emptyset$ if and only if $\R(s) \neq \emptyset$.
These correspond to the site events in Fortune's algorithm~\cite[Section~7.2]{ref:deBergCo08}.

\item \emph{Voronoi vertex events.}
Let $v$ be a Voronoi vertex of $\vd(S)$ and let $s \in S$ be a site associated with $v$.
A Voronoi vertex event occurs when $v$ first appears as a Voronoi vertex in $\vd_t(S)$, i.e., when $d_s(v) = d(v^*, v) + t$.
Thus, we define $t_v = d_s(v) - d(v^*, v)$ as the Voronoi vertex event of $v$.

Lemma~\ref{lem:events} shows that $v$ is a Voronoi vertex of $\vd_t(S)$ if and only if $t \ge t_v$.
These correspond to circle events in Fortune's algorithm~\cite[Section~7.2]{ref:deBergCo08}.

\item \emph{Voronoi terminal events.}
Let $u$ be a Voronoi terminal of $\vd(S)$ and let $s \in S$ be a site associated with $u$.
A Voronoi terminal event occurs when $u$ first becomes a Voronoi terminal in $\vd_t(S)$, i.e., when $d_s(u) = d(v^*, u) + t$.
We define $t_u = d_s(u) - d(v^*, u)$ as the Voronoi terminal event of $u$.

Lemma~\ref{lem:events} shows that $u$ is a Voronoi terminal of $\vd_t(S)$ if and only if $t \ge t_u$.

\item \emph{Pocket events.}
Let $v$ be a vertex of $P$ and $s \in S$ a site.
Suppose that for some value $t$, the bisector $B(s, v^*)$ contains $v$ and bounds a nonempty region $R$ such that for all $p \in R$, we have $d_s(p) = d_{v^*}(p)$ and both $\pi(p,s)$ and $\pi(p,v^*)$ contain $v$ (see Figure~\ref{fig:bis_deg2}).
This occurs when $d_s(v) = d(v^*, v) + t$, i.e., $t = d_s(v) - d(v^*, v)$.
We define $t_{sv} = d_s(v) - d(v^*, v)$ as the pocket event of $(s, v)$.

We call $R$ the \emph{pocket} of $B(v^*, s)$ and denote it by $pocket(B(v^*, s))$; the vertex $v$ is called its \emph{apex}.

Following this definition, for any vertex $v'$ of $P$ inside $R$ and $v' \neq v$, we have $d_s(v') = d(v^*, v') + t$ for $t = t_{sv}$ (which is also equal to $t_{sv'}$ following the same definition as for $t_{sv}$). However, we consider this event to be for $v$ but not for $v'$. That is, a pocket event is only for the apex of the pocket but not for any other vertex in the pocket.

By Lemma~\ref{lem:sweeping}, if $t_{sv}$ does not coincide with any other event, then $\W_t$ and $\vd_t(S)$ do not change combinatorially at $t = t_{sv}$, and the event can be ignored.
Under our general position assumptions (to be discussed later), pocket events can only coincide with Voronoi vertex or terminal events.
Accordingly, when processing a Voronoi vertex or terminal event, we will first check if it coincides with a pocket event. If so, the pocket event will be processed, and the Voronoi vertex or terminal event will be processed simultaneously.

\item \emph{Final event.}
At some point, the influence of $v^*$ disappears, i.e., $\R_t(v^*)$ becomes empty and $\vd_t(S) = \vd(S)$.
This occurs when $d_{v^*}(v^*) = t = \min_{s \in S} d_s(v^*)$.
We define $t_{v^*} = \min_{s \in S} d_s(v^*)$ as the final event.

Lemma~\ref{lem:events} shows that $\R_t(v^*) \neq \emptyset$ if and only if $t \le t_{v^*}$.
This corresponds to the point at which the event queue becomes empty in Fortune's algorithm~\cite[Section~7.2]{ref:deBergCo08}.
\end{enumerate}

The following lemma justifies the definitions of the events.

\begin{lemma} \label{lem:events}
\begin{enumerate}
\item Let $t_s$ be the site event of a site $s \in S$.
If $t < t_s$, then $\R_t(s) = \emptyset$; otherwise, $\R_t(s) \neq \emptyset$ if and only if $\R(s) \neq \emptyset$.

\item Let $t_v$ be the Voronoi vertex event of a Voronoi vertex $v$.
If $t < t_v$, then $v$ is not a Voronoi vertex of $\vd_t(S)$; otherwise, $v$ is a Voronoi vertex of $\vd_t(S)$.

\item Let $t_u$ be the Voronoi terminal event of a Voronoi terminal $u$.
If $t < t_u$, then $u$ is not a Voronoi terminal of $\vd_t(S)$; otherwise, $u$ is a Voronoi terminal of $\vd_t(S)$.

\item Let $t_{v^*}$ be the final event time.
If $t \le t_{v^*}$, then $\R_t(v^*) \neq \emptyset$; otherwise, $\R_t(v^*) = \emptyset$.
\end{enumerate}
\end{lemma}

\begin{proof}
\leavevmode
\begin{enumerate}
\item If $t < t_s$, then
$d_{v^*}(s) = d(v^*, s) + t < d(v^*, s) + t_s = d_s(s)$.
Thus, $s \notin \R_t(s)$, and by Observation~\ref{obser:nonempty}, $\R_t(s) = \emptyset$.

Now assume that $t \ge t_s$.
By Observation~\ref{obser:nonempty}, $\R_t(s) \neq \emptyset$ if and only if $s \in \R_t(s)$, and $\R(s) \neq \emptyset$ if and only if $s \in \R(s)$.
It therefore suffices to show that $s \in \R_t(s)$ if and only if $s \in \R(s)$.

If $s \notin \R(s)$, then clearly $s \notin \R_t(s)$.
On the other hand, if $s \in \R(s)$, then $d_s(s) \le d_{s'}(s)$ for all $s' \in S$.
Since $t \ge t_s$, we have
$d_{v^*}(s) = d(v^*, s) + t \ge d(v^*, s) + t_s = d_s(s)$.
Thus, $d_s(s) \le d_{s'}(s)$ for all $s' \in S \cup \{v^*\}$, implying $s \in \R_t(s)$.

\item Let $s \in S$ be a site associated with the Voronoi vertex $v$.

If $t < t_v$, then
$d_{v^*}(v) = d(v^*, v) + t < d(v^*, v) + t_v = d_s(v)$.
Since $v$ is a Voronoi vertex of $\vd(S)$, we have $d_s(v) \le d_{s'}(v)$ for all $s' \in S$.
Hence $d_{v^*}(v) < d_{s'}(v)$ for all $s' \in S$, which implies that $v$ lies in the interior of $\R_t(v^*)$ and therefore is not a Voronoi vertex of $\vd_t(S)$.

If $t \ge t_v$, then
$d_{v^*}(v) = d(v^*, v) + t \ge d(v^*, v) + t_v = d_s(v)$.
Thus, the sites associated with $v$ in $\vd(S)$ remain the nearest sites to $v$ in $\vd_t(S)$, and hence $v$ is a Voronoi vertex of $\vd_t(S)$.

\item The argument for Voronoi terminal events is analogous to that for Voronoi vertex events and is omitted.

\item If $t \le t_{v^*}$, then
$d_{v^*}(v^*) = t \le t_{v^*} = \min_{s \in S} d_s(v^*)$.
Hence $v^* \in \R_t(v^*)$, and thus $\R_t(v^*) \neq \emptyset$.

If $t > t_{v^*}$, then
$d_{v^*}(v^*) = t > t_{v^*} = \min_{s \in S} d_s(v^*)$.
Thus $v^* \notin \R_t(v^*)$, and by Observation~\ref{obser:nonempty}, $\R_t(v^*) = \emptyset$.
\end{enumerate}

This completes the proof.
\end{proof}

The following lemma, proved in Section~\ref{sec:correctness}, shows that it suffices to handle events and thus establishes the correctness of our algorithm.

\begin{restatable}{lemma}{sweeping} \label{lem:sweeping}
As $t$ varies from $-\infty$ to $\infty$, the combinatorial structures of $\W_t$ and $\vd_t(S)$ remain unchanged except at events.
If $t_{sv}$ is a pocket event that does not coincide with any other event, then the combinatorial structures of $\W_t$ and $\vd_t(S)$ do not change at $t = t_{sv}$.

More specifically, as $t$ changes, each vertex $v$ of $\W_t$ traces along a Voronoi edge or a $\partial P$-portion of $\vd(S)$, moving monotonically from the point where it is created by an event to the point where it is removed by another event.
During this motion, $v$ moves continuously except at pocket events, where it may make a discontinuous jump, but always in the same direction as its prior motion.
\end{restatable}

We describe how to process events (i.e., how to update $\W_t$ and $\vd_t(S)$) in Sections~\ref{sec:site_events} and~\ref{sec:nonsite_events}.
At the beginning of the algorithm, we can compute the final event and all site events.
Although all pocket events could also be computed, doing so would take $\Omega(nm)$ time, as there are $\Omega(nm)$ pairs $(s, v)$.
Instead, by Lemma~\ref{lem:sweeping}, we handle a pocket event only when it coincides with another event.

We cannot compute all Voronoi vertex or terminal events in advance, since the Voronoi vertices and terminals of $\vd(S)$ are not known a priori.
Therefore, our event queue maintains \emph{candidate} Voronoi vertex and terminal events.
Any candidate event that is not a true event will eventually be removed from the queue before being processed.

\subsection{General position assumptions}
\label{sec:genpos}
We make several general position assumptions to simplify the exposition.
These assumptions can be removed without affecting the results, e.g., by standard perturbation techniques~\cite{ref:EdelsbrunnerSi90,ref:YapA90} or by slightly perturbing the locations of the sites or vertices of $P$ (similar assumptions were also used in the previous work~\cite{ref:AronovOn89,ref:AronovTh93,ref:OhOp19,ref:OhVo20}).

First, for any two sites $s_1, s_2 \in S$, we assume that
(1) $d_{s_1}(s_1) \neq d_{s_2}(s_1)$ and $d_{s_1}(s_2) \neq d_{s_2}(s_2)$, and
(2) for any vertex $v$ of $P$, $d_{s_1}(v) \neq d_{s_2}(v)$.
The second condition ensures that the bisector $B(s_1, s_2)$ does not pass through any vertex of $P$, and hence is a smooth curve connecting two points on $\partial P$ by Lemma~\ref{lem:bisector}.
These assumptions apply only to sites in $S$ and not to $v^*$.
Indeed, the case $d_{v^*}(s) = d_s(s)$ for $s \in S$ corresponds to the site event $t_s$, and the case $d_{v^*}(v) = d_s(v)$ for $v \in V_P$ and $s \in S$ corresponds to the pocket event $t_{sv}$.

We further assume that no point in $P$ is equidistant (in weighted distance) from four sites in $S$, ensuring that each Voronoi vertex of $\vd(S)$ is associated with exactly three sites.
We also assume that no point on $\partial P$ is equidistant from three sites in $S$, so that each Voronoi terminal is associated with exactly two sites.
Finally, we assume that no three points in $S \cup V_P$ are collinear; in particular, for any three sites $s_1, s_2, s_3 \in S \cup \{v^*\}$, $s_2 \notin \pi(s_1, s_3)$.

To simplify the treatment of pocket events, we assume that for any $s \in S$ and $v \in V_P$, when $t = t_{sv}$, the region $pocket(B(v^*, s))$ has at most one connected component.
(Otherwise, it may consist of two components on opposite sides of $\pi(v^*, s)$, each with its own apex.)

Finally, we make assumptions on event timing.
We partition Voronoi vertex events into \emph{pocket} and \emph{non-pocket} types: a Voronoi vertex event $t_v$ is a pocket Voronoi vertex event if $v$ lies in some $pocket(B(v^*, s))$ for a site $s$ associated with $v$, and a non-pocket event otherwise.
Voronoi terminal events are partitioned analogously.
We assume that no two events among site events, non-pocket Voronoi vertex and terminal events, pocket events, and the final event occur at the same time.
Under this assumption, a pocket event can coincide only with a pocket Voronoi vertex event or a pocket Voronoi terminal event.

\subsection{Global preprocessing for $\boldsymbol{P}$}
\label{sec:global}
Before running our algorithm to construct $\vd(S)$, we assume that the following preprocessing has been performed on $P$ before $S$ is given.
We refer to this as \emph{global preprocessing}, which takes $O(m)$ time in total.

First, we construct the GH data structure and a ray-shooting data structure as described in Section~\ref{sec:pre}.
Second, we select a convex vertex $v^*$ of $P$ and compute a list of the edges of $P$ sorted in counterclockwise order, starting from $v^*$, and assign each edge an index according to this order.
Third, we build a point location data structure on $\partial P$ so that, given a point $p \in \partial P$, we can determine in $O(\log m)$ time the edge of $P$ that contains $p$.
We refer to this as the \emph{polygon boundary point location data structure}.

All three steps can be carried out in $O(m)$ time.
In particular, the third step can be implemented by first triangulating $P$~\cite{ref:ChazelleTr91} and then constructing a point location data structure on the triangulation~\cite{ref:KirkpatrickOp83,ref:EdelsbrunnerOp86}.
Although the GH data structure already supports such queries, we state this explicitly for clarity.

With this global preprocessing, many operations on $S$ can be performed in time polylogarithmic in $m$.
In the remainder of the paper, we use these data structures without further mention.

\section{Geometric primitives and sweeping status} \label{sec:basics}

In Section~\ref{sec:bisectors}, we establish several properties of bisectors and describe operations on them that are used in our algorithm.
In Section~\ref{sec:struct}, we discuss the combinatorial structures of $\vd_t(S)$ and $\W_t$, and describe the data structures that maintain them, which together constitute the sweeping status.

\subsection{Bisector properties and operations} \label{sec:bisectors}

For convenience, we call a value of $t$ \emph{general} if it is neither a site event nor a pocket event.
We begin with the following observation, due to Aronov~\cite{ref:AronovOn89}.

\begin{observation}[{Aronov~\cite{ref:AronovOn89}}]\label{obs:star_shaped}
Voronoi regions are star-shaped with respect to their sites.
That is, for any $s \in S$ (or $s = v^*$ for a general value of $t$) and any $p \in \R_t(s)$, we have $\pi(s, p) \subseteq \R_t(s)$ and $\pi(s, p) \setminus \{p\}$ does not intersect any Voronoi edge of $\R_t(s)$.
For any two sites $s_1, s_2 \in S$ (or in $S \cup \{v^*\}$ for a general value of $t$) and any point $p \in P$, the intersection $\pi(s_1, p) \cap B(s_1, s_2)$ contains at most one point.
\end{observation}

\begin{proof}
Aronov~\cite{ref:AronovOn89} proved the statement for the unweighted case.
The argument extends to the additively weighted setting as discussed in~\cite[Lemma~29]{ref:BrewerSh26}.
It relies on general position assumptions that $d_{s_1}(s_2) \neq d_{s_2}(s_2)$, $d_{s_1}(s_1) \neq d_{s_2}(s_1)$, and that no vertex $v$ of $P$ satisfies $d_{s_1}(v) = d_{s_2}(v)$.
For sites in $S$, these are included in our general position assumptions.
When $v^*$ is involved, degeneracies occur precisely at site events and pocket events.
\end{proof}

For any two sites $s_1, s_2 \in S$, Observation~\ref{obs:star_shaped} implies that $\pi(s_1, s_2)$ intersects $B(s_1, s_2)$ exactly once.
This intersection partitions $B(s_1, s_2)$ into two curves, called \emph{half-bisectors}.
One lies to the right of $\pi(s_1, s_2)$ and is called the \emph{right half-bisector}, denoted $B_r(s_1, s_2)$; the other is the \emph{left half-bisector}, denoted $B_l(s_1, s_2)$.
Similarly, for any site $s \in S$, the bisector $B(s, v^*)$ also consists of two half-bisectors for a general value of $t$.
Each half-bisector intersects $\partial P$ exactly once.

With the global preprocessing for $P$, we have the following result.

\begin{lemma}[{\cite{ref:OhVo20,ref:BrewerSh26}}]\label{lem:bpp}
For any $s_1, s_2 \in S$ (or in $S \cup \{v^*\}$ for a general value of $t$), we can compute
$B_l(s_1, s_2) \cap \partial P$ and $B_r(s_1, s_2) \cap \partial P$ in $O(\log^2 m)$ time.
\end{lemma}
\begin{proof}
Oh and Ahn~\cite[Lemma~5.7]{ref:OhVo20} gave an algorithm for the unweighted case assuming the GH and ray-shooting data structures have been constructed. The algorithm was extended to the weighted case by Brewer and Wang~\cite[Lemmas~31 and~34]{ref:BrewerSh26}.
We briefly discuss the idea below.

Let $\seg{s_2 s_2'}$ be the last-edge extension of $\pi(s_1, s_2)$. Similarly, let $\seg{s_1 s_1'}$ be the last-edge extension of $\pi(s_2, s_1)$. Let $\gamma$ be the portion of $\partial P$ from $s_1'$ to $s_2'$ counterclockwise. Thus, the intersection $q = B_r(s_1, s_2) \cap \partial P$ lies in $\gamma$. Hence, we search for $q$ on $\gamma$ using binary search. In each iteration, we pick the middle vertex $v$ in the remaining portion of $\gamma$ and compare $d_{s_1}(v)$ with $d_{s_2}(v)$ (by computing their values in $O(\log m)$ time using the GH data structure). If $d_{s_1}(v) < d_{s_2}(v)$, then we search $\gamma$ toward the $s_2'$ side; otherwise, we search toward the $s_1'$ side. Since each iteration takes $O(\log m)$ time and there are $O(\log m)$ iterations in the binary search, the above algorithm will eventually find an edge $e_q$ of $\partial P$ that contains $q$ in $O(\log^2 m)$ time. After that, $q$ can be found on $e_q$ in an additional $O(\log^2 m)$ time.

Similarly, $B_l(s_1, s_2) \cap \partial P$ can be computed in $O(\log^2 m)$ time.

\paragraph{Remark.}
This procedure is a key bottleneck of our entire algorithm. It seems challenging to improve the time to $O(\log m)$. More specifically, the challenge is to compute $e_q$ in $O(\log m)$ time. Once $e_q$ is found, the last step of finding $q$ on $e_q$ may be solved in $O(\log m)$ time. For this, Lemma~\ref{lem:gendiagonalintersection} in Section~\ref{sec:diagonalintersection} gives an $O(\log m)$-time algorithm for the unweighted case. It should be possible to extend it to the weighted case by combining the weighted techniques from~\cite[Lemmas~31 and~34]{ref:BrewerSh26}. However, computing the edge $e_q$ in $O(\log m)$ time poses a tremendous challenge. Indeed, we feel that a nested binary search might be necessary, implying that $O(\log^2 m)$ time might be the best possible.
\end{proof}

For any three sites $s_1, s_2, s_3 \in S$, it is known that in the unweighted case, there is at most one point in $P$ equidistant from all three sites; equivalently, the bisectors $B(s_1, s_2)$ and $B(s_2, s_3)$ intersect at most once~\cite{ref:AronovTh93}.
This property no longer holds in the weighted case: the two bisectors may intersect twice even in the Euclidean plane~\cite{ref:SharirIn85}.
Our algorithm requires an efficient subroutine for computing such intersections.
To this end, we first prove the following lemma.

\begin{lemma}\label{lem:halfintersect}
For any three sites $s_1, s_2, s_3 \in S$ (or in $S \cup \{v^*\}$ for a general value of $t$), each intersection point $v \in B(s_1, s_2) \cap B(s_2, s_3)$ is the unique intersection of two half-bisectors of $B(s_1, s_2)$ and $B(s_2, s_3)$.
\end{lemma}

\begin{proof}
Let $S' = \{s_1, s_2, s_3\}$. For each $s \in S'$, let $\R(s, S')$ denote the Voronoi region of $s$ in $\vd(S')$.

If $B(s_1, s_2) \cap B(s_2, s_3)$ has a single intersection point, then the statement holds trivially.
Otherwise, the Voronoi region of one site, say $s_2$, in $\vd(S')$ is closed, in the sense that its boundary does not intersect $\partial P$.
In this case, the two intersection points of $B(s_1, s_3)$ with $\partial P$ are Voronoi terminals of $\vd(S')$.
Let $u_1$ and $u_2$ be these two points.
There is a Voronoi edge connecting $u_1$ to a Voronoi vertex $v_1$ (see Figure~\ref{fig:halfbisector}), and similarly a Voronoi edge connecting $u_2$ to a Voronoi vertex $v_2$.
The boundary of $\R(s_2, S')$ consists of two Voronoi edges with endpoints $v_1$ and $v_2$.
One of them, denoted by $e_1$, lies on $B(s_1, s_2)$ and separates $\R(s_1, S')$ from $\R(s_2, S')$; the other, denoted by $e_2$, lies on $B(s_2, s_3)$ and separates $\R(s_2, S')$ from $\R(s_3, S')$.

\begin{figure}
    \centering
    \includegraphics[height=2.0in]{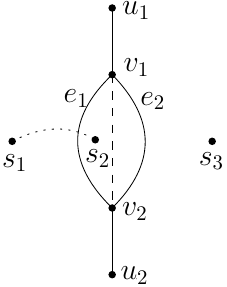}
    \caption{Illustrating the proof of Lemma~\ref{lem:halfintersect}. The dashed segment represents $B'$.} \label{fig:halfbisector}
\end{figure}

The bisector $B(s_1, s_3)$ partitions $P$ into two subpolygons: one, denoted by $P_1$, containing $s_1$, and the other, denoted by $P_3$, containing $s_3$.
Without loss of generality, assume that $s_2 \in P_1$ (including the case where $s_2 \in B(s_1, s_3)$).
By Observation~\ref{obs:star_shaped}, $\pi(s_1, s_2) \setminus \{s_2\}$ does not intersect $B(s_1, s_3)$.

Let $B'$ denote the portion of $B(s_1, s_3)$ between $v_1$ and $v_2$, excluding the endpoints.
By definition, $B'$ lies in the interior of $\R(s_2, S')$ and partitions $\R(s_2, S')$ into two subregions.
Let $R_1$ be the subregion contained in $P_1$.
Since $s_2 \in P_1$, we have $s_2 \in R_1$, and $R_1$ is bounded by $B'$ and $e_1$.
As $s_1 \notin \R(s_2, S')$, we have $s_1 \notin R_1$.
Therefore, $\pi(s_1, s_2)$ must intersect $\partial R_1$, and hence must intersect either $e_1$ or $B'$.

We claim that $\pi(s_1, s_2)$ must intersect $e_1$.
We consider two cases.

\begin{itemize}
\item If $s_2 \notin B'$, then $s_2$ lies in the interior of $R_1$ (since $s_2 \notin e_1$ by the general position assumption).
Thus, $\pi(s_1, s_2) \setminus \{s_2\}$ must intersect either $e_1$ or $B'$.
Since it does not intersect $B(s_1, s_3)$, it cannot intersect $B'$, and therefore must intersect $e_1$.

\item If $s_2 \in B'$, suppose, for contradiction, that $\pi(s_1, s_2)$ does not intersect $e_1$.
Since $s_2$ lies in the interior of $\R(s_2, S')$, which is bounded by $e_1$ and $e_2$, the path $\pi(s_1, s_2) \setminus \{s_2\}$ must intersect $e_2$ and also intersect the portion of $B(s_1, s_3)$ between $u_i$ and $v_i$ for some $i \in \{1,2\}$.
Hence, $\pi(s_1, s_2)$ intersects $B(s_1, s_3)$ at least twice, contradicting Observation~\ref{obs:star_shaped}.
\end{itemize}

This proves that $\pi(s_1, s_2)$ intersects $e_1$.

Since $v_1$ and $v_2$ are the endpoints of $e_1 \subseteq B(s_1, s_2)$, the above claim implies that they lie on different half-bisectors of $B(s_1, s_2)$.
Without loss of generality, assume that $v_1 \in B_l(s_1, s_2)$ and $v_2 \in B_r(s_1, s_2)$.

Now consider the bisector $B(s_2, s_3)$. Since $B_l(s_1, s_2)$ contains $v_1$ but does not contain $v_2$, $v_1$ is the unique intersection between $B_l(s_1, s_2)$ and $B(s_2, s_3)$, which further implies that $v_1$ is the unique intersection between $B_l(s_1, s_2)$ and the half-bisector of $B(s_2, s_3)$ containing $v_1$. By similar reasoning, $v_2$ is the unique intersection between $B_r(s_1, s_2)$ and the half-bisector of $B(s_2, s_3)$ containing $v_2$. This completes the proof.
\end{proof}

Based on Lemma~\ref{lem:halfintersect}, the following lemma gives an algorithm for computing the intersections of two bisectors among three sites. The algorithm extends that of Wang~\cite{ref:WangAn23DCG} for the unweighted case.

\begin{lemma} \label{lem:bis_branch_int}
For any three sites $s_1, s_2, s_3 \in S$ (or in $S \cup \{v^*\}$ for a general value of $t$), given $B(s_1, s_2) \cap \partial P$,
we can compute $B(s_1, s_2) \cap B(s_2, s_3)$ in $O(\log m)$ time.
\end{lemma}
\begin{proof}
Consider a half-bisector of $B(s_1, s_2)$, say $B_r(s_1, s_2)$, and a half-bisector of $B(s_2, s_3)$, say $B_r(s_2, s_3)$, and suppose they have a unique intersection point $\alpha$.
To compute $\alpha$, we can adapt Wang's algorithm~\cite[Lemma~3.1]{ref:WangAn23DCG} for the unweighted case.

The algorithm requires a special point $p_1 \in B_r(s_1, s_2)$ such that the portion of $B_r(s_1, s_2)$ between $p$ and $p_1$ contains $\alpha$, where $p$ is the endpoint of $B_r(s_1, s_2)$ that is not on $\partial P$.
For our problem, we can take $p_1$ to be the intersection point $B_r(s_1, s_2) \cap \partial P$.
Consequently, we can compute $\alpha$ by applying Wang's algorithm using weighted distances.
The algorithm, based on the tentative prune-and-search technique of Kirkpatrick and Snoeyink~\cite{ref:KirkpatrickTe95}, runs in $O(\log m)$ time assuming the GH data structure is available.

To handle the case where the two half-bisectors do not intersect, we add a verification step: we run the algorithm to obtain a candidate point $\alpha$, and then check whether $\alpha$ is equidistant from the three sites. If not, we discard $\alpha$.

Finally, we consider all $O(1)$ pairs of half-bisectors of $B(s_1, s_2)$ and $B(s_2, s_3)$.
For each pair, we apply the above procedure.
By Lemma~\ref{lem:halfintersect}, this suffices to compute all intersection points in $B(s_1, s_2) \cap B(s_2, s_3)$ in $O(\log m)$ total time.
\end{proof}

Combining Lemmas~\ref{lem:bpp} and~\ref{lem:bis_branch_int}, we obtain the following subroutine, which will be used frequently in our algorithm.

\begin{lemma} \label{lem:vc_compute}
Let $v$ be a Voronoi vertex or Voronoi terminal of $\vd(S)$ (or $\vd_t(S)$ for a general value of $t$).
Suppose that the cyclic order of the incident edges around $v$ and the sites associated with those edges are known.
Then the coordinates of $v$ can be computed in $O(\log^2 m)$ time.
\end{lemma}

\begin{proof}
If $v$ is a Voronoi terminal, then exactly one of its three incident edges is a Voronoi edge; let $s_1$ and $s_2$ be the two sites associated with that edge.
Then $v$ is either $B_l(s_1, s_2) \cap \partial P$ or $B_r(s_1, s_2) \cap \partial P$, which can be determined from the cyclic order of incident edges.
In either case, $v$ can be computed in $O(\log^2 m)$ time using Lemma~\ref{lem:bpp}.

If $v$ is a Voronoi vertex, then it is incident to three Voronoi edges and associated with three sites $s_1, s_2, s_3$.
From the cyclic order of edges around $v$, we obtain the cyclic order of these sites; assume without loss of generality that they are in counterclockwise order.

We first compute $B(s_1, s_2) \cap B(s_2, s_3)$, which contains at most two points, in $O(\log^2 m)$ time using Lemmas~\ref{lem:bpp} and~\ref{lem:bis_branch_int}.
For each candidate intersection point $u$, we compute the anchors $u_i$ of $u$ on $\pi(u, s_i)$ for $i = 1, 2, 3$ in $O(\log m)$ time using the GH data structure.
If $u_1, u_2, u_3$ appear in counterclockwise order around $u$, then $u$ is the desired vertex $v$.

Thus, $v$ can be computed in $O(\log^2 m)$ time.
\end{proof}

The following lemma provides another operation required in our algorithm.

\begin{lemma} \label{lem:x_voronoi}
For any two sites $s, s' \in S$, except during the site event $t_{s'}$ or a pocket event associated with $s$ or $s'$, we can compute $B(v^*, s') \cap \seg{s x}$ in $O(\log m)$ time, where $\seg{s x}$ is the last-edge extension of $\pi(v^*, s)$.
\end{lemma}

\begin{proof}
As discussed in Section~\ref{sec:pre}, the segment $\seg{s x}$ can be computed in $O(\log m)$ time using the GH data structure and ray shooting.
Note that $\pi(v^*, x) = \pi(v^*, s) \cup \overline{s x}$.

By Observation~\ref{obs:star_shaped}, $B(v^*, s')$ intersects $\seg{s x}$ in at most one point, since $\seg{s x} \subseteq \pi(v^*, x)$.
If $d_{s'}(s) < d_{v^*}(s)$ or $d_{v^*}(x) < d_{s'}(x)$ (which can be checked in $O(\log m)$ time using the GH data structure), then $B(v^*, s') \cap \seg{s x} = \emptyset$.
Otherwise, let $q = B(v^*, s') \cap \seg{s x}$. To compute $q$, it suffices to find the anchor $a_q$ of $q$ in $\pi(s', q)$ (see Figure~\ref{fig:x_voronoi}).

Using the GH data structure, we first compute the junction vertex $b$ of $\pi(s', s)$ and $\pi(s', x)$.
Let $F$ be the concatenation of $\pi(s, b)$ and $\pi(b, x)$; both are concave chains.
Then $a_q$ lies on $F$.
We perform a binary search on the vertices of $F$ to locate $a_q$.

Specifically, for a vertex $v$ of $F$, we construct a line through $v$ tangent to the chain of $F$ containing $v$ (if $v = b$, the line is tangent to both chains).
This line intersects $\overline{s x}$ at a point $v'$.
Then $\pi(s', v') = \pi(s', v) \cup \overline{v v'}$, and hence $d_{s'}(v') = d_{s'}(v) + |\overline{v v'}|$.

If $d_{v^*}(v') = d_{s'}(v')$, then $a_q = v$ (and $q = v'$) and we are done.
If $d_{v^*}(v') < d_{s'}(v')$, we continue the search on the portion of $F$ toward $x$; otherwise, we search toward $s$.

\begin{figure}
    \centering
    \includegraphics{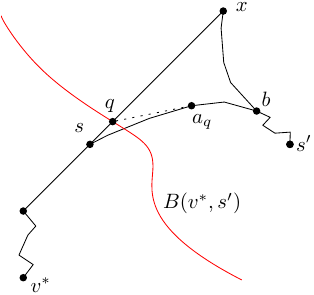}
    \caption{Illustration for the proof of Lemma~\ref{lem:x_voronoi}.}
    \label{fig:x_voronoi}
\end{figure}

Note that both $d_{v^*}(v')$ and $d_{s'}(v')$ can be computed in $O(1)$ time.
Indeed, $d_{v^*}(v') = d_{v^*}(s) + |\overline{s v'}|$, and $d_{v^*}(s)$ can be precomputed.
Similarly, $d_{s'}(v') = d_{s'}(v) + |\overline{v v'}|$, and the value $d_{s'}(v)$ is available during the binary search using the GH data structure.

Thus, the binary search takes $O(\log m)$ time, completing the proof.
\end{proof}

\subsection{Structures and representations of $\boldsymbol{\vd_t(S)}$ and $\boldsymbol{\W_t}$} \label{sec:struct}

We describe the structures and representations used for the beach line $\W_t$ and the Voronoi diagram $\vd_t(S)$.

\begin{figure}
    \centering
    \begin{subfigure}[b]{\textwidth}
        \centering
        \includegraphics[width=0.3\linewidth]{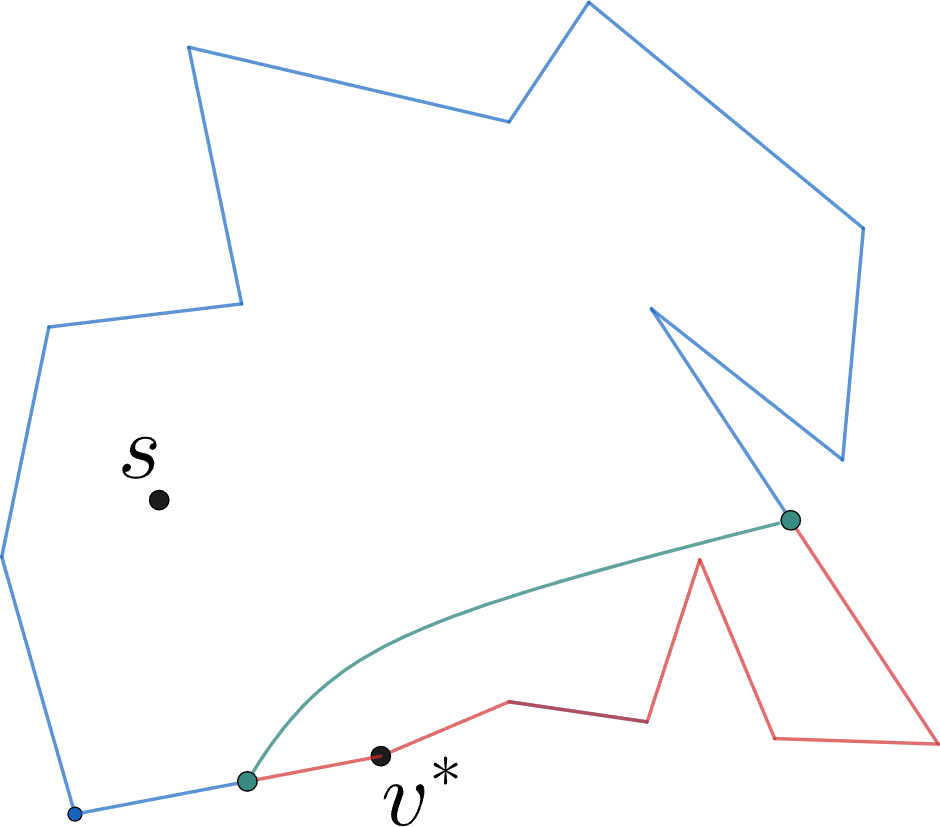}
    \end{subfigure}
    \caption{Illustration of $\W_t\sa$ (red boundary portions) and $\W_t\ca$ (green curve).}
    \label{fig:ellsaellca}
\end{figure}

The beach line $\W_t$ is the boundary of $\R_t(v^*)$ in $\vd_t(S)$, and it consists of Voronoi edges and $\partial P$-portions.
Define $\W_t\sa = \W_t \cap \partial P$, and let $\W_t\ca$ be the closure of $\W_t \setminus \W_t\sa$.
In other words, $\W_t\sa$ consists of the points of $\W_t$ lying on $\partial P$, while $\W_t\ca$ consists of the points of $\W_t$ lying on Voronoi edges of $\vd_t(S)$; see Figure~\ref{fig:ellsaellca}.

In Fortune's algorithm~\cite[Section~7.2]{ref:deBergCo08}, the beach line is $x$-monotone due to the horizontal sweep line.
This monotonicity is crucial to the efficiency of the algorithm.
In our setting, we generalize this idea by defining an order on the beach line $\W_t$.
We will also introduce two additional orders that will be used later; these three orders are mutually consistent.

\subsubsection{Three orders}

We introduce a total order $\lw$ on $\W_t$, a partial order $\la$ on the points of $P$, and a total order $\lp$ on $\partial P$.

\paragraph{A total order $\boldsymbol{\lw}$.}
For a general value of $t$, Voronoi regions are star-shaped and their boundaries consist of hyperbolic arcs~\cite{ref:AronovOn89,ref:BrewerSh26}.
Since $v^*$ is a vertex of $P$, we have $v^* \in \partial \R_t(v^*) = \W_t$.
Thus, we can view $\W_t$ as a closed curve traversed counterclockwise starting and ending at $v^*$.

For any two points $p, q \in \W_t$, we define $p \lw q$ if $p$ appears before $q$ along this traversal.
In particular, we define $p \lw v^*$ for any $p \in \W_t \setminus \{v^*\}$, i.e., $v^*$ is treated as the last point.
This defines a total order on $\W_t$.

\paragraph{A partial order $\boldsymbol{\la}$.}
Consider two distinct points $p$ and $q$ in $P$.
If $p \in \pi(v^*, q)$ or $q \in \pi(v^*, p)$, then $p$ and $q$ are not $\la$-comparable.
Otherwise, they are $\la$-comparable, and we define their order as follows.

Let $b$ be the junction vertex of $\pi(v^*, p)$ and $\pi(v^*, q)$.
We distinguish two cases.

\begin{figure}
        \centering
        \begin{subfigure}[b]{.45\textwidth}
            \centering
            \includegraphics[width=0.6\linewidth]{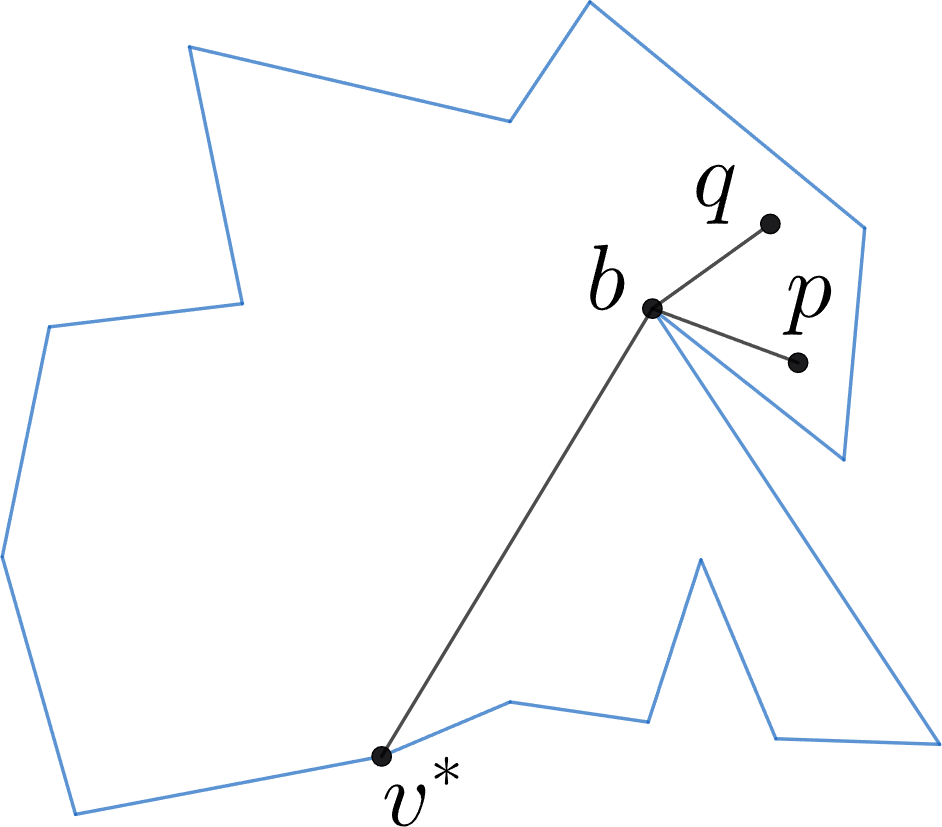}
            \subcaption{The case $b \neq v^*$.}
            \label{fig:leqvs}
        \end{subfigure}
        \hfill
        \begin{subfigure}[b]{.45\textwidth}
            \centering
            \includegraphics[width=0.6\linewidth]{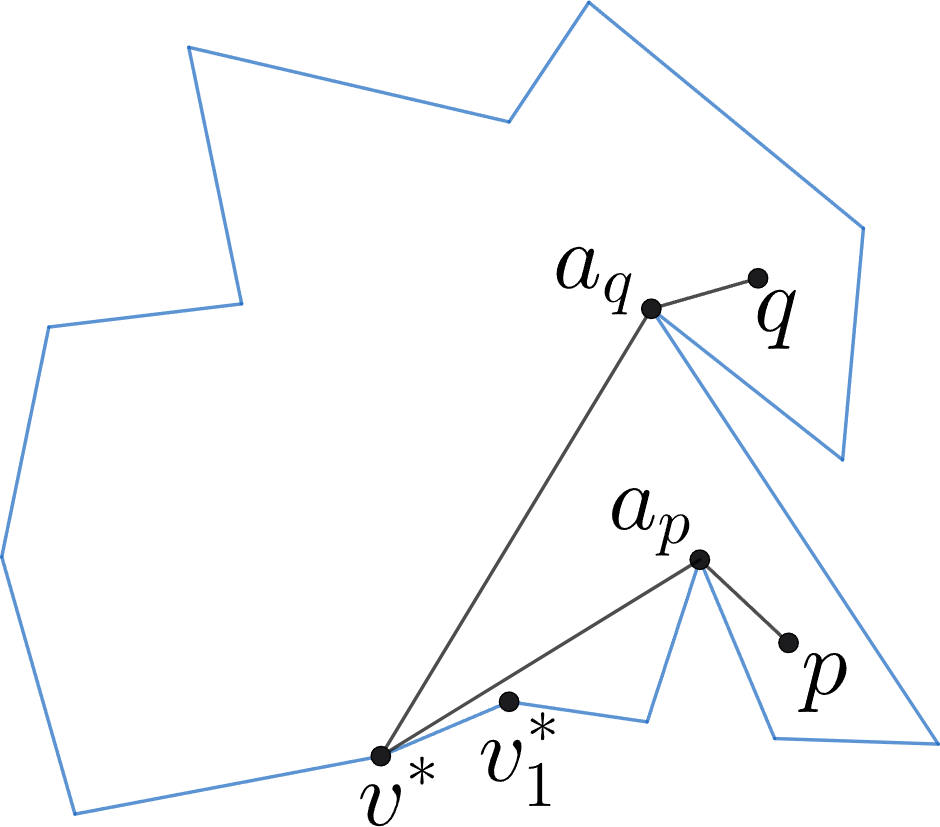}
            \subcaption{The case of $b=v^*$.}
            \label{fig:leqvs2}
        \end{subfigure}
        \caption{Illustration of the definition of $p \la q$.}
    \end{figure}

\begin{itemize}
\item If $b \neq v^*$, then we define $p \la q$ if the anchors of $b$ in the three paths $\pi(b, v^*)$, $\pi(b, p)$, and $\pi(b, q)$ appear in counterclockwise order around $b$ (see Figure~\ref{fig:leqvs}).

\item If $b = v^*$, let $a_p$ and $a_q$ be the anchors of $v^*$ in $\pi(v^*, p)$ and $\pi(v^*, q)$, respectively.
Let $v^*_1$ be the counterclockwise neighbor of $v^*$ on $\partial P$.
Define $\angle v^*_1 v^* p$ (resp., $\angle v^*_1 v^* q$) as the counterclockwise angle from $\overline{v^* v^*_1}$ to $\overline{v^* a_p}$ (resp., $\overline{v^* a_q}$); see Figure~\ref{fig:leqvs2}.
We define $p \la q$ if $\angle v^*_1 v^* p < \angle v^*_1 v^* q$.
\end{itemize}

We also define $p \ea q$ if $p = q$.
We have the following lemma.

\begin{lemma} \label{lem:la_compute}
The relation $\la$ defines a partial order on the points of $P$.
Given any two points in $P$, their order under $\la$ can be determined in $O(\log m)$ time using the GH data structure.
\end{lemma}

\begin{proof}
It suffices to prove that $\la$ is transitive.
Let $p, q, r \in P$ such that $p \la q$ and $q \la r$. We show that $p \la r$.

Let $b_{pq}$ be the junction vertex of $\pi(v^*, p)$ and $\pi(v^*, q)$, and define $b_{pr}$ and $b_{qr}$ analogously.
Since $b_{pq}, b_{qr} \in \pi(v^*, q)$, there are two cases (see Figure~\ref{fig:partial_order}):

\begin{itemize}
\item If $b_{qr} \in \pi(v^*, b_{pq})$, then $b_{pr} = b_{qr}$.
In this case, the ordering around $b_{pr}$ implied by $q \la r$ yields $p \la r$, regardless of whether $b_{pr} = v^*$.

\item If $b_{pq} \in \pi(v^*, b_{qr})$, then $b_{pr} = b_{pq}$.
In this case, the ordering around $b_{pr}$ implied by $p \la q$ yields $p \la r$, again regardless of whether $b_{pr} = v^*$.
\end{itemize}

\begin{figure}
    \centering
    \begin{subfigure}[b]{.45\textwidth}
        \centering
        \includegraphics[scale=0.5]{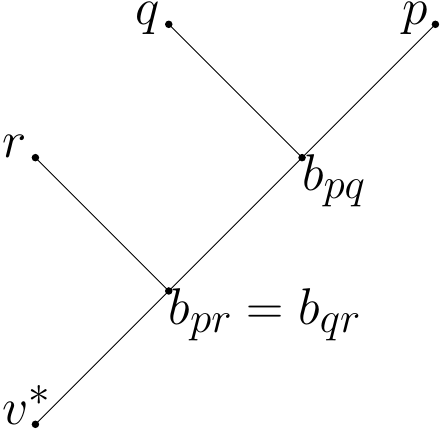}
        \subcaption{$b_{qr} \in \pi(v^*, b_{pq})$.}
        \label{fig:cqr_in_cpq}
    \end{subfigure}
    \hfill
    \begin{subfigure}[b]{.45\textwidth}
        \centering
        \includegraphics[scale=0.5]{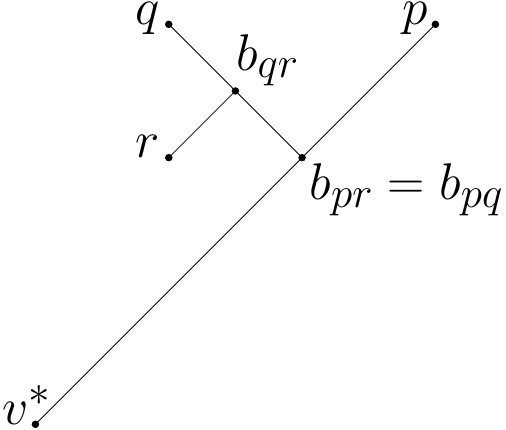}
        \subcaption{$b_{pq} \in \pi(v^*, b_{qr})$.}
        \label{fig:cpq_in_cqr}
    \end{subfigure}
    \caption{Illustration for the proof of Lemma~\ref{lem:la_compute}.}
    \label{fig:partial_order}
\end{figure}

This proves transitivity.

To determine the order between two points $p, q \in P$, we proceed as follows.
If $p = q$, then $p \ea q$.
Otherwise, we check whether $p \in \pi(v^*, q)$ or $q \in \pi(v^*, p)$; if so, the two points are not $\la$-comparable.
If they are comparable, we compute the junction vertex $b_{pq}$ and the anchors of $v^*$ in the paths $\pi(b_{pq}, v^*)$, $\pi(b_{pq}, p)$, and $\pi(b_{pq}, q)$.
The order between $p$ and $q$ under $\la$ can then be determined accordingly.
All these operations can be performed in $O(\log m)$ time using the GH data structure.
\end{proof}

We remark that in Fortune's algorithm, the order $\la$ corresponds to the $x$-coordinate order of points in the plane; points with the same $x$-coordinate are not comparable. Thus, the $x$-coordinate order is also a partial order.

The following lemma shows that the orders $\la$ and $\lw$ are consistent on $\W_t$.

\begin{lemma} \label{lem:vs_monotone}
Under a general value of $t$, for any two points $p, q \in \W_t$ that are $\la$-comparable, we have $p \la q$ if and only if $p \lw q$.
Moreover, $\la$ is a total order on $\W_t\ca$.
\end{lemma}

\begin{proof}
Suppose $p \la q$; we show that $p \lw q$.
Since $p$ and $q$ are $\la$-comparable, we have $p \notin \pi(v^*, q)$ and $q \notin \pi(v^*, p)$.

The path $\pi(v^*, p)$ partitions $\R_t(v^*)$ into two regions.
Let $R_1$ be the region to the right of $\pi(v^*, p)$ and $R_2$ the other region.
Let $\W_t^1 = \W_t \cap R_1$ and $\W_t^2 = \W_t \cap R_2$.
By definition, for any $p_1 \in \W_t^1 \setminus \{p\}$ and $p_2 \in \W_t^2 \setminus \{p\}$, we have $p_1 \lw p \lw p_2$.

Since $p \la q$, the definition of $\la$ implies that $q$ lies to the left of $\pi(v^*, p)$, and hence $q \in \W_t^2$.
Therefore, $p \lw q$.

Conversely, suppose $p \lw q$.
If $p$ and $q$ are $\la$-comparable, assume for contradiction that $q \la p$.
By the above argument, this would imply $q \lw p$, contradicting $p \lw q$.
Hence $p \la q$.

Finally, we show that $\la$ is a total order on $\W_t\ca$.
Let $p, q \in \W_t\ca$ be distinct points.
Since both lie on Voronoi edges of $\partial \R_t(v^*)$ and $\R_t(v^*)$ is star-shaped (for general $t$),  $\pi(v^*, p) \setminus \{p\}$ does not intersect any Voronoi edge of $\R_t(v^*)$ by Observation~\ref{obs:star_shaped}.
Since $q \in \W_t\ca \subseteq \W_t = \partial \R_t(v^*)$, $q$ is on a Voronoi edge of $\R_t(v^*)$. We thus obtain $q \notin \pi(v^*, p)$.
Similarly, $p \notin \pi(v^*, q)$.
Hence $p$ and $q$ are $\la$-comparable, proving that $\la$ is total on $\W_t\ca$.
\end{proof}

\paragraph{A total order $\boldsymbol{\lp}$.}
We define a total order $\lp$ on the points of $\partial P$.
For two points $p, q \in \partial P \setminus \{v^*\}$, we define $p \lp q$ if $v^*, p, q$ appear in counterclockwise order along $\partial P$.
For any $p \in \partial P \setminus \{v^*\}$, we define $p \lp v^*$.
This defines a total order on $\partial P$.

The following lemma shows that $\la$ and $\lp$ are consistent.

\begin{lemma} \label{lem:pvconsistent}
For any two points $p, q \in \partial P$ that are $\la$-comparable, we have $p \la q$ if and only if $p \lp q$.
Moreover, given any two points $p, q \in \partial P$, we can determine their order under $\lp$ in $O(\log m)$ time.
\end{lemma}

\begin{proof}
We first prove that $p \la q \implies p \lp q$.
Suppose $p \la q$.
Since $p$ and $q$ are $\la$-comparable, we have $p \notin \pi(v^*, q)$ and $q \notin \pi(v^*, p)$.

The path $\pi(v^*, p)$ partitions $P$ into two regions.
Let $R_1$ be the region to the right of $\pi(v^*, p)$ and $R_2$ the other region. We assume that neither region contains $\pi(v^*, p)$.
Hence, $R_1$, $R_2$, and $\pi(v^*, p)$ form a partition of $P$.
By definition of $\lp$, for any $p_1 \in (\partial P \cap R_1) \setminus \{p\}$ and $p_2 \in (\partial P \cap R_2) \setminus \{p\}$, we have $p_1 \lp p \lp p_2$.

Since $p \la q$, the definition of $\la$ implies that $q$ lies to the left of $\pi(v^*, p)$ and $q \notin \pi(v^*, p)$, and hence $q \in \partial P \cap R_2$.
Therefore, $p \lp q$.

The converse direction $p \lp q \implies p \la q$ follows by an argument analogous to that in the proof of Lemma~\ref{lem:vs_monotone}, and we omit the details.

To determine the order under $\lp$, given two points $p, q \in \partial P$, we proceed as follows.
Using the polygon boundary point location data structure, we find in $O(\log m)$ time the edges $e_p$ and $e_q$ of $P$ that contain $p$ and $q$, respectively.
Using the indices of $e_p$ and $e_q$ in the counterclockwise ordering of $\partial P$ (established during global preprocessing), we can determine the order of $p$ and $q$ under $\lp$.
The total time is $O(\log m)$.
\end{proof}

If $p\lp q$, then we also write $q>_{\partial P} p$. We use $p\leq_{\partial P} q$ to mean either $p=q$ or $p\lp q$.
We do the same for the other two orders $\la$ and $\lw$.

\subsubsection{The representation of $\boldsymbol{\W_t}$}
\label{sec:representation}

We store the Voronoi edges and $\partial P$-portions of $\W_t$ in a balanced binary search tree, ordered according to $\lw$.

For each Voronoi edge $e$ of $\W_t$, the edge lies on a bisector $B(v^*, s)$ for some site $s \in S$; we store this site and denote it by $s_e$, instead of storing $e$ explicitly.
For each $\partial P$-portion, we store only a label indicating that it is a $\partial P$-portion.
Note that no two $\partial P$-portions of $\W_t$ are adjacent, except possibly the two incident to $v^*$ (in which case $v^*$ is considered to lie between them).

We use the term \emph{edge} to refer to either a Voronoi edge or a $\partial P$-portion of $\W_t$.
In addition to the binary search tree, we maintain a doubly linked list over the edges of $\W_t$, so that the neighbors of each edge can be accessed in $O(1)$ time.

Furthermore, for every pair of adjacent edges $(e, f)$ in $\W_t$ that are both Voronoi edges, we store the intersection points
$B(s_e, s_f) \cap \partial P$. In addition, for every triple of consecutive edges $(e, f, g)$ in $\W_t$ that are all Voronoi edges, we store $B(s_e, s_g) \cap \partial P$. These values are needed when invoking the algorithm of Lemma~\ref{lem:bis_branch_int}.

\begin{lemma} \label{lem:updateWt}
Inserting or removing an edge $e$ from $\W_t$ can be done in $O(\log n + \log^2 m)$ time, assuming that the neighbors of $e$ in $\W_t$ are known.
\end{lemma}

\begin{proof}
We will show in Lemma~\ref{lem:vd_size} that $\vd_t(S)$ has size $O(n)$.
Since every edge of $\W_t$ is also an edge of $\vd_t(S)$, it follows that $|\W_t| = O(n)$.

Maintaining $\W_t$ in a balanced binary search tree therefore supports insertion and deletion in $O(\log n)$ time.
Updating the doubly linked list of neighbors requires only $O(1)$ time, since each update affects at most a constant number of edges.

For each triple of consecutive Voronoi edges $(e, f, g)$, we maintain the intersection points
$B(s_e, s_f) \cap \partial P$, $B(s_e, s_g) \cap \partial P$, and $B(s_f, s_g) \cap \partial P$.
Each such intersection can be computed in $O(\log^2 m)$ time by Lemma~\ref{lem:bpp}.
Since inserting or removing an edge affects only $O(1)$ such triples, maintaining these values requires $O(\log^2 m)$ time in total.

Combining the above bounds, the total update time is $O(\log n + \log^2 m)$.
\end{proof}

Although the vertices of $\W_t$ are not stored explicitly, each can be computed in $O(\log^2 m)$ time using Lemma~\ref{lem:vc_compute}.

\subsubsection{The representation of $\boldsymbol{\vd_t(S)}$}

We represent $\vd_t(S)$ implicitly using a doubly-connected edge list (DCEL).
The vertices of $\vd_t(S)$ are Voronoi vertices and Voronoi terminals, the edges are Voronoi edges and $\partial P$-portions between Voronoi terminals, and the faces correspond to Voronoi regions.

Instead of storing each vertex $v$ explicitly, we store only the set of sites associated with $v$.
Similarly, instead of storing each edge $e$ explicitly, we store the identifiers of its two incident vertices together with the sites associated with $e$.

Since $\vd_t(S)$ is represented implicitly, its size is $O(n)$ rather than $O(n + m)$, as shown in Lemma~\ref{lem:vd_size}.
Although vertex coordinates are not stored explicitly, they can be computed in $O(\log^2 m)$ time using Lemma~\ref{lem:vc_compute} when needed.

\begin{lemma} \label{lem:vd_size}
The size of $\vd_t(S)$ is $O(n)$.
\end{lemma}

\begin{proof}
Since each Voronoi region of $\vd_t(S)$ is defined by a site in $S \cup \{v^*\}$ and $|S| = n$, $\vd_t(S)$ has $O(n)$ faces.

Under our general position assumptions, each Voronoi vertex is incident to exactly three Voronoi edges, and each Voronoi terminal is incident to one Voronoi edge and two $\partial P$-portions.
In addition, the vertex $v^*$ is incident to two $\partial P$-portions.
Therefore, every vertex of $\vd_t(S)$ except $v^*$ has degree three.

Applying Euler's formula to the planar graph underlying $\vd_t(S)$, we conclude that both the number of edges and the number of vertices are $O(n)$.
\end{proof}

\section{Processing the site events}
\label{sec:site_events}

In this section, we describe how to process site events.
Consider a site event $t_s$ of a site $s \in S$.
To process this event, we need to update both $\vd_t(S)$ and $\W_t$.
For convenience, we assume that $\vd_{t_s - \epsilon}(S)$ and $\W_{t_s - \epsilon}$ are given for an arbitrarily small $\epsilon > 0$.
Our goal is to update them to obtain $\vd_{t_s + \epsilon}(S)$ and $\W_{t_s + \epsilon}$.

By Lemmas~\ref{lem:events} and~\ref{lem:sweeping}, we can view this update as inserting the region $\R_{t_s + \epsilon}(s)$ into $\vd_{t_s - \epsilon}(S)$.
Based on this observation, we will first compute $\R_{t_s}(s)$; from this, $\R_{t_s + \epsilon}(s)$ can be determined easily.
Then, we will insert $\R_{t_s + \epsilon}(s)$ into $\vd_{t_s - \epsilon}(S)$.

In the following, we first discuss the degenerate configuration at $t = t_s$.

\subsection{The degenerate configuration at $\boldsymbol{t = t_s}$}

By definition, $d_{v^*}(s) = d_s(s)$ at $t = t_s$.
By the triangle inequality, for every point $p \in P$,
\[
d_{v^*}(p) \le d_{v^*}(s) + d(s, p) = d_s(s) + d(s, p) = d_s(p).
\]
Hence, $\R_{t_s}(s) \subseteq \R_{t_s}(v^*)$.

This implies that at $t = t_s$, the Voronoi region $\R_{t_s}(v^*)$ is the same as in $\vd_{t_s}(S \setminus \{s\})$.
Therefore, although degeneracies may occur at $t = t_s$, the region $\R_{t_s}(v^*)$ remains star-shaped (as in Observation~\ref{obs:star_shaped}), and $\W_t$ remains a well-defined curve with the properties in Lemma~\ref{lem:vs_monotone}.
For reference, we summarize this in the following lemma.

\begin{lemma} \label{lem:vs_monotone_site_event}
At $t = t_s$, $\R_t(v^*)$ is star-shaped and $\W_t$ is a curve starting and ending at $v^*$.
For any two points $p, q \in \W_t$ that are $\la$-comparable, we have $p \la q \iff p \lw q$.
Moreover, $\la$ is a total order on $\W_t\ca$.
\end{lemma}

Let $\seg{s s'}$ be the last-edge extension of $\pi(v^*, s)$.
The following lemma characterizes $\R_{t_s}(s)$.
This situation is analogous to the site event in Fortune's algorithm (see~\cite[Figure~7.2]{ref:deBergCo08}).

\begin{lemma} \label{lem:site_event_bisector}
At $t = t_s$, the following hold:
\begin{enumerate}
\item $B(v^*, s) = \seg{s s'}$.
\item $\R_{t_s}(s) \neq \emptyset$ if and only if $\seg{s s'} \cap \W_{t_s} \neq \emptyset$.
\item If $\seg{s s'} \cap \W_{t_s} \neq \emptyset$, then $\seg{s s'} \cap \W_{t_s}$ consists of a single point $z$ lying on a single edge $e$ of $\W_{t_s}$, and $\R_{t_s}(s) = \seg{s z}$.
\end{enumerate}
\end{lemma}

\begin{proof}
\leavevmode
\begin{enumerate}
\item At $t = t_s$, we have
\[
d_{v^*}(s) = d(v^*, s) + t_s = d(v^*, s) + d_s(s) - d(v^*, s) = d_s(s).
\]
Thus $s \in B(v^*, s)$.
By the triangle inequality, $B(v^*, s)$ extends from $s$ along the ray opposite to the direction of the anchor of $s$ in $\pi(v^*, s)$, until it reaches $\partial P$.
Therefore, $B(v^*, s) = \seg{s s'}$.

\item If $\R_{t_s}(s) \neq \emptyset$, then by Lemma~\ref{lem:events}, $\R(s) \neq \emptyset$, and thus $s \in \R(s)$ by Observation~\ref{obser:nonempty}.
Hence $d_s(s) = d_S(s)$, and since $d_{v^*}(s) = d_s(s)$, we obtain $d_{v^*}(s) = d_S(s)$, implying $s \in \R_{t_s}(v^*)$.
Because $\seg{s s'}$ connects $s \in \R_{t_s}(v^*)$ to $s' \in \partial P$, it must intersect $\partial \R_{t_s}(v^*) = \W_{t_s}$.

Conversely, if $\seg{s s'} \cap \W_{t_s} \neq \emptyset$, let $z$ be a point in the intersection.
Since $z \in \W_{t_s}$, we have $d_{v^*}(z) = d_{S \cup \{v^*\}}(z)$.
Because $z \in B(v^*, s)$, we also have $d_s(z) = d_{v^*}(z)$, and hence $d_s(z) = d_{S \cup \{v^*\}}(z)$, implying $z \in \R_{t_s}(s)$.
Thus $\R_{t_s}(s) \neq \emptyset$.

\item First, since $\pi(s, s') = \overline{s s'}$, points on $\overline{s s'}$ are not $\la$-comparable.
Since $\la$ is a total order on $\W_{t_s}\ca$ by Lemma~\ref{lem:vs_monotone_site_event}, the segment $\overline{s s'}$ can intersect $\W_{t_s}\ca$ at most once.
Moreover, $\overline{s s'}$ can intersect $\W_{t_s}\sa$ only at $s'$.

Suppose for contradiction that $\seg{s s'} \cap \W_{t_s}$ contains two distinct points.
Then $\seg{s s'}$ must intersect $\W_{t_s}\ca$ at some point $p$, and also contain $s' \in \W_{t_s}\sa$.
Let $q$ be a point on $\seg{p s'}$ strictly between $p$ and $s'$.
Then
\[
d_S(q) \le d_S(p) + |\overline{p q}| = d_{v^*}(p) + |\overline{p q}| = d_{v^*}(q).
\]
Since $\overline{s s'}$ intersects $\W_{t_s}\ca$ only at $p$ and $q \neq p$, we have $q \notin \W_{t_s}\ca$. As $q \neq s'$, we also have $q \notin \W_{t_s}\sa$. Therefore, $q \notin \W_{t_s}$ and $d_S(q) \neq d_{v^*}(q)$.
Consequently, we obtain $d_S(q) < d_{v^*}(q)$.
Thus,
\[
d_S(s') \le d_S(q) + |\overline{q s'}| < d_{v^*}(q) + |\overline{q s'}| = d_{v^*}(s'),
\]
which implies $s' \notin \W_{t_s}$, a contradiction.
Hence $\seg{s s'}$ intersects $\W_{t_s}$ at most once.

If the intersection is nonempty, it consists of a single point $z$.
We claim that $z$ lies on a single edge of $\W_{t_s}$.
Otherwise, $z$ would be a Voronoi vertex or terminal of $\vd_{t_s}(S)$, and since $d_{v^*}(z) = d_s(z)$, it would also be a Voronoi vertex or terminal of $\vd(S)$ associated with $s$.
This implies $t_s = t_z$, contradicting the general position assumption that site events do not coincide with Voronoi vertex or terminal events.

Finally, by an argument similar to part (1), we obtain
\[
\R_{t_s}(s) = \seg{s s'} \cap \R_{t_s}(v^*) = \seg{s z}.
\]
\end{enumerate}

This completes the proof.
\end{proof}

By Lemma~\ref{lem:site_event_bisector}, our algorithm for handling the site event at $t = t_s$ proceeds as follows.
First, we determine whether $\R_{t_s}(s) = \emptyset$; if not, we compute the edge $e$.
Second, we update our data structures for $\vd_{t_s - \epsilon}(S)$ and $\W_{t_s - \epsilon}$ to obtain $\vd_{t_s + \epsilon}(S)$ and $\W_{t_s + \epsilon}$ using $e$.
The following lemma describes the second step; the algorithm for computing $e$ is given in Section~\ref{sec:computeedge}.

\begin{lemma} \label{lem:site_event_updates}
Given the edge $e$, we can update the data structures for $\vd_t(S)$ and $\W_t$ from $t = t_s - \epsilon$ to $t = t_s + \epsilon$ in $O(\log n + \log^2 m)$ time.
\end{lemma}

\begin{proof}
Let $v_1$ and $v_2$ be the endpoints of $e$ with $v_1 <_{\W_{t_s}} v_2$.
We distinguish two cases depending on whether $e$ is a Voronoi edge or a $\partial P$-portion.

\begin{itemize}
\item If $e$ is a Voronoi edge (see
    Figure~\ref{fig:evor}), then we create two new vertices $w_1$ and $w_2$ and remove $e$ from $\vd_t(S)$.
Further, we add edges $(v_1, w_1)$ and $(w_2, v_2)$, both Voronoi edges associated with $v^*$ and $s_e$ (where $s_e$ is the site associated with $e$ other than $v^*$).
We also add two edges between $w_1$ and $w_2$: one Voronoi edge associated with $s$ and $s_e$, and one associated with $v^*$ and $s$.
These two edges form the boundary of the new Voronoi region $\R_t(s)$.
All updates to $\vd_t(S)$ take $O(1)$ time.

For $\W_t$, we replace $e$ with three edges in order: $e_1$, $e_2$, and $e_3$, where $e_1$ and $e_3$ are Voronoi edges associated with $v^*$ and $s_e$, and $e_2$ is a Voronoi edge associated with $v^*$ and $s$.
Updating $\W_t$ takes $O(\log n + \log^2 m)$ time by Lemma~\ref{lem:updateWt}.

\begin{figure}
    \centering
    \begin{subfigure}[b]{.45\textwidth}
        \centering
        \includegraphics[width=0.8\linewidth]{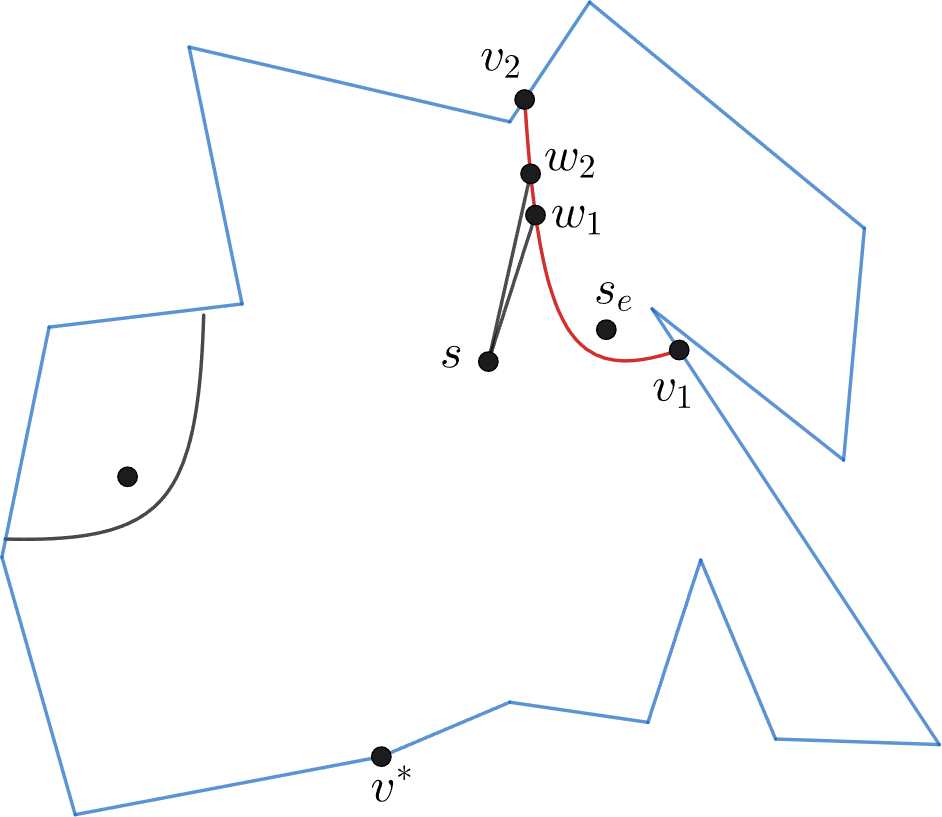}
        \subcaption{$e$ is a Voronoi edge.}
        \label{fig:evor}
    \end{subfigure}
    \hfill
    \begin{subfigure}[b]{.45\textwidth}
        \centering
        \includegraphics[width=0.8\linewidth]{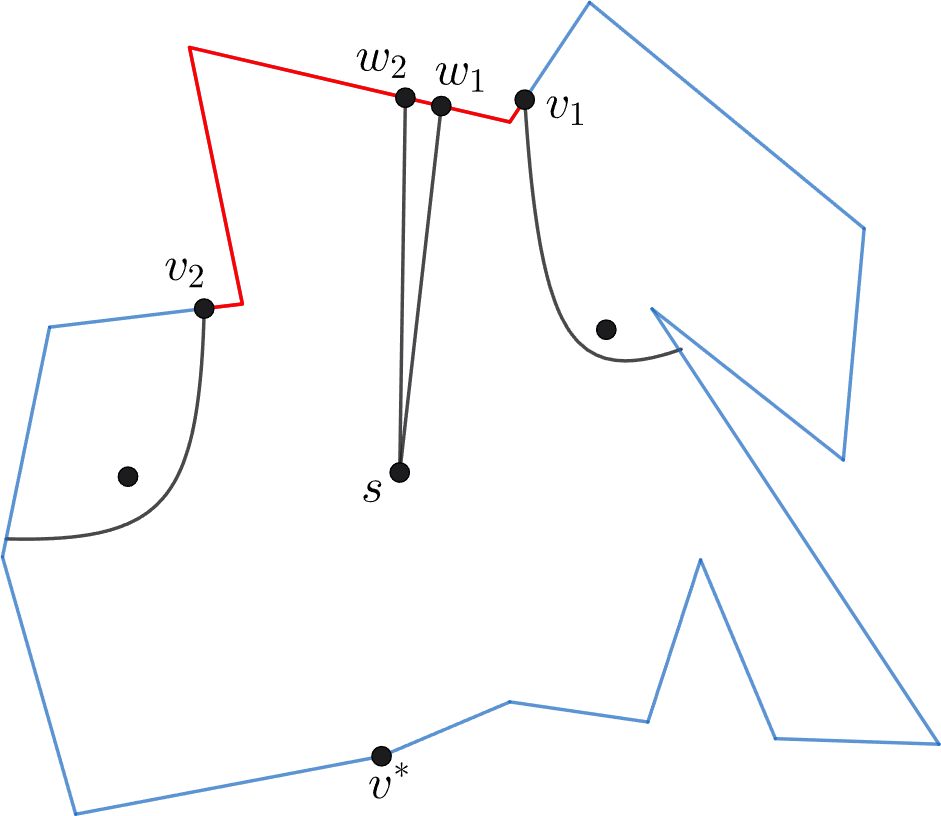}
        \subcaption{$e$ is a $\partial P$-portion.}
        \label{fig:epp}
    \end{subfigure}
    \caption{Illustrations of the proof of Lemma~\ref{lem:site_event_updates}.}
    \label{fig:site_event}
\end{figure}

\item If $e$ is a $\partial P$-portion (see Figure~\ref{fig:epp}), then we create two vertices $w_1$ and $w_2$ and remove $e$ from $\vd_t(S)$. Further, we add edges $(v_1, w_1)$, $(w_1, w_2)$, and $(w_2, v_2)$, all of which are $\partial P$-portions.
In addition, we add a Voronoi edge between $w_1$ and $w_2$ associated with $v^*$ and $s$.
The two edges between $w_1$ and $w_2$ together define the boundary of the new region $\R_t(s)$.
These updates to $\vd_t(S)$ take $O(1)$ time.

For $\W_t$, we replace $e$ with three edges in order: $e_1$, $e_2$, and $e_3$, where $e_1$ and $e_3$ are $\partial P$-portions and $e_2$ is a Voronoi edge associated with $v^*$ and $s$.
Updating $\W_t$ again takes $O(\log n + \log^2 m)$ time by Lemma~\ref{lem:updateWt}.
\end{itemize}

This completes the proof.
\end{proof}

\subsection{Computing the edge $\boldsymbol{e}$}
\label{sec:computeedge}

We now describe the first main step of the algorithm, namely determining whether $\R_{t_s}(s) = \emptyset$, and if not, computing the edge $e$ (and the intersection point $z$).
We will show that this can be done in $O(\log n \log m + \log^2 m)$ time.

We first present an algorithm for computing $e$ under the assumption that $\R_{t_s}(s) \neq \emptyset$.
We will later explain how to test whether $\R_{t_s}(s) = \emptyset$ and how to compute $z$.

Let $f$ and $g$ be two adjacent edges along $\W_{t_s}$ such that $f$ precedes $g$, and let $v$ be their common endpoint.
We describe an $O(\log m)$-time procedure to determine whether $v <_{\W_{t_s}} z$.
Using this procedure, we can perform a binary search on $\W_{t_s}$ to locate $e$.
Since $\W_{t_s}$ is stored in a balanced binary search tree of height $O(\log n)$, this yields an $O(\log n \log m)$-time algorithm for computing $e$.

A straightforward approach to determining whether $v <_{\W_{t_s}} z$ would be to compute the coordinates of $v$.
If $v$ is a Voronoi vertex, this can be done in $O(\log m)$ time by Lemma~\ref{lem:bis_branch_int}.
However, if $v$ is a Voronoi terminal, computing its coordinates takes $O(\log^2 m)$ time by Lemma~\ref{lem:bpp}, which would exceed our target bound.
Therefore, we must avoid explicitly computing $v$ in this case.
Achieving an $O(\log m)$-time test for $v <_{\W_{t_s}} z$ is crucial for obtaining the overall $O(n \log n + n \log^2 m)$ running time.

\paragraph{Algorithm description.}
Algorithm~\ref{algo:vx} gives the pseudocode.
If $v = v^*$, then by definition $z <_{\W_{t_s}} v^*$, so we assume $v \neq v^*$.
Hence, at least one of $f$ and $g$ must be a Voronoi edge.

If both $f$ and $g$ are Voronoi edges, we first compute the coordinates of $v$.
We then determine whether $v \la s$.
If so, we return \texttt{true} (i.e., $v <_{\W_{t_s}} z$); otherwise, we return \texttt{false}.

If exactly one of $f$ and $g$ is a Voronoi edge, assume without loss of generality that $f$ is a Voronoi edge.
Let $s_f$ be the site associated with $f$ other than $v^*$.
We first compute the last-edge extension $\seg{s s'}$ of $\pi(v^*, s)$.
We then test whether $d_{s_f}(s') \le d_{v^*}(s')$.
If so, we return \texttt{false}.
Otherwise, we test whether $s_f \la s$.
If so, we return \texttt{true}; otherwise, we return \texttt{false}.

\paragraph{Running time.}
Since $B(s_f, s_g) \cap \partial P$ are already stored as discussed in Section~\ref{sec:representation},
Line~\ref{ln:computev} can be executed in $O(\log m)$ time by Lemma~\ref{lem:bis_branch_int}.
Lines~\ref{ln:vsorder} and~\ref{ln:sfsorder} take $O(\log m)$ time by Lemma~\ref{lem:la_compute}.
Lines~\ref{ln:sprime} and~\ref{ln:distance} can each be implemented in $O(\log m)$ time using the GH and ray-shooting data structures.
Therefore, the total running time of the procedure is $O(\log m)$.

\begin{algorithm}
    \caption{Decide whether $v <_{\W_{t_s}} z$.} \label{algo:vx}
    \KwIn{$s, f, g$.}
    \KwOut{Whether $v <_{\W_{t_s}} z$.}
    \If{$f$ and $g$ are Voronoi edges}{
        Compute $v$\;\label{ln:computev}
        \If{$v \la s$ \label{ln:vsorder}}
        {
        \Return{true\;}
        }
        \Else{
            \Return{false\;}
        }
    }
    \If{$f$ is a Voronoi edge and $g$ is a $\partial P$-portion}{
        Compute $s'$\;\label{ln:sprime}
        \If{$d_{s_f}(s') \leq d_{v^*}(s')$\label{ln:distance}}
        {
            \Return{false\;}
        }
        \Else{
        \If{$s_f \la s$\label{ln:sfsorder}}
        {
        \Return{true\;}
        }
        \Else{
            \Return{false\;}
        }
        }
    }
    \If{$f$ is a $\partial P$-portion and $g$ is a Voronoi edge}{
        \tcp{Symmetric to the above case}
    }
\end{algorithm}

\paragraph{Algorithm correctness.}
The correctness of the algorithm is nontrivial. We now give a rigorous proof.
We begin with two lemmas that will be used repeatedly.

\begin{lemma} \label{lem:ext_comp2}
For any three distinct points $p, q, r \in P$ such that $p \in \pi(v^*, q)$, if $p$ and $r$ are $\la$-comparable, then $p \la r$ if and only if $q \la r$.
\end{lemma}

\begin{proof}
Since $p \in \pi(v^*, q)$ and $p$ and $r$ are $\la$-comparable, it follows that $q$ and $r$ are also $\la$-comparable.

\begin{figure}
    \centering
    \includegraphics[scale=0.5]{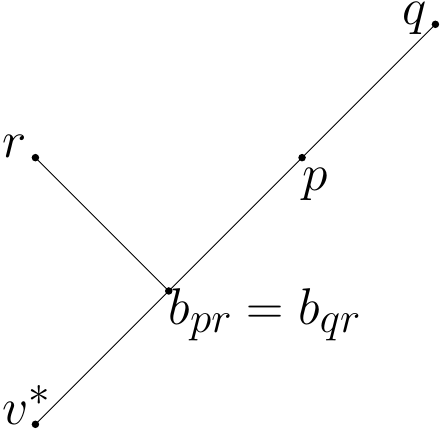}
    \caption{Illustration for the proof of Lemma~\ref{lem:ext_comp2}.}
    \label{fig:ext_comp}
\end{figure}

Let $b_{pr}$ be the junction vertex of $\pi(v^*, p)$ and $\pi(v^*, r)$, and define $b_{qr}$ analogously (see Figure~\ref{fig:ext_comp}).
Since $\pi(v^*, p)$ is a prefix of $\pi(v^*, q)$ and $p \notin \pi(v^*, r)$, we must have $b_{pr} = b_{qr} \in \pi(v^*, p) \setminus \{p\}$.
This implies $p \la r \iff q \la r$.
\end{proof}

\begin{lemma} \label{lem:v_comparison}
We have $s' \notin \pi(v^*, v)$ and $v \notin \pi(v^*, s')$.
In particular, $v$ is $\la$-comparable with each of $s$, $z$, and $s'$.
Furthermore, $s \la v \iff z \la v \iff s' \la v$.
\end{lemma}

\begin{proof}
By Lemma~\ref{lem:site_event_bisector}, $\overline{s s'} \cap \W_{t_s}$ consists of a single point $z$ lying on a single edge of $\W_{t_s}$.
Since $v$ lies on two edges of $\W_{t_s}$, we have $z \neq v$ and hence $v \notin \seg{s s'}$.

We first show that $s' \notin \pi(v^*, v)$.
Suppose for contradiction that $s' \in \pi(v^*, v)$.
Then $\pi(v^*, v) = \pi(v^*, s') \cup \pi(s', v)$.
Note that $\pi(v^*, s') = \pi(v^*, s) \cup \seg{s s'}$.
This implies that $s'$ is a vertex of $\partial P$, and $s'$, $s$, and the anchor of $s$ in $\pi(v^*, s)$ are collinear points in $S \cup V_P$, contradicting our general position assumption.

Next, we show that $v \notin \pi(v^*, s')$.
Suppose for contradiction that $v \in \pi(v^*, s')$.
Since $v \notin \seg{s s'}$, we must have $v \in \pi(v^*, s)$.
Let $s_v \in S$ be a site associated with $v$.
Then
\[
d_{s_v}(z) \le d_{s_v}(v) + d(v, z).
\]
Because $v \in \W_{t_s}$, we have $d_{v^*}(v) = d_{s_v}(v)$ at $t = t_s$.
Since $v \in \pi(v^*, s')$ and $z \in \seg{s s'}$, it follows that
\[
d_{v^*}(z) = d_{v^*}(v) + d(v, z).
\]
Thus,
\[
d_{s_v}(z) \le d_{v^*}(z).
\]
On the other hand, since $z \in \W_{t_s}$, we also have $d_{v^*}(z) \le d_{s_v}(z)$.
Hence $d_{v^*}(z) = d_{s_v}(z)$.

By Lemma~\ref{lem:site_event_bisector}, $B(v^*, s) \cap \W_{t_s} = \{z\}$, so we also have $d_s(z) = d_{v^*}(z) = d_{s_v}(z)$.
Since $s \neq s_v$ (otherwise $v$ would lie on $B(v^*, s) = \seg{s s'}$, contradicting $v \notin \seg{s s'}$),
it follows that $z$ lies on a Voronoi edge associated with $v^*$ and $s_v$, and hence $z \in \W_{t_s}\ca$.
Since at least one of $f$ and $g$ is a Voronoi edge, we also have $v \in \W_{t_s}\ca$.
By Lemma~\ref{lem:vs_monotone_site_event}, $\la$ is a total order on $\W_{t_s}\ca$, so $z$ and $v$ are $\la$-comparable.
This implies $v \notin \pi(v^*, z)$, contradicting $v \in \pi(v^*, s)$ because $\pi(v^*, s) \subseteq \pi(v^*, z)$.

Thus $v \notin \pi(v^*, s')$.

Since $s' \notin \pi(v^*, v)$, $v \notin \pi(v^*, s')$, and $\pi(v^*, s') = \pi(v^*, s) \cup \overline{s s'}$, the point $v$ is $\la$-comparable with $s$, $z$, and $s'$.
Applying Lemma~\ref{lem:ext_comp2}, we obtain
\[
s \la v \iff z \la v \iff s' \la v,
\]
as claimed.
\end{proof}

We now prove the correctness of the algorithm.
There are two cases depending on whether both $f$ and $g$ are Voronoi edges.

If both $f$ and $g$ are Voronoi edges, then
\[
v \la s \overset{(1)}{\implies} v \la z \overset{(2)}{\implies} v <_{\W_{t_s}} z,
\]
where (1) follows from Lemma~\ref{lem:v_comparison} and (2) follows from Lemma~\ref{lem:vs_monotone_site_event}.

We next consider the case where exactly one of $f$ and $g$ is a Voronoi edge.
Without loss of generality, assume that $f$ is a Voronoi edge and $g$ is a $\partial P$-portion.
Depending on whether $d_{s_f}(s') \le d_{v^*}(s')$, we distinguish two subcases.

\begin{lemma}\label{lem:correct10}
If $d_{s_f}(s') \le d_{v^*}(s')$, then $v >_{\W_{t_s}} z$.
\end{lemma}

\begin{proof}
See Figure~\ref{fig:sfspleqvssp} for an illustration.
We first consider the case $d_{s_f}(s') < d_{v^*}(s')$; the equality case will be handled later.

Since $f$ is a Voronoi edge associated with $s_f$ and $v^*$, we have $f \subseteq B(v^*, s_f)$, which is a curve with endpoints on $\partial P$. Since $v \in \partial P$, $v$ must be an endpoint of $B(v^*, s_f)$.
Let $w$ be the other endpoint of $B(v^*, s_f)$. Note that $w \in \partial P$. Since $f \subseteq B(v^*, s_f)$, $f$ precedes $g$ along $\W_{t_s}$, $v$ is the common endpoint of $f$ and $g$, and $g$ is a $\partial P$-portion, $w$ must precede $v$ under the order $\lp$. That is, $w <_{\partial P} v$.

\begin{figure}
    \centering
    \includegraphics[width=0.4\linewidth]{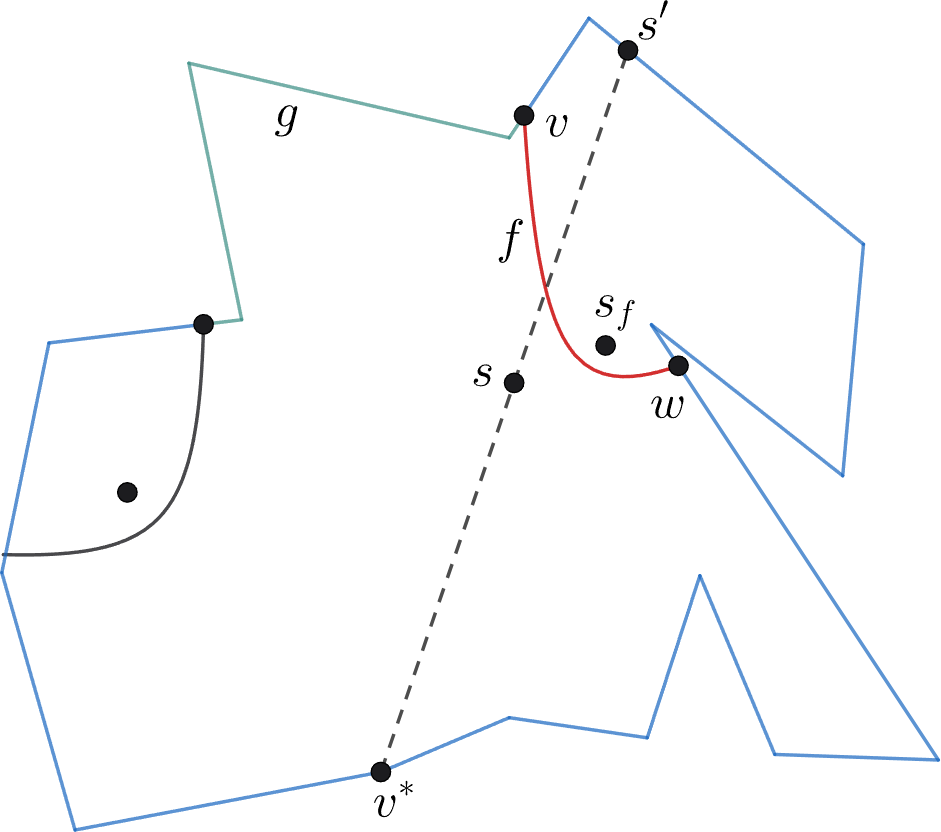}
    \caption{Illustration for the proof of Lemma~\ref{lem:correct10}.}
    \label{fig:sfspleqvssp}
\end{figure}

Since $B(v^*, s_f)$ separates the boundary $\partial P$ into two parts and $w <_{\partial P} v$, all points $p \in \partial P$ satisfying $d_{s_f}(p) < d_{v^*}(p)$ must lie before $v$ in the order $<_{\partial P}$.
Applying this to $p = s'$, we obtain $s' <_{\partial P} v$.
Hence,
\[
s' <_{\partial P} v \overset{(1)}{\implies} s' \la v \overset{(2)}{\implies} z \la v \overset{(3)}{\implies} z <_{\W_{t_s}} v,
\]
where (1) follows from Lemma~\ref{lem:pvconsistent} (using that $v$ and $s'$ are $\la$-comparable by Lemma~\ref{lem:v_comparison}),
(2) follows from Lemma~\ref{lem:v_comparison} and $z \in \pi(v^*, s')$, and
(3) follows from Lemma~\ref{lem:vs_monotone_site_event}.
Thus, $v >_{\W_{t_s}} z$.

\paragraph{The case $\boldsymbol{d_{s_f}(s') = d_{v^*}(s')}$.}
If $d_{s_f}(s') = d_{v^*}(s')$, then since $s' \in \partial P$, it must be an endpoint of $B(s_f, v^*)$.
By Lemma~\ref{lem:v_comparison}, $s' \neq v$, and hence $s' = w$.
From the above argument, we still have $s' <_{\partial P} v$, and therefore
\[
s' <_{\partial P} v \implies s' \la v \implies z \la v \implies z <_{\W_{t_s}} v.
\]
This completes the proof.
\end{proof}

\begin{lemma}\label{lem:correct20}
If $d_{s_f}(s') > d_{v^*}(s')$, then $s_f \la s$ if and only if $v <_{\W_{t_s}} z$.
\end{lemma}

\begin{proof}
Let $\seg{s_f s_f'}$ be the last-edge extension of $\pi(v^*, s_f)$.
By our general position assumption, for any three sites $s_1, s_2, s_3 \in S \cup \{v^*\}$, $s_2 \notin \pi(s_1, s_3)$. This implies that $s_f$ and $s$ are $\la$-comparable. Hence, either $s_f \la s$ or $s \la s_f$ holds. We discuss the two cases below.

\paragraph{Case 1: $s_f \la s$.}
We derive (see Figure~\ref{fig:sfpleqsp}):
\[
s_f \la s
\overset{(1)}{\implies} s_f \la s'
\overset{(2)}{\implies} s_f' \la s'
\overset{(3)}{\implies} s_f' <_{\partial P} s'
\]
\[
\overset{(4)}{\implies} v <_{\partial P} s'
\overset{(5)}{\implies} v \la s'
\overset{(6)}{\implies} v \la z
\overset{(7)}{\implies} v <_{\W_{t_s}} z.
\]

Here:
\begin{itemize}
\item (1) follows from Lemma~\ref{lem:ext_comp2}, since $s \in \pi(v^*, s')$.
\item (2) follows from Lemma~\ref{lem:ext_comp2}, since $s_f \in \pi(v^*, s_f')$.
\item (3) follows from Lemma~\ref{lem:pvconsistent}.
\item (4) follows by contradiction: if $s_f' <_{\partial P} s' \le_{\partial P} v$, then for all $p$ with $s_f' <_{\partial P} p <_{\partial P} v$, we have $d_{s_f}(p) \le d_{v^*}(p)$; applying this to $p=s'$ contradicts $d_{s_f}(s') > d_{v^*}(s')$.
\item (5) follows from Lemma~\ref{lem:pvconsistent} and the fact that by Lemma~\ref{lem:v_comparison}, $v$ and $s'$ are $\la$-comparable.
\item (6) follows from Lemma~\ref{lem:v_comparison}.
\item (7) follows from Lemma~\ref{lem:vs_monotone_site_event}.
\end{itemize}

\begin{figure}
    \centering
    \begin{subfigure}[b]{.45\textwidth}
        \centering
        \includegraphics[width=0.8\linewidth]{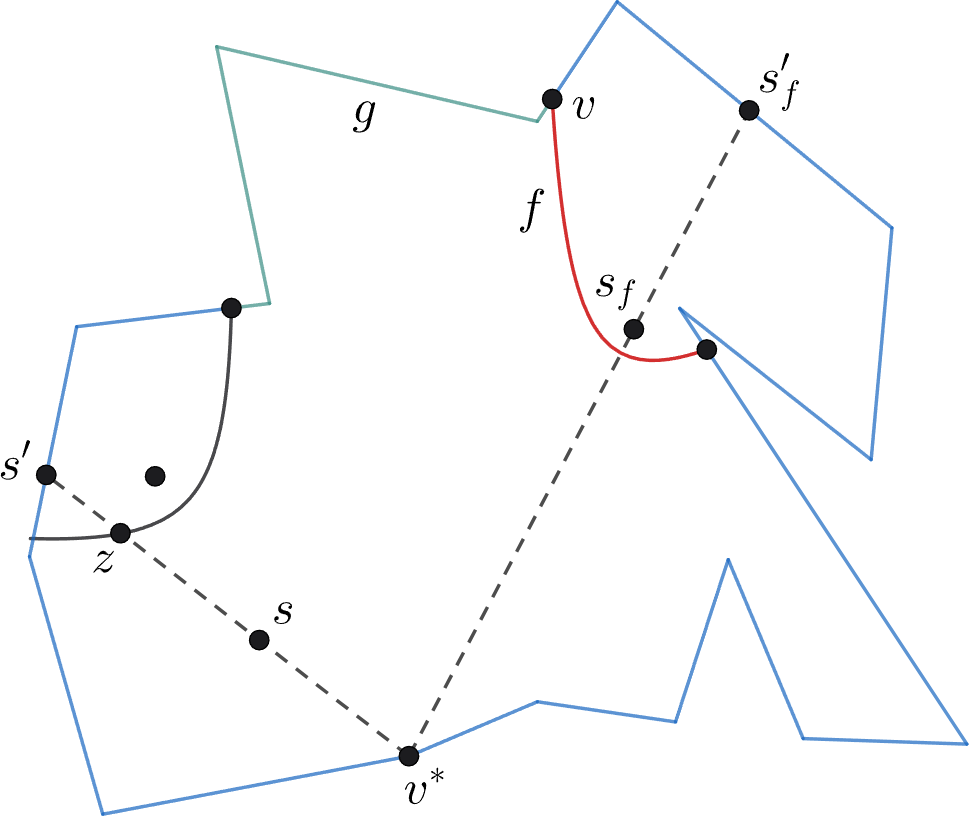}
        \subcaption{The case $s_f \la s$.}
        \label{fig:sfpleqsp}
    \end{subfigure}
    \hfill
    \begin{subfigure}[b]{.45\textwidth}
        \centering
        \includegraphics[width=0.8\linewidth]{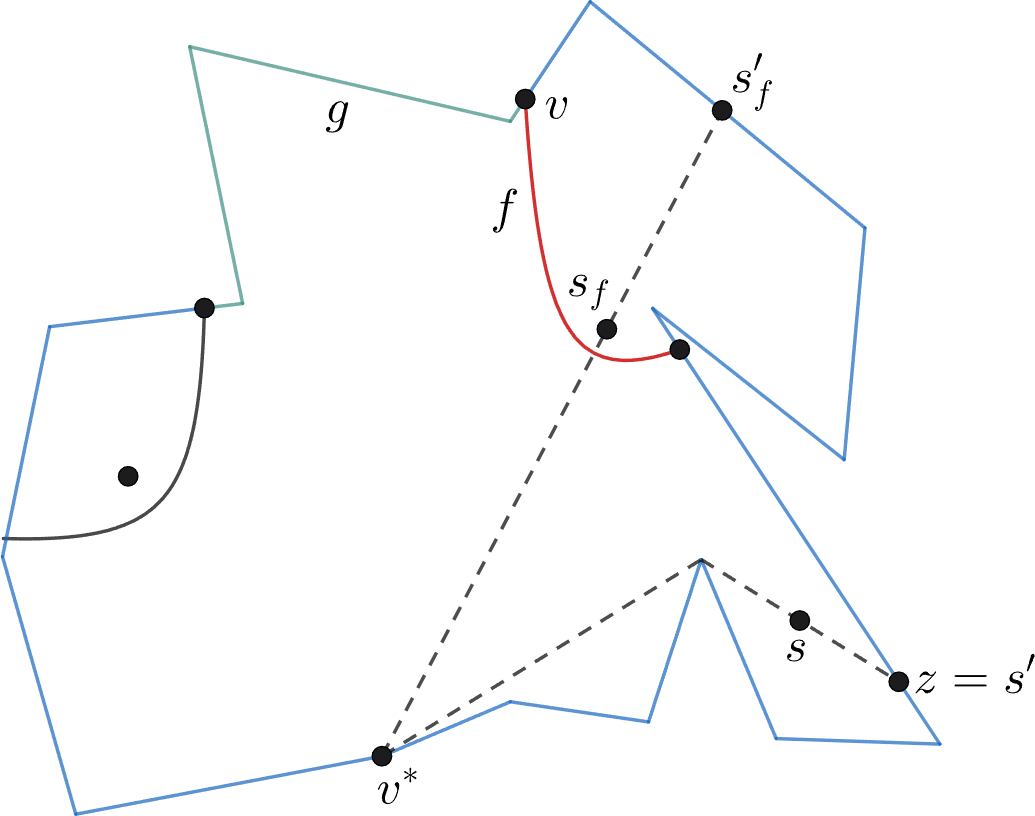}
        \subcaption{The case $s \la s_f$.}
        \label{fig:sfpgeqsp}
    \end{subfigure}
    \caption{Illustration for the proof of Lemma~\ref{lem:correct20}.}
\end{figure}

\paragraph{Case 2: $s \la s_f$.}
In this case, we can obtain $ z \lw v$ by a symmetric argument.
Specifically, we derive  (see Figure~\ref{fig:sfpgeqsp})

\[
s \la s_f
\os{1} \implies s' \la s_f
\os{2} \implies s' \la s_f'
\os{3} \implies s' \lp s_f' \]
\[\os{4} \implies s' \lp v
\os{5} \implies s' \la v
\os{6} \implies z \la v
\os{7} \implies z \lw v.\]

Here:
\begin{itemize}
\item (1) follows from Lemma~\ref{lem:ext_comp2}.
\item (2) follows from Lemma~\ref{lem:ext_comp2}.
\item (3) follows from Lemma~\ref{lem:pvconsistent}.
\item (4) follows because $d_{s_f}(s_f') < d_{v^*}(s_f')$, and all points $p \in \partial P$ with $d_{s_f}(p) < d_{v^*}(p)$ satisfy $p <_{\partial P} v$, implying $s_f' <_{\partial P} v$.
\item (5) follows from Lemma~\ref{lem:pvconsistent} and the fact that by Lemma~\ref{lem:v_comparison}, $v$ and $s'$ are $\la$-comparable.
\item (6) follows from Lemma~\ref{lem:v_comparison}.
\item (7) follows from Lemma~\ref{lem:vs_monotone_site_event}.
\end{itemize}

Combining the two cases completes the proof.
\end{proof}

This establishes the correctness of the algorithm. We therefore obtain the following result.

\begin{lemma}
If $\R(s) \neq \emptyset$, then we can find $e$ in $O(\log n \log m)$ time.
\end{lemma}

\paragraph{The case $\boldsymbol{\R(s) = \emptyset}$ and computing $\boldsymbol{z \in e}$.}

The above algorithm computes $e$ under the assumption that $\R(s) \neq \emptyset$.
We now describe how to determine whether $\R(s) = \emptyset$ and, if not, how to compute the point $z \in e$.

To test whether $\R(s) = \emptyset$, we run the above algorithm, which returns an edge $e'$ of $\W_{t_s}$ in $O(\log n \log m)$ time.
In what follows, we show how to compute the intersection $\seg{s s'} \cap e'$ in $O(\log^2 m)$ time.
If the intersection is nonempty, then the intersection point is $z$ and $e = e'$; otherwise, we conclude that $\R(s) = \emptyset$.

Let $v_1$ and $v_2$ be the endpoints of $e'$ with $v_1 <_{\W_{t_s}} v_2$.
Both $v_1$ and $v_2$ can be computed in $O(\log^2 m)$ time by Lemma~\ref{lem:vc_compute}.
We distinguish two cases depending on whether $e'$ is a Voronoi edge.

\begin{itemize}
\item
If $e'$ is a Voronoi edge, then $e' \subseteq B(v^*, s_{e'})$.
We compute $z' = \seg{s s'} \cap B(v^*, s_{e'})$ in $O(\log m)$ time using Lemma~\ref{lem:x_voronoi}.
If $v_1 \la z' \la v_2$ (which can be checked in $O(\log m)$ time by Lemma~\ref{lem:la_compute}), then $\seg{s s'} \cap e' = \{z'\}$; otherwise, $\seg{s s'} \cap e' = \emptyset$.

\item
If $e'$ is a $\partial P$-portion, then we compute $s'$ in $O(\log m)$ time using the GH data structure and ray shooting.
If $v_1 <_{\partial P} s' <_{\partial P} v_2$ (which can be checked in $O(\log m)$ time by Lemma~\ref{lem:pvconsistent}), then $\seg{s s'} \cap e' = \{s'\}$; otherwise, the intersection is empty.
\end{itemize}

In summary, we can determine whether $\R(s) = \emptyset$, and if not, compute both $e$ and $z$, in $O(\log n \log m + \log^2 m)$ time.
Together with Lemma~\ref{lem:site_event_updates}, this yields the following result.

\begin{lemma} \label{lem:handle_site_event}
Processing each site event takes $O(\log n \log m + \log^2 m)$ time.
\end{lemma}

\section{Processing other (non-site) events} \label{sec:nonsite_events}

In this section, we describe how to handle the remaining events, including non-pocket Voronoi vertex events (Section~\ref{sec:voronoi_vertex_events}), non-pocket Voronoi terminal events (Section~\ref{sec:voronoi_terminal_events}), pocket events (Section~\ref{sec:pocket_events}), and the final event (Section~\ref{sec:final_event}).

\subsection{Voronoi vertex events} \label{sec:voronoi_vertex_events}

To process a non-pocket Voronoi vertex event at time $t_v$ for a Voronoi vertex $v$, we are given $\vd_{t_v - \epsilon}(S)$ and $\W_{t_v - \epsilon}$, and our goal is to update them to obtain $\vd_{t_v + \epsilon}(S)$ and $\W_{t_v + \epsilon}$.

\begin{figure}[t]
    \centering
    \begin{subfigure}[b]{.38\textwidth}
        \centering
        \includegraphics[width=\linewidth]{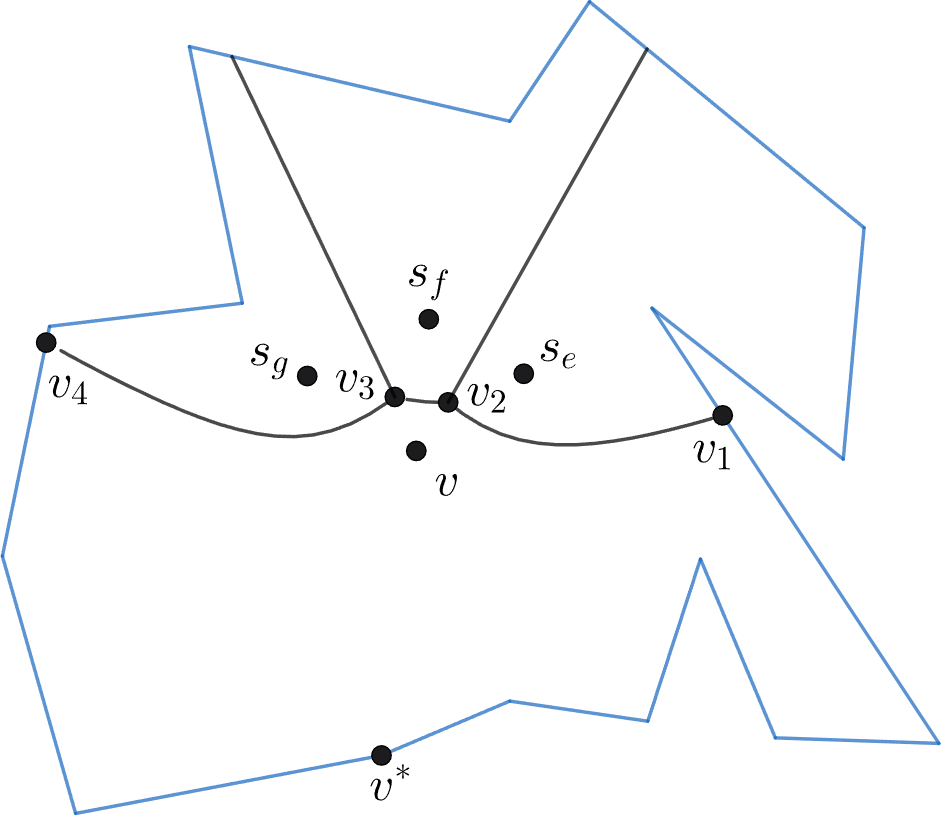}
        \subcaption{$t = t_v - \epsilon$.}
        \label{fig:vv_event_0}
    \end{subfigure}
    \hfill
    \begin{subfigure}[b]{.38\textwidth}
        \centering
        \includegraphics[width=\linewidth]{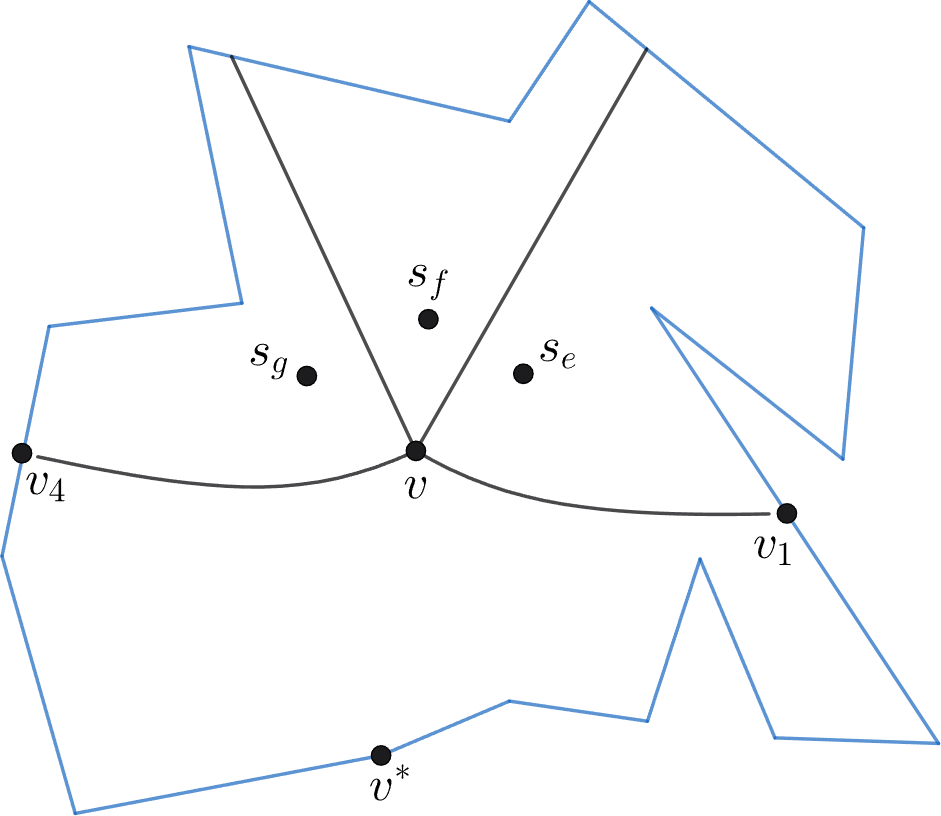}
        \subcaption{$t = t_v$.}
        \label{fig:vv_event_1}
    \end{subfigure}
    \hfill
    \begin{subfigure}[b]{.38\textwidth}
        \centering
        \includegraphics[width=\linewidth]{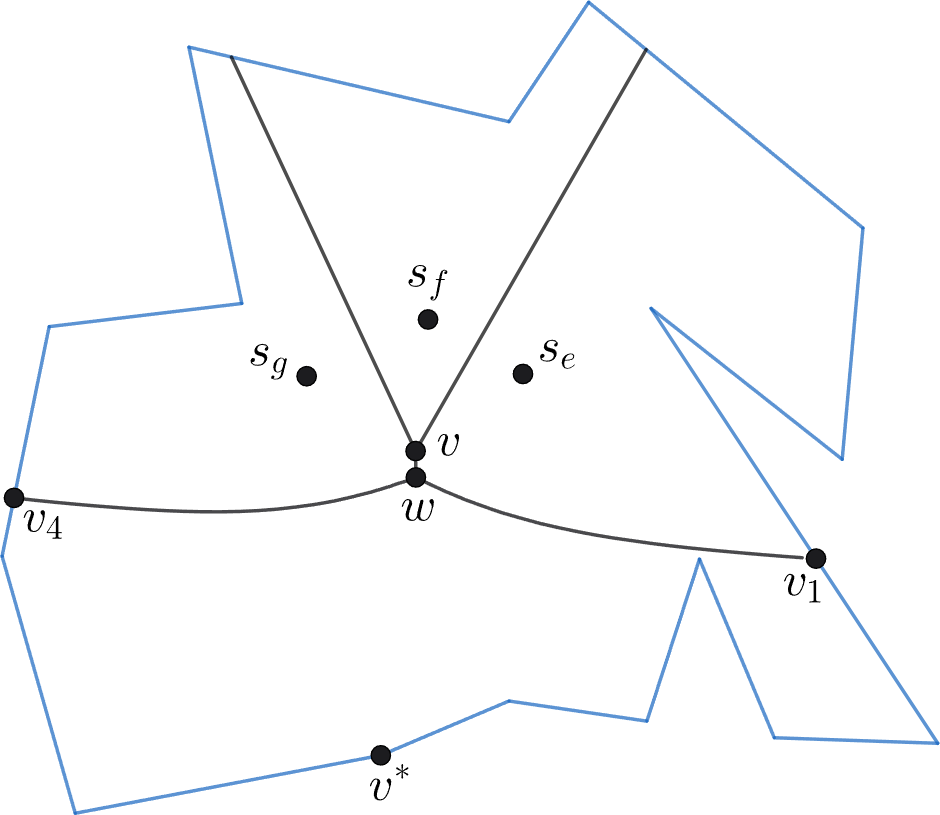}
        \subcaption{$t = t_v + \epsilon$.}
        \label{fig:vv_event_2}
    \end{subfigure}
    \caption{Illustration of a non-pocket Voronoi vertex event.}
    \label{fig:vv_event}
\end{figure}

Let the edges along $\W_{t_v - \epsilon}$ be $e$, $f$, and $g$ in order, such that $v$ is associated with the sites $s_e$, $s_f$, and $s_g$.
Let $v_i$ for $1 \le i \le 4$ denote the endpoints of $e$, $f$, and $g$ in order along $\W_{t_v - \epsilon}$ (see Figure~\ref{fig:vv_event}; note that $v_1$ and $v_4$ are shown as Voronoi terminals in the figure, but they may also be Voronoi vertices).
In particular, $v_2$ and $v_3$ must be Voronoi vertices of $\vd_{t_v - \epsilon}(S)$.

To update $\vd_t(S)$ from $t = t_v - \epsilon$ to $t = t_v + \epsilon$, we remove $f$, $v_2$, and $v_3$, and introduce two new vertices $v$ and $w$.
We update the endpoints of $e$ and $g$ by replacing $v_2$ and $v_3$ with $w$, respectively.
For the Voronoi edge associated with $s_e$ and $s_f$, we replace its endpoint $v_2$ with $v$; similarly, for the Voronoi edge associated with $s_f$ and $s_g$, we replace its endpoint $v_3$ with $v$.
Finally, we add a new Voronoi edge between $v$ and $w$, associated with $s_e$ and $s_g$.
These updates to $\vd_t(S)$ take $O(1)$ time.

To update $\W_t$, we simply remove the edge $f$, which takes $O(\log n + \log^2 m)$ time by Lemma~\ref{lem:updateWt}.
This yields the following result.

\begin{lemma} \label{lem:handle_voronoi_vertex_event}
We can handle each non-pocket Voronoi vertex event in $O(\log n + \log^2 m)$ time.
\end{lemma}

\subsection{Voronoi terminal events} \label{sec:voronoi_terminal_events}

To process a non-pocket Voronoi terminal event $t_u$ of a Voronoi terminal $u$, we are given $\vd_{t_u - \epsilon}(S)$ and $\W_{t_u - \epsilon}$, and our goal is to update them to obtain $\vd_{t_u + \epsilon}(S)$ and $\W_{t_u + \epsilon}$.

The event occurs when an edge $f$ of $\vd_{t_u - \epsilon}(S)$ shrinks to a point on $\partial P$, namely $u$.
Let $e$ and $g$ be the edges immediately before and after $f$ in $\vd_{t_u - \epsilon}(S)$ (see Figure~\ref{fig:c1_event}(a)).
Let $v_i$, for $1 \le i \le 4$, denote the endpoints of $e$, $f$, and $g$ in order along $\W_{t_u - \epsilon}$ under $<_{\W_{t_u - \epsilon}}$.
Note that at least one of the three edges must be a $\partial P$-portion.
We distinguish two cases depending on whether $f$ is a $\partial P$-portion.

\begin{itemize}
\item If $f$ is a $\partial P$-portion, then we refer to the event as a \emph{type-1} non-pocket Voronoi terminal event (see Figure~\ref{fig:c1_event}).
In this case, both $e$ and $g$ are Voronoi edges, and thus $v_2$ and $v_3$ are Voronoi terminals.
The point $u$ is the intersection $B(s_e, s_g) \cap \partial P$.

To update $\vd_t(S)$ from $t = t_u - \epsilon$ to $t = t_u + \epsilon$, we remove $f$, $v_2$, and $v_3$, and introduce two new vertices $u$ and $w$.
We update the endpoints of $e$ and $g$ by replacing $v_2$ and $v_3$ with $w$, respectively.
We then add a new Voronoi edge between $u$ and $w$, associated with $s_e$ and $s_g$.
In addition, there is another edge (other than $f$ and $g$) whose endpoint was $v_3$, and we update that endpoint to $u$.
Similarly, there is another edge (other than $f$ and $e$) whose endpoint was $v_2$, and we update that endpoint to $u$.
These updates take $O(1)$ time.

For $\W_t$, we remove the edge $f$, which takes $O(\log n + \log^2 m)$ time by Lemma~\ref{lem:updateWt}.
Thus, a type-1 event can be handled in $O(\log n + \log^2 m)$ time.

\begin{figure}[t]
    \centering
    \begin{subfigure}[b]{.38\textwidth}
        \centering
        \includegraphics[width=\linewidth]{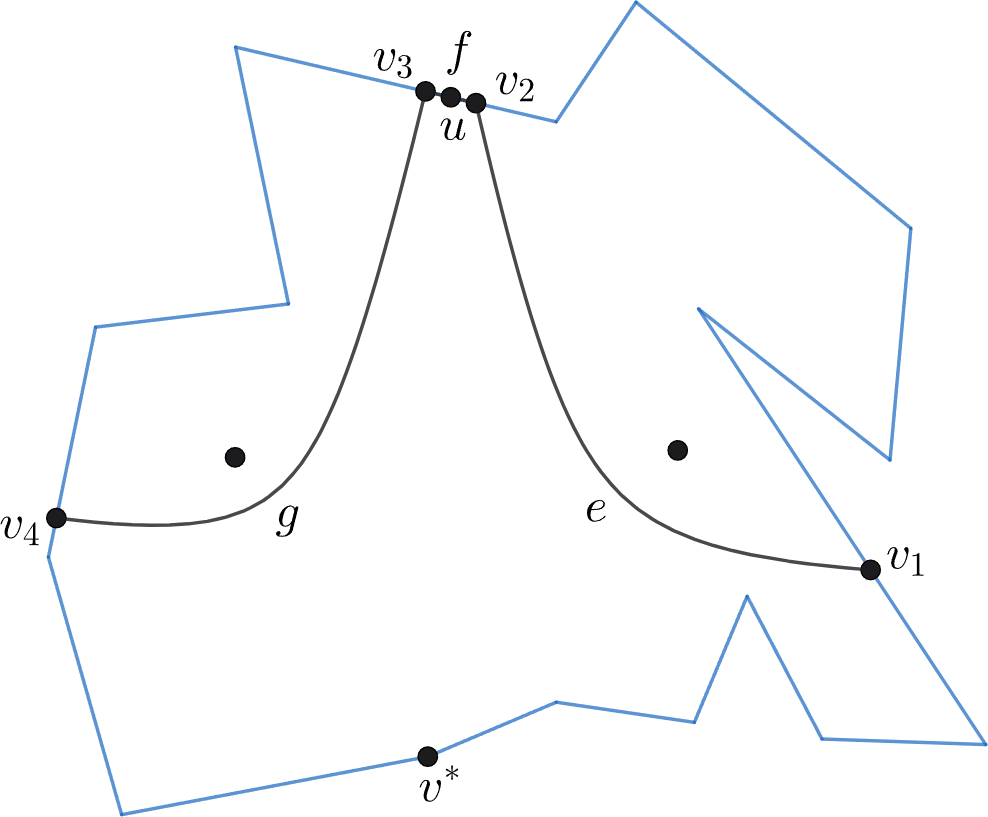}
        \subcaption{$t = t_u - \epsilon$.}
        \label{fig:c1_event_0}
    \end{subfigure}
    \hfill
    \begin{subfigure}[b]{.38\textwidth}
        \centering
        \includegraphics[width=\linewidth]{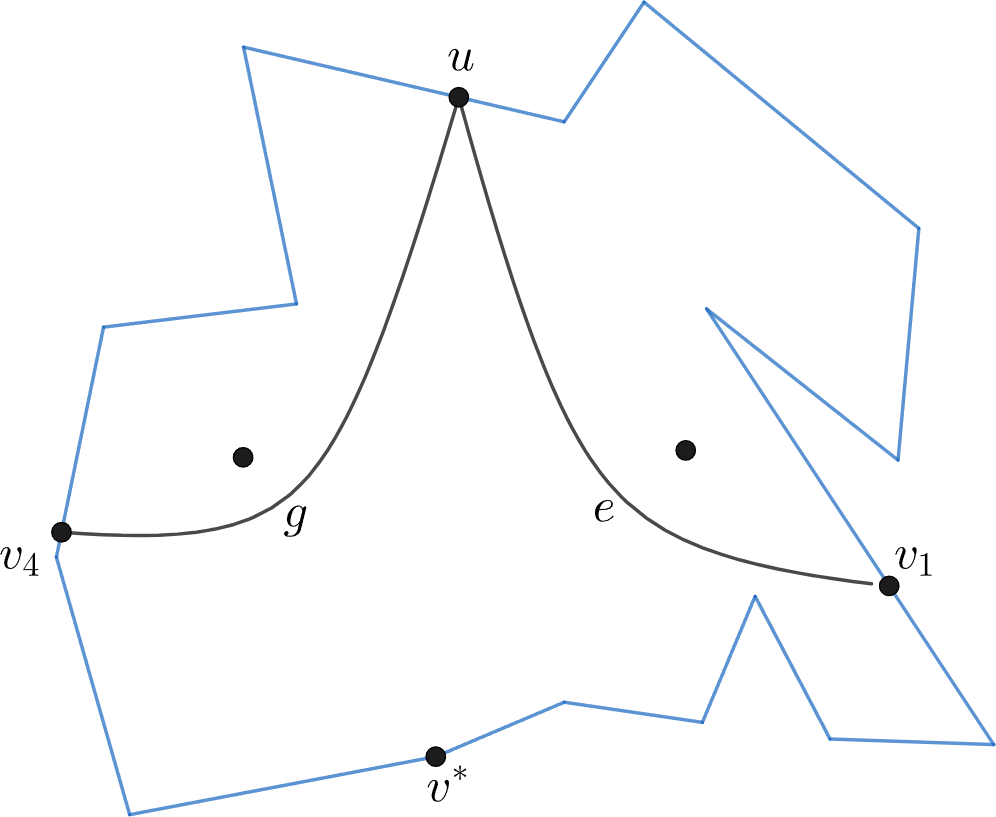}
        \subcaption{$t = t_u$.}
        \label{fig:c1_event_1}
    \end{subfigure}
    \hfill
    \begin{subfigure}[b]{.38\textwidth}
        \centering
        \includegraphics[width=\linewidth]{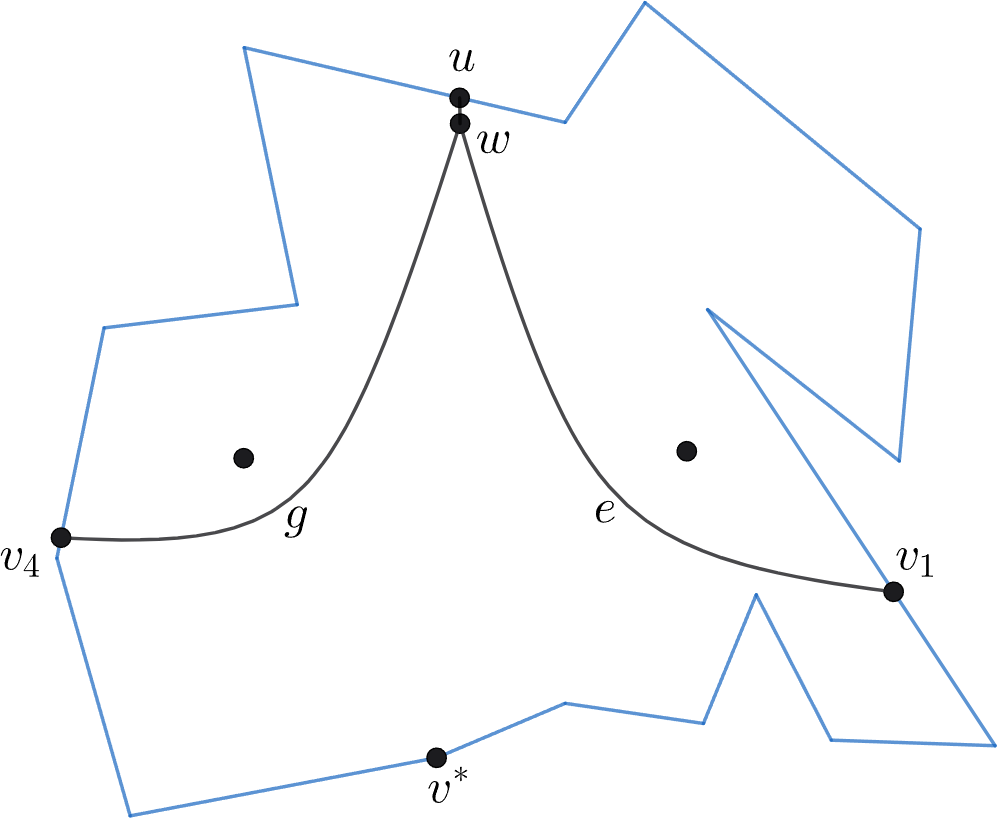}
        \subcaption{$t = t_u + \epsilon$.}
        \label{fig:c1_event_2}
    \end{subfigure}
    \caption{Illustration of a type-1 non-pocket Voronoi terminal event.}
    \label{fig:c1_event}
\end{figure}

\item If $f$ is not a $\partial P$-portion, then we refer to the event as a \emph{type-2} non-pocket Voronoi terminal event.
In this case, exactly one of $e$ and $g$ is a Voronoi edge.
Indeed, if both were Voronoi edges, the event would be a Voronoi vertex event; if neither were, $f$ would not disappear.
There are two symmetric subcases; we describe the case where $e$ is a Voronoi edge and $g$ is a $\partial P$-portion (see Figure~\ref{fig:c2_event}).

In this case, we have $u = B(s_e, s_f) \cap \partial P$.
We update $\vd_t(S)$ by removing $f$, $v_2$, and $v_3$, and introducing two new vertices $u$ and $w$.
We update the endpoints of $e$ and $g$ by replacing $v_2$ and $v_3$ with $w$, respectively.
We then add a new edge between $u$ and $w$, which is a $\partial P$-portion. In addition, there is another edge (other than $f$ and $g$) whose endpoint was $v_3$, and we update that endpoint to $u$. Similarly, there is another edge (other than $f$ and $g$) whose endpoint was $v_2$, and we update that endpoint to $u$.
These updates take $O(1)$ time.

For $\W_t$, we remove $f$, which takes $O(\log n + \log^2 m)$ time by Lemma~\ref{lem:updateWt}.

\begin{figure}[t]
    \centering
    \begin{subfigure}[b]{.38\textwidth}
        \centering
        \includegraphics[width=\linewidth]{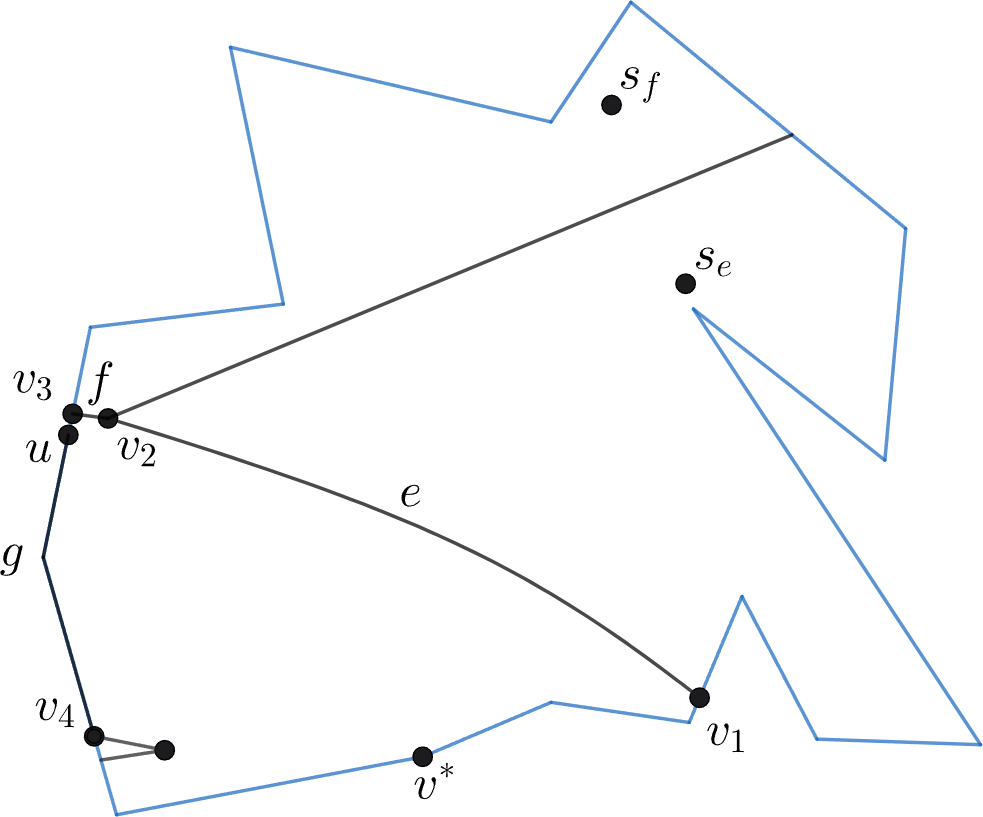}
        \subcaption{$t = t_u - \epsilon$.}
        \label{fig:c2_event_0}
    \end{subfigure}
    \hfill
    \begin{subfigure}[b]{.38\textwidth}
        \centering
        \includegraphics[width=\linewidth]{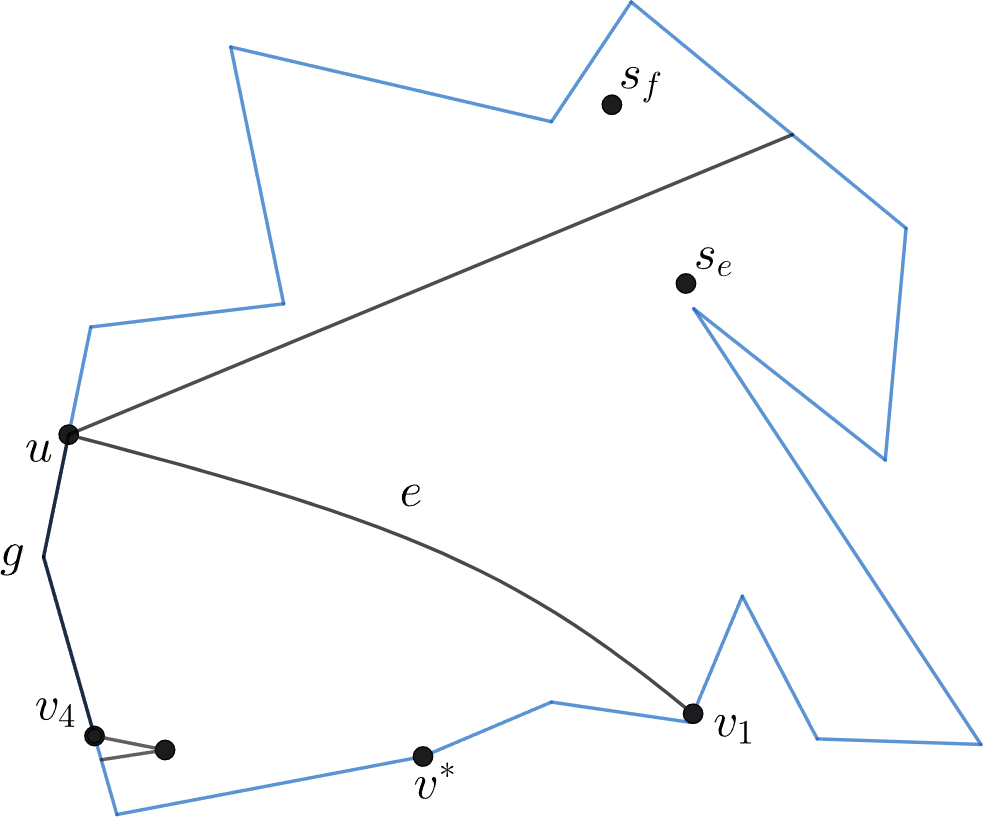}
        \subcaption{$t = t_u$.}
        \label{fig:c2_event_1}
    \end{subfigure}
    \hfill
    \begin{subfigure}[b]{.38\textwidth}
        \centering
        \includegraphics[width=\linewidth]{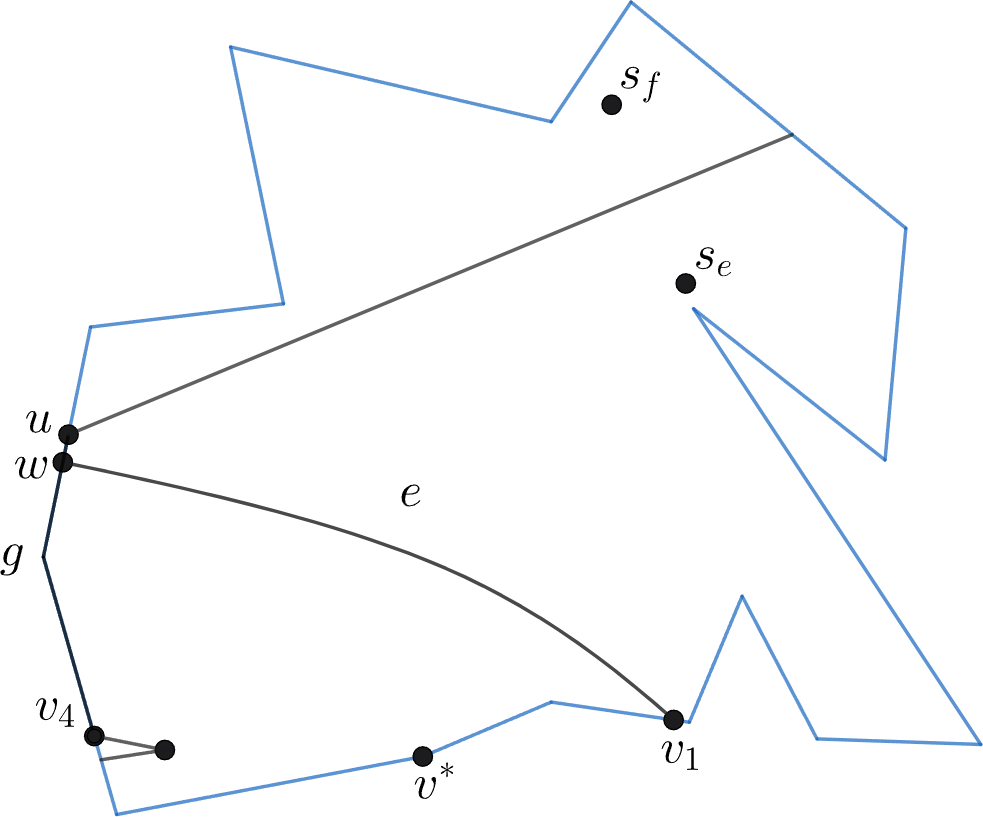}
        \subcaption{$t = t_u + \epsilon$.}
        \label{fig:c2_event_2}
    \end{subfigure}
    \caption{Illustration of a type-2 non-pocket Voronoi terminal event.}
    \label{fig:c2_event}
\end{figure}

\end{itemize}

We therefore obtain the following result.

\begin{lemma} \label{lem:handle_cap_events}
We can handle each non-pocket Voronoi terminal event in $O(\log n + \log^2 m)$ time.
\end{lemma}

\subsection{Pocket events} \label{sec:pocket_events}

We now describe how to handle pocket events.
As discussed in Section~\ref{sec:genpos}, we only need to handle a pocket event when it coincides with a Voronoi vertex or terminal event.
Accordingly, whenever such an event occurs, we first check whether it coincides with a pocket event.
If not, we process it as described in the previous subsections.
Otherwise, we handle the pocket event.
As we will see, once the pocket event is processed, all coincident Voronoi vertex or terminal events are handled simultaneously.

Consider an event $t_u$ of a Voronoi vertex or terminal $u$.
Let $S_u$ be the set of sites associated with $u$.
Note that $|S_u| = 2$ if $u$ is a Voronoi terminal and $|S_u| = 3$ if $u$ is a Voronoi vertex.

We first describe how to detect whether $t_u$ coincides with a pocket event.
For each site $s \in S_u$, we compute the junction vertex $b_u$ of $\pi(u, s)$ and $\pi(u, v^*)$ using the GH data structure.
If $b_u \neq u$, then at $t = t_u$, we have $d_{v^*}(b_u) = d_s(b_u)$, implying that $t_u = t_{s b_u}$ is a pocket event (with $b_u$ as the apex of $pocket(B(v^*, s))$).
If $b_u = u$ for all $s \in S_u$, then $t_u$ does not coincide with any pocket event.
This yields the following lemma.

\begin{lemma} \label{lem:elevate}
Given a Voronoi vertex or terminal event, we can determine in $O(\log m)$ time whether it coincides with a pocket event.
\end{lemma}

Assume that $t_u$ coincides with a pocket event $t_{sv}$ for some $s \in S_u$ and a vertex $v$ of $P$.
We now describe how to process the event.
Our goal is to update $\vd_t(S)$ and $\W_t$ from $t = t_{sv} - \epsilon$ to $t = t_{sv} + \epsilon$.

\begin{figure}[t]
    \centering
    \begin{subfigure}[b]{.38\textwidth}
        \centering
        \includegraphics[width=\linewidth]{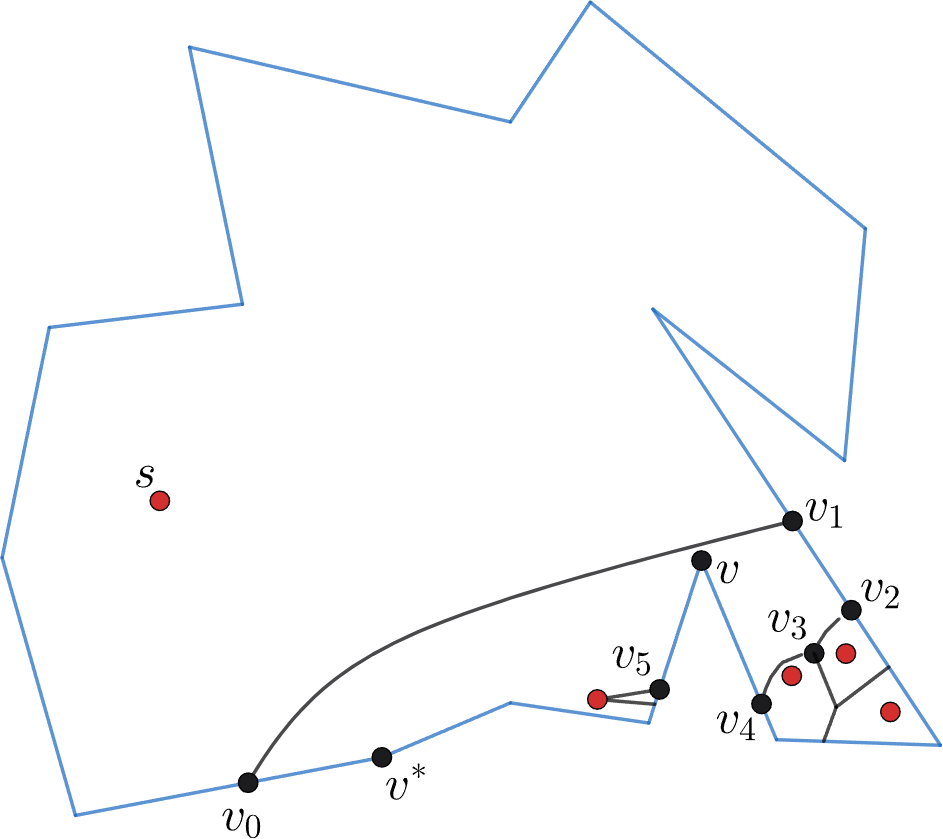}
        \subcaption{$t = t_{sv} - \epsilon$.}
        \label{fig:p_event_0}
    \end{subfigure}
    \hfill
    \begin{subfigure}[b]{.38\textwidth}
        \centering
        \includegraphics[width=\linewidth]{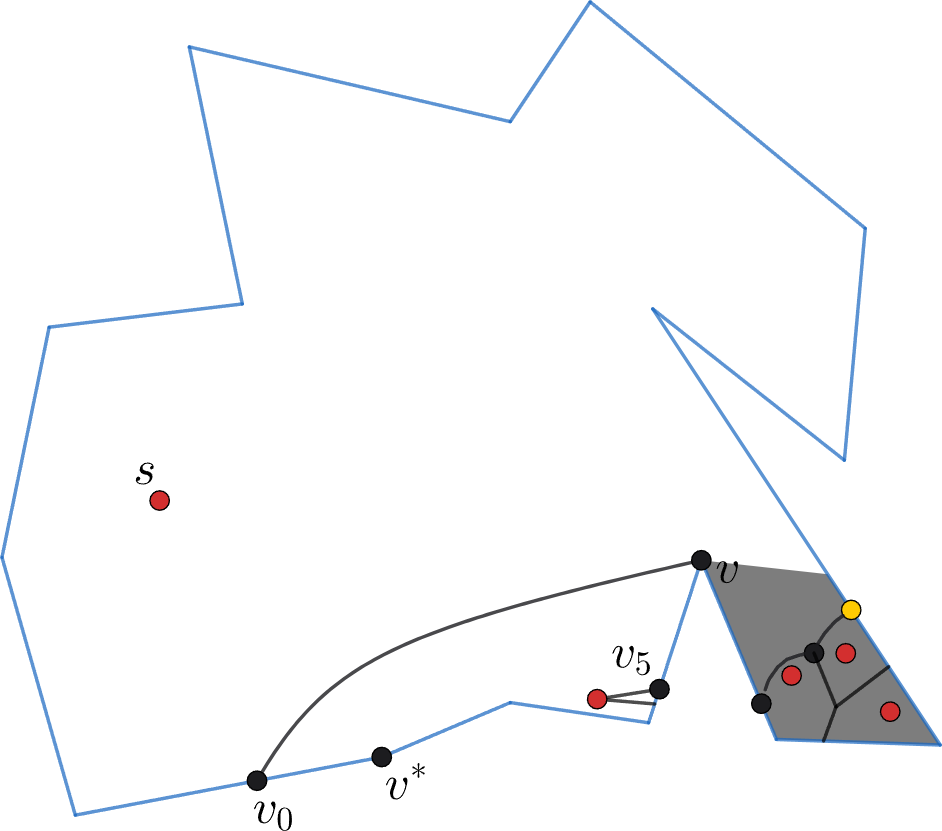}
        \subcaption{$t = t_{sv}$.}
        \label{fig:p_event_1}
    \end{subfigure}
    \hfill
    \begin{subfigure}[b]{.38\textwidth}
        \centering
        \includegraphics[width=\linewidth]{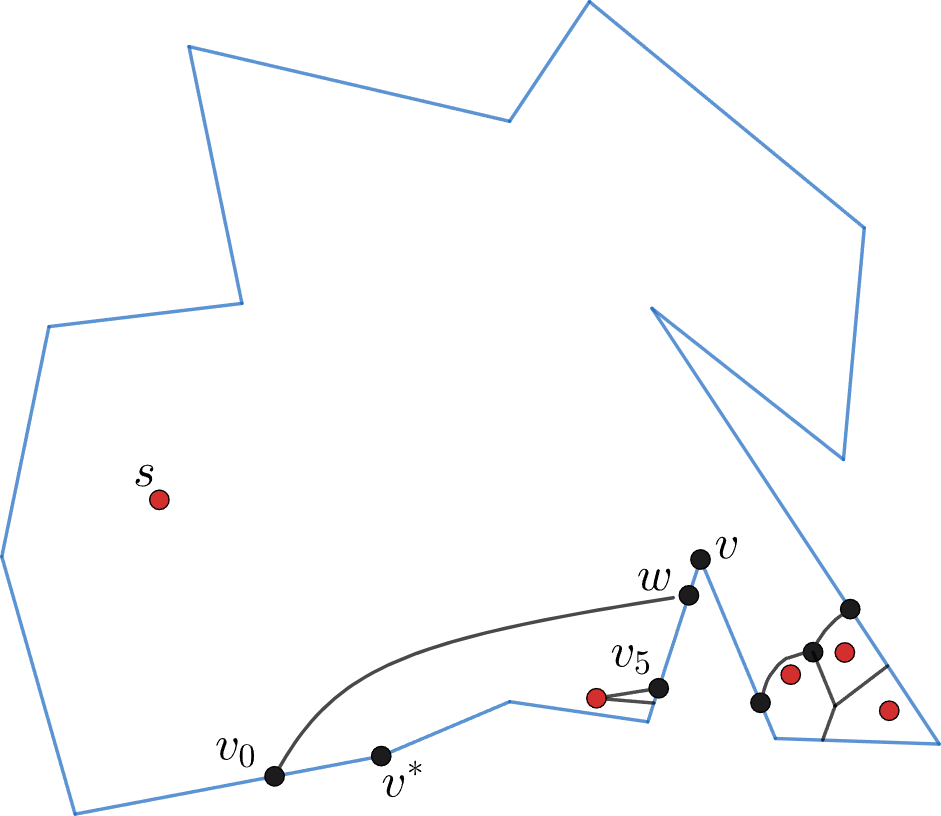}
        \subcaption{$t = t_{sv} + \epsilon$.}
        \label{fig:p_event_2}
    \end{subfigure}
        \caption{Illustration of a pocket event that coincides with a Voronoi terminal event.
    The Voronoi terminal $u$ is colored gold in (b), which results from ``merging'' $v_1$ and $v_2$ from (a), and it
    is the Voronoi terminal whose event coincides with the pocket event.}
    \label{fig:p_event}
\end{figure}

\begin{figure}[t]
    \centering
    \begin{subfigure}[b]{.38\textwidth}
        \centering
        \includegraphics[width=\linewidth]{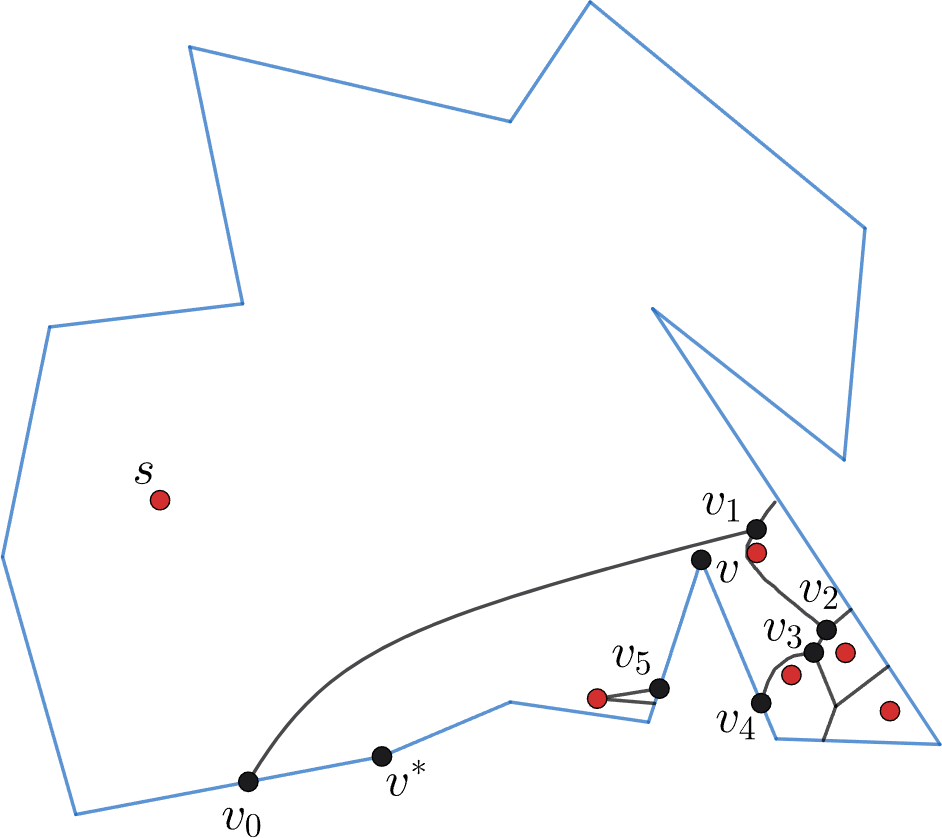}
        \subcaption{$t = t_{sv} - \epsilon$.}
        \label{fig:p_event_t2_0}
    \end{subfigure}
    \hfill
    \begin{subfigure}[b]{.38\textwidth}
        \centering
        \includegraphics[width=\linewidth]{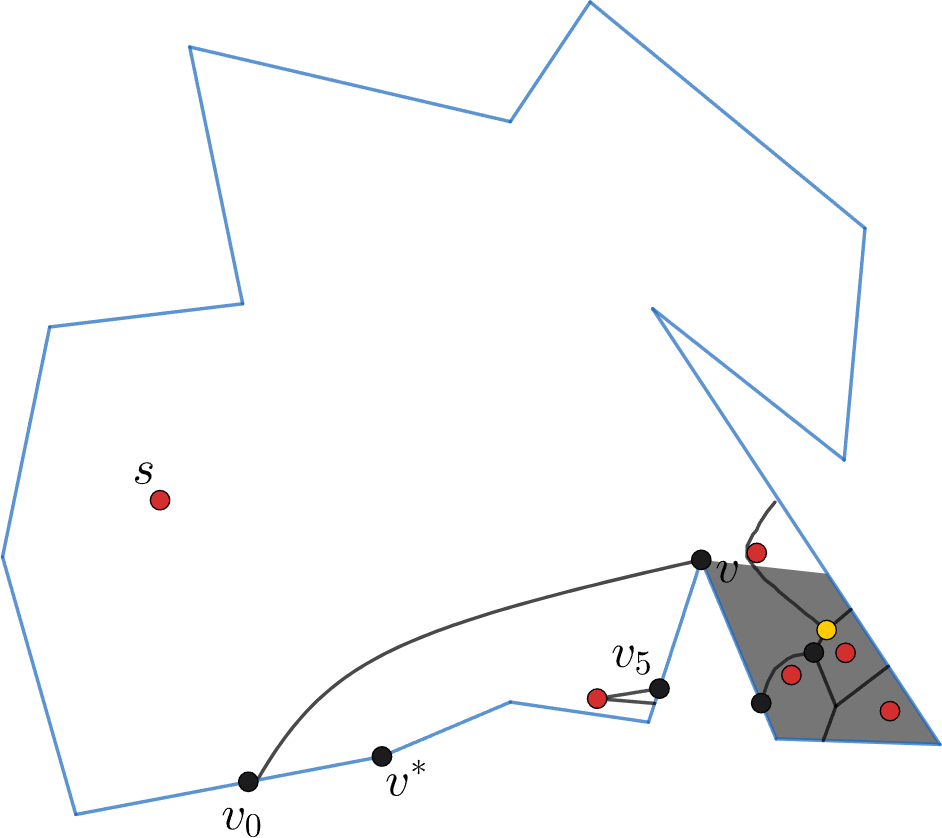}
        \subcaption{$t = t_{sv}$.}
        \label{fig:p_event_t2_1}
    \end{subfigure}
    \hfill
    \begin{subfigure}[b]{.38\textwidth}
        \centering
        \includegraphics[width=\linewidth]{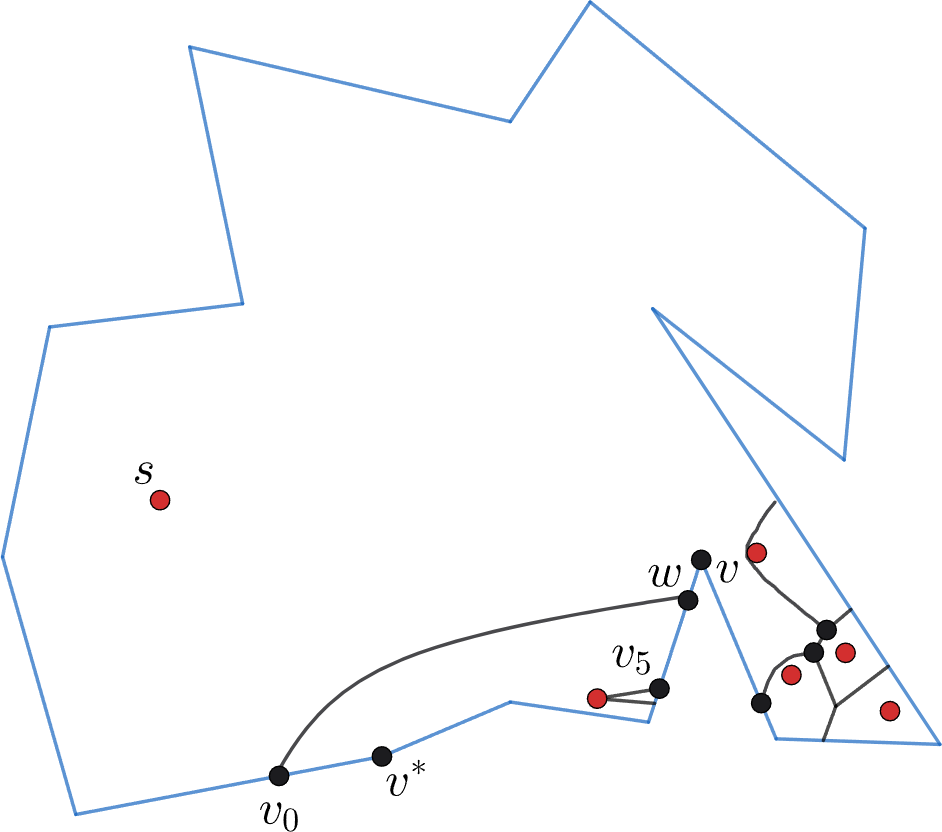}
        \subcaption{$t = t_{sv} + \epsilon$.}
        \label{fig:p_event_t2_2}
    \end{subfigure}
    \caption{Illustration of a pocket event that coincides with a Voronoi vertex event (the golden point in (b)). This is a ``jump to'' case.}
    \label{fig:p_event_t2}
\end{figure}

We describe the case where $u$ is a Voronoi terminal (see Figure~\ref{fig:p_event}); the Voronoi vertex case is analogous (see Figure~\ref{fig:p_event_t2}).

\begin{figure}[t]
    \centering
    \begin{subfigure}[b]{.38\textwidth}
        \centering
        \includegraphics[width=\linewidth]{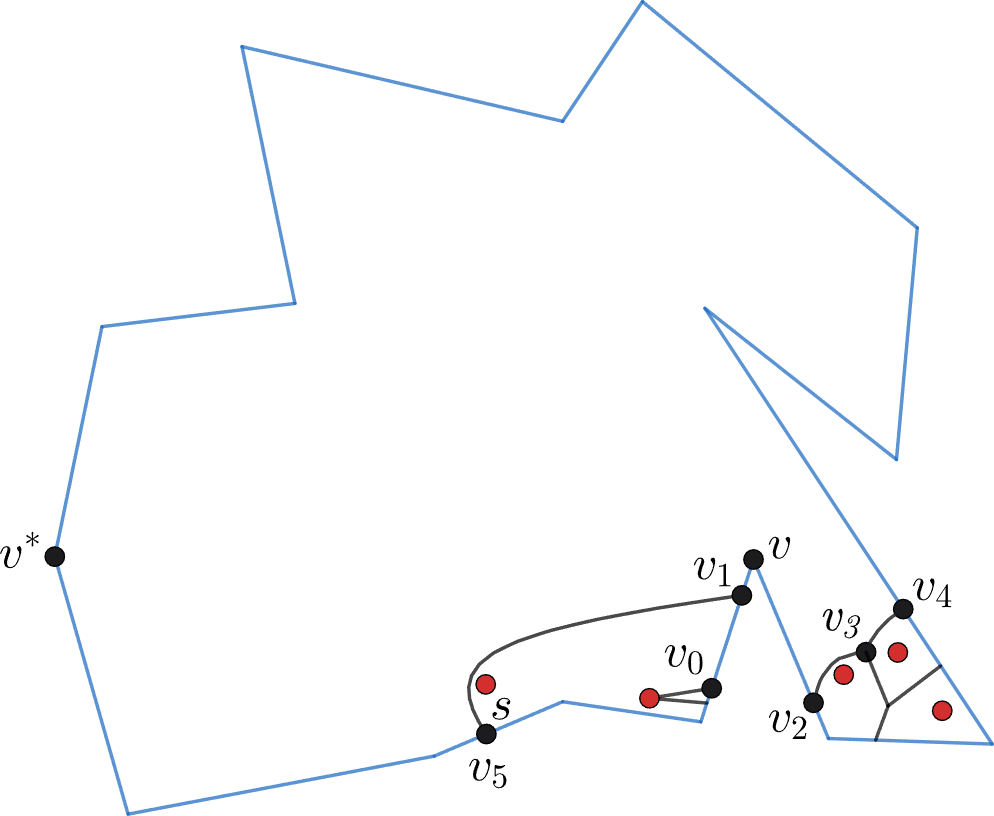}
        \subcaption{$t = t_{sv} - \epsilon$.}
        \label{fig:p_event_t3_0}
    \end{subfigure}
    \hfill
    \begin{subfigure}[b]{.38\textwidth}
        \centering
        \includegraphics[width=\linewidth]{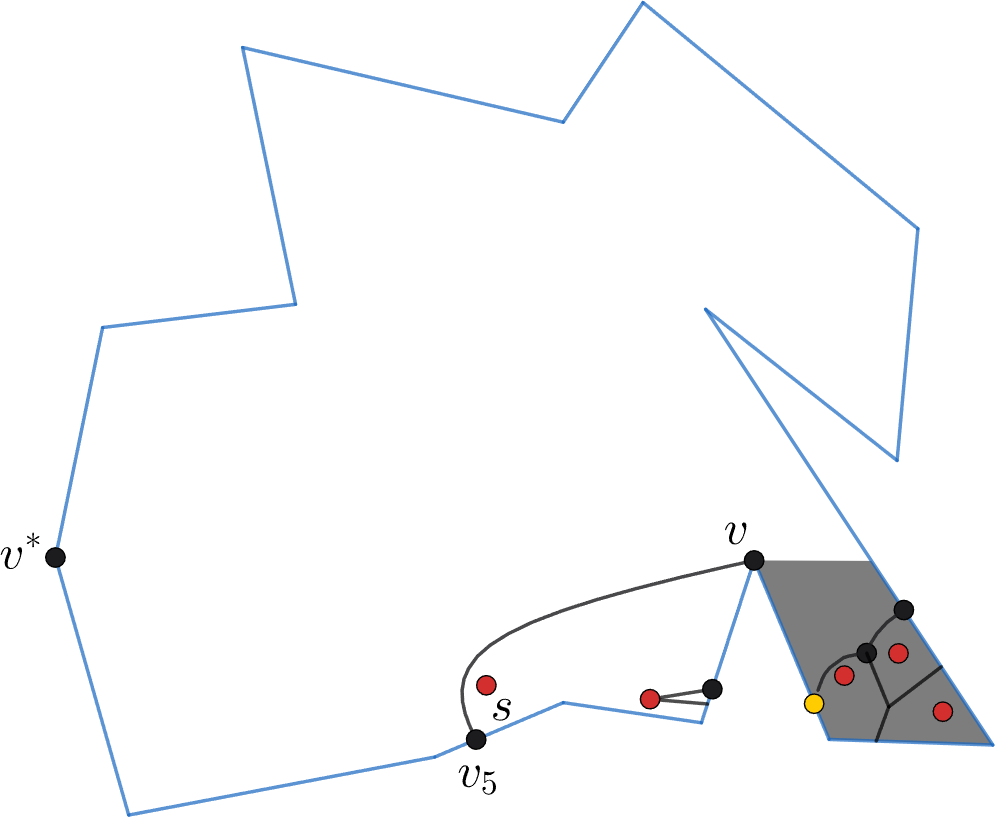}
        \subcaption{$t = t_{sv}$.}
        \label{fig:p_event_t3_1}
    \end{subfigure}
    \hfill
    \begin{subfigure}[b]{.38\textwidth}
        \centering
        \includegraphics[width=\linewidth]{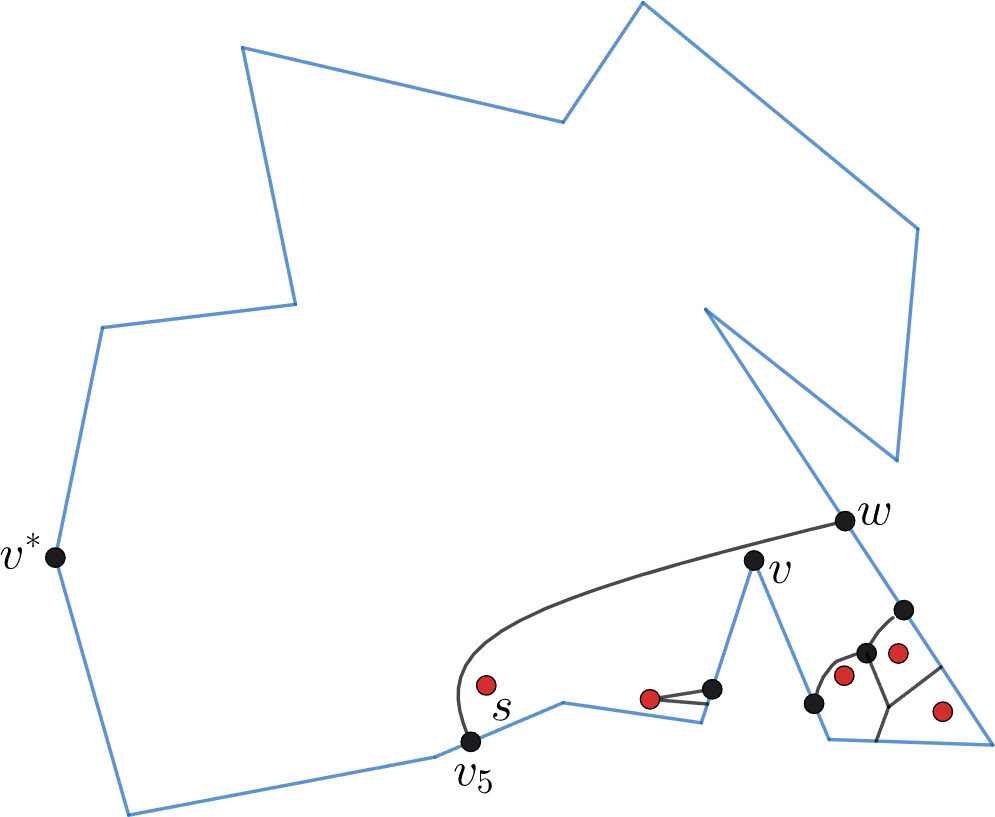}
        \subcaption{$t = t_{sv} + \epsilon$.}
        \label{fig:p_event_t3_2}
    \end{subfigure}
    \caption{Illustration of a pocket event that coincides with a Voronoi terminal event. This is a ``jump from'' case.}
    \label{fig:p_event_t3}
\end{figure}

At $t = t_{sv}$, the bisector $B(v^*, s)$ passes through $v$, which is the apex of $pocket(B(v^*, s))$.
As $t$ increases from $t_{sv} - \epsilon$ to $t_{sv} + \epsilon$, the bisector may ``jump to'' $v$ (see Figure~\ref{fig:p_event}) or ``jump from'' $v$ (see Figure~\ref{fig:p_event_t3}).
We describe the former case; the latter is symmetric.

As $t_{sv}$ also corresponds to a Voronoi terminal event $t_u$, the event occurs because an edge $f$ in $\W_{t_u - \epsilon}$ shrinks to $u$.
Let $e$ and $g$ be the predecessor and successor of $f$ in $\W_{t_u - \epsilon}$ (in clockwise order), and let the endpoints of $e, f, g$ be $v_0, v_1, v_2, v_3$ in clockwise order (see Figure~\ref{fig:p_event}(a)).

Starting from $v_2$, we walk clockwise along $\W_{t_u - \epsilon}$.
For each visited vertex $v_i$, we compute its coordinates at $t = t_{sv}$ in $O(\log^2 m)$ time using Lemma~\ref{lem:vc_compute}.
Since we are not at a site or pocket event for sites associated with $v_i$, the lemma applies.

We test whether $v_i$ lies in $pocket(B(v^*, s))$ by checking whether $v \in \pi(v^*, v_i)$ and $v \in \pi(s, v_i)$, which can be done in $O(\log m)$ time using the GH data structure.
We continue the walk while vertices lie in the pocket.
Let $v_k$ be the first vertex outside the pocket ($k = 5$ in Figure~\ref{fig:p_event}).
The walk takes $O(n_{sv} \log^2 m)$ time, where $n_{sv}$ is the number of Voronoi vertices and terminals inside the pocket.

We then update $\vd_t(S)$ as follows.
We remove $v_1$ and merge the two $\partial P$-portions incident to it.
All vertices and edges between $v_2$ and $v_{k - 1}$ (inclusive), which were associated with $v^*$, are updated to be associated with $s$.
Let $e'$ be the edge between $v_{k - 1}$ and $v_k$; it is a $\partial P$-portion.
We split $e'$ by introducing a new vertex $w$, creating two $\partial P$-portions $(v_{k - 1}, w)$ and $(w, v_k)$.
Finally, we update the endpoint of $e$ to $w$.
These updates take $O(n_{sv})$ time.

To update $\W_t$, we remove all edges between $v_1$ and $v_{k - 1}$ and replace $e'$ with the new edge $(w, v_k)$.
By Lemma~\ref{lem:updateWt}, this takes $O(n_{sv} (\log n + \log^2 m))$ time.

We thus obtain the following result.

\begin{lemma} \label{lem:handle_pocket_event}
Each pocket event can be processed in $O(n_{sv} \cdot (\log n + \log^2 m))$ time, where $n_{sv}$ is the number of Voronoi vertices and terminals inside the pocket.
\end{lemma}

\subsection{The final event} \label{sec:final_event}

To process the final event, we are given $\vd_{t_{v^*} - \epsilon}(S)$ and $\W_{t_{v^*} - \epsilon}$, and our goal is to update them to obtain $\vd_{t_{v^*} + \epsilon}(S)$ and $\W_{t_{v^*} + \epsilon}$.
Note that $\vd_{t_{v^*} + \epsilon}(S) = \vd(S)$ and $\W_{t_{v^*} + \epsilon} = \emptyset$.

\begin{figure}[t]
    \centering
    \begin{subfigure}[b]{.38\textwidth}
        \centering
        \includegraphics[width=\linewidth]{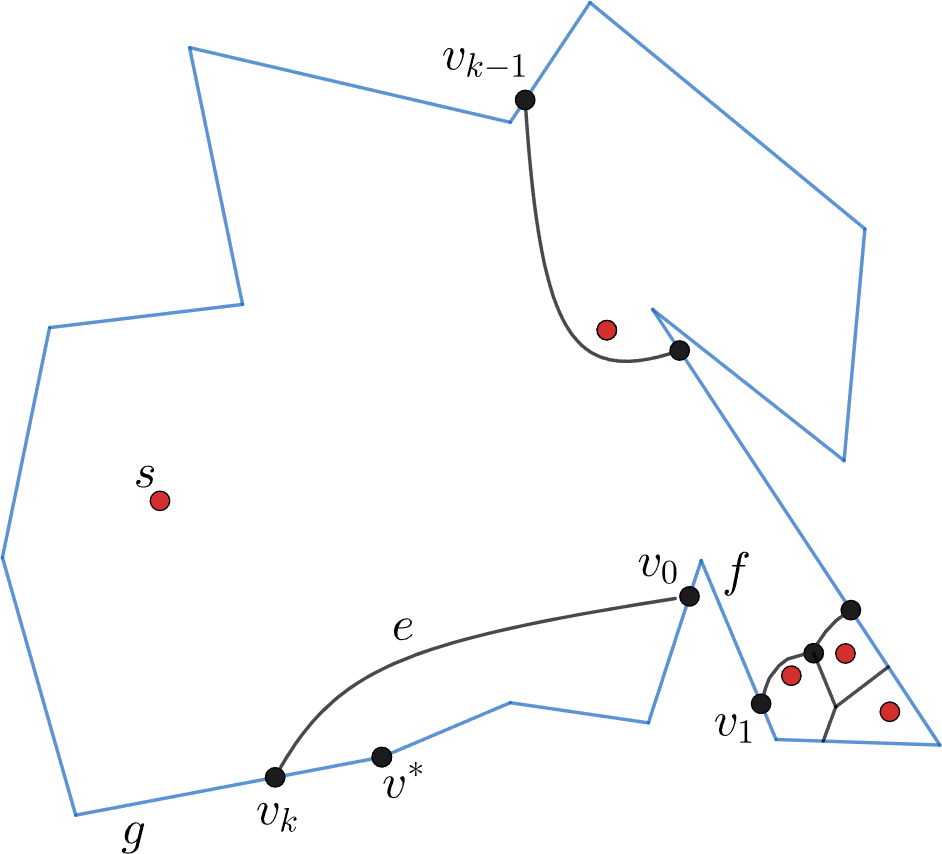}
        \subcaption{$t = t_{v^*} - \epsilon$.}
        \label{fig:f_event_0}
    \end{subfigure}
    \hfill
    \begin{subfigure}[b]{.38\textwidth}
        \centering
        \includegraphics[width=\linewidth]{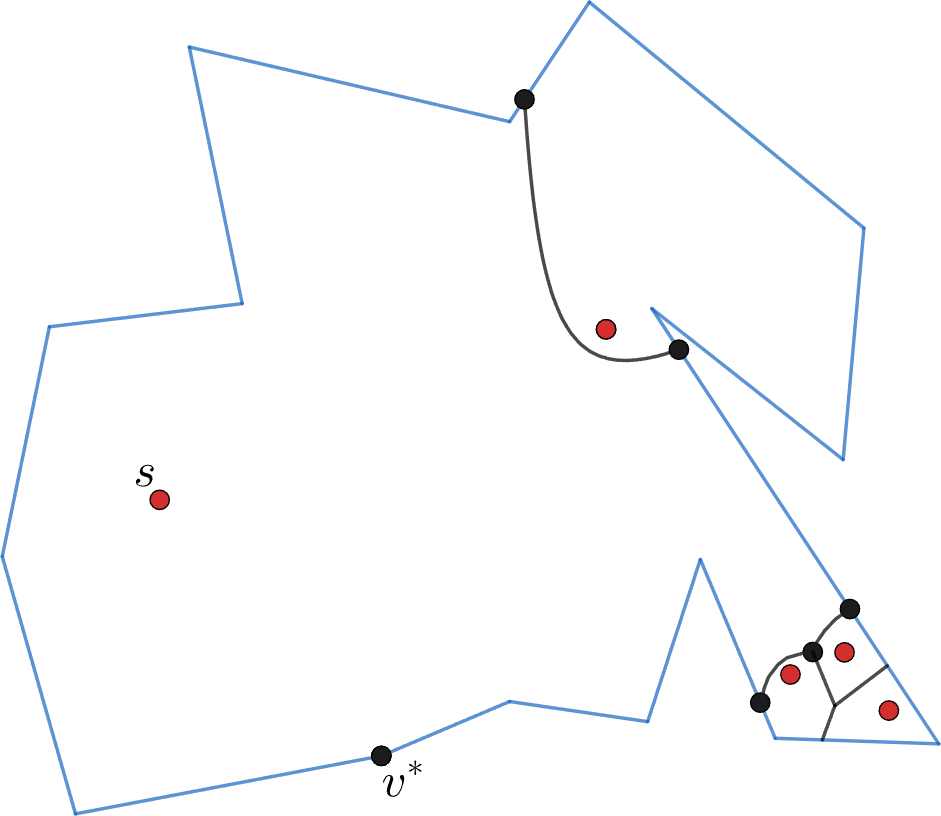}
        \subcaption{$t = t_{v^*}$.}
        \label{fig:f_event_1}
    \end{subfigure}
    \hfill
    \begin{subfigure}[b]{.38\textwidth}
        \centering
        \includegraphics[width=\linewidth]{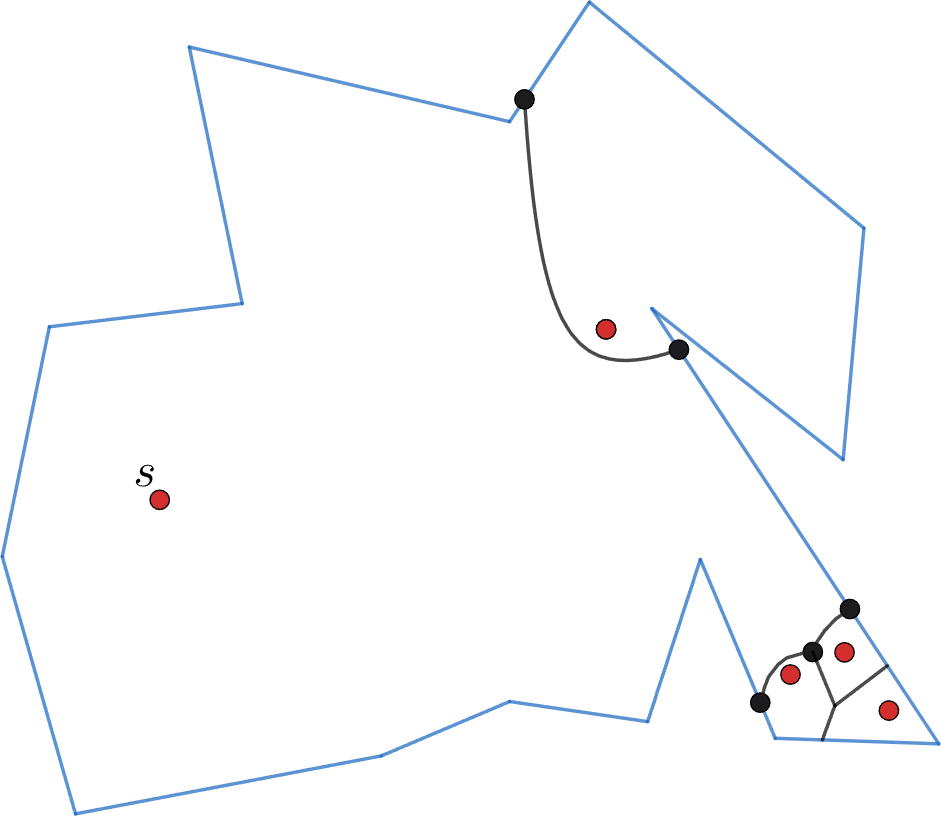}
        \subcaption{$t = t_{v^*} + \epsilon$.}
        \label{fig:f_event_2}
    \end{subfigure}
    \caption{Illustration of the final event.}
    \label{fig:f_event}
\end{figure}

Let $s \in S$ be the site associated with the final event (see Figure~\ref{fig:f_event}), i.e.,
$s = \arg\min_{s' \in S} d_{s'}(v^*)$.
At $t = t_{v^*} - \epsilon$, the region $\R_{t_{v^*} - \epsilon}(s)$ is the only neighbor of $\R_{t_{v^*} - \epsilon}(v^*)$ in $\vd_{t_{v^*} - \epsilon}(S)$.
Thus, $\W_{t_{v^*} - \epsilon}$ consists of three edges: a $\partial P$-portion starting at $v^*$, a Voronoi edge $e$ associated with $s$, and a $\partial P$-portion ending at $v^*$.

Let $f$ and $g$ be the neighbors of $e$ in $\R_{t_{v^*} - \epsilon}(s)$.
Both $f$ and $g$ are $\partial P$-portions, and it is possible that $f = g$.
We update $\vd_t(S)$ as follows.

If $f \neq g$, we delete the region $\R_t(v^*)$ together with all its incident edges and vertices, including $e$.
We then simply merge $f$ and $g$ as a single edge.

If $f = g$, then $\R(s)$ is the only nonempty Voronoi region in $\vd(S) = \vd_{t_{v^*} + \epsilon}(S)$.
In this case, $\vd(S)$ consists of a single region, namely $\R(s) = P$.

In both cases, we set $\W_{t_{v^*} + \epsilon} = \emptyset$.
All updates take $O(1)$ time.

\begin{lemma} \label{lem:handle_final_event}
The final event can be processed in $O(1)$ time.
\end{lemma}

\section{Time analysis and algorithm correctness} \label{sec:wrap_up}

We analyze the running time in Section~\ref{sec:time} and prove correctness in Section~\ref{sec:correctness}.

\subsection{Time analysis} \label{sec:time}

We maintain all events in a priority queue $Q$.
We will show that $|Q| = O(n)$ throughout the algorithm, so each insertion or deletion takes $O(\log n)$ time.

\paragraph{Site events.}
At the beginning of the algorithm, we compute all site events in $O(n \log m)$ time using the GH data structure and insert them into $Q$.

\paragraph{Voronoi vertex events.}
As discussed in Section~\ref{sec:voronoi_vertex_events}, each Voronoi vertex event $t_v$ corresponds to three adjacent edges $e, f, g$ in $\W_t$ at time $t = t_v - \epsilon$, with distinct associated sites $s_e, s_f, s_g$.
We detect such events as follows.

Whenever $\W_t$ is updated, for each new triple of adjacent Voronoi edges $e, f, g$ in $\W_t$ (in order under $\lw$), we compute the intersections $B(s_e, s_f) \cap B(s_f, s_g)$.
Since $B(s_e, s_f) \cap \partial P$, $B(s_e, s_g) \cap \partial P$, and $B(s_f, s_g) \cap \partial P$ are already stored as discussed in Section~\ref{sec:representation}, we can compute these intersections in $O(\log m)$ time using Lemma~\ref{lem:bis_branch_int} (in fact, here $O(\log^2 m)$ suffices).

Since there may be up to two intersection points, we identify the eligible one as follows.
For each candidate point $v$, we compute the anchors $a_e, a_f, a_g$ of $v$ in $\pi(v, s_e)$, $\pi(v, s_f)$, and $\pi(v, s_g)$ in $O(\log m)$ time each.
If these anchors appear in counterclockwise order around $v$, then $v$ is eligible, and we insert $t_v = d_{s_e}(v) - d(v^*, v)$ into $Q$ as a candidate Voronoi vertex event.
At most one intersection point is eligible.

However, if before $t_v$ is processed, the neighboring status of $e$, $f$, and $g$ changes during the algorithm (e.g., one of them is deleted due to the processing of other events), then $t_v$ is not a true event, and we simply remove $t_v$ from $Q$.

For the analysis, note that removing a candidate event from $Q$ is triggered by a true event.
Each site, Voronoi vertex, or terminal event affects only $O(1)$ triples of adjacent edges, and thus causes $O(1)$ candidate removals.
For pocket events, the number of affected triples is proportional to the number of pocket Voronoi vertices and terminals in the pocket.
Since the total number of Voronoi vertices and terminals is $O(n)$, the total number of candidate vertex events generated is $O(n)$.

Hence, there are $O(n)$ candidate Voronoi vertex events overall, and each takes $O(\log m)$ time to compute and $O(\log n)$ time to insert/delete from $Q$.

\paragraph{Voronoi terminal events.}
As discussed in Section~\ref{sec:voronoi_terminal_events}, each Voronoi terminal event $t_u$ corresponds to three adjacent edges $e, f, g$ in $\W_t$ at time $t = t_u - \epsilon$.
We distinguish the two cases depending on whether $t_u$ is a type-1 or type-2 event.

\begin{itemize}
\item
In type-1 events, $f$ is a $\partial P$-portion, and $e$ and $g$ are Voronoi edges.
We detect such events as follows.

Whenever $\W_t$ is updated, we identify new triples of edges $e, f, g$ in $\W_t$ satisfying the following conditions:
(1) they are adjacent and ordered according to $\lw$;
(2) $f$ is a $\partial P$-portion, while $e$ and $g$ are Voronoi edges; and
(3) $s_e \neq s_g$.
For each such triple, we compute the intersection point $u$ between the right half-bisector $B_r(s_e, s_g)$ and $\partial P$, which can be done in $O(\log^2 m)$ time by Lemma~\ref{lem:bpp}.
We then insert $t_u = d_{s_e}(u) - d(v^*, u)$ into $Q$ as a candidate Voronoi terminal event.

As in the Voronoi vertex event case, we remove $t_u$ from $Q$ whenever the adjacency of $e, f, g$ changes during the algorithm.

\item
In type-2 events, $f$ is a Voronoi edge and exactly one of $e$ and $g$ is a Voronoi edge.
We describe the subcase where $e$ is a Voronoi edge and $g$ is a $\partial P$-portion; the other subcase is symmetric.

Whenever $\W_t$ is updated, we identify new triples of edges $e, f, g$ in $\W_t$ satisfying the following conditions:
(1) they are adjacent and ordered according to $\lw$;
(2) $e$ and $f$ are Voronoi edges, while $g$ is a $\partial P$-portion; and
(3) $s_e \neq s_f$.
For each such triple, we compute the intersection of the left half-bisector $B_l(s_e, s_f)$ with $\partial P$, which can be done in $O(\log^2 m)$ time by Lemma~\ref{lem:bpp}.
We then insert $t_u = d_{s_e}(u) - d(v^*, u)$ into $Q$ as a candidate Voronoi terminal event.

As before, we remove $t_u$ from $Q$ whenever the adjacency of $e, f, g$ changes during the algorithm.
\end{itemize}

For the time analysis, by an argument similar to that for Voronoi vertex events, the total number of candidate Voronoi terminal events throughout the algorithm is $O(n)$.
Each event can be computed in $O(\log^2 m)$ time, and inserting or removing an event from $Q$ takes $O(\log n)$ time.

\paragraph{Putting it all together.}
According to the above discussion, there are $O(n)$ events in total and thus $|Q| = O(n)$.
Combining the above with Lemmas~\ref{lem:handle_site_event}, \ref{lem:handle_voronoi_vertex_event}, \ref{lem:handle_cap_events}, \ref{lem:handle_pocket_event}, and~\ref{lem:handle_final_event}, the total running time is
\[
O\bigl(n (\log n \log m + \log^2 m)\bigr).
\]
This simplifies to $O(n \log^2 m)$ when $n < m$.

If $n > m$, we can instead construct $\vd(S)$ explicitly in $O((n+m)\log(n+m)) = O(n \log n)$ time using the algorithm in~\cite{ref:HershbergerAn99}.
Combining both cases yields an overall construction time of
\[
O(n \log n + n \log^2 m),
\]
as stated in Theorem~\ref{thm:mainthm}.

\paragraph{Computing $\boldsymbol{\vd(S)}$ explicitly.}
Given the implicit representation of $\vd(S)$, we can construct it explicitly in an additional $O(m + n \log m)$ time.
For each Voronoi edge $e$, as discussed in~\cite{ref:OhVo20} for a similar problem, we can compute it explicitly in $O(\log m + k_e)$ time using the GH data structure, where $k_e$ is its combinatorial complexity.
Since there are $O(n)$ edges and $\sum_e k_e = O(m + n)$, the total time is $O(m + n \log m)$.
Thus, we can compute $\vd(S)$ explicitly in $O(m + n \log n + n \log^2 m)$
time overall.

\subsection{Correctness} \label{sec:correctness}

We now prove the correctness of our algorithm.
In particular, we prove Lemma~\ref{lem:sweeping}.
We first give an overview of the proof.

\subsubsection{Proof overview}
\label{sec:proofoverview}

Let $e$ be an edge of $\vd(S)$ that is a $\partial P$-portion, and let $s$ be the site associated with $e$ (i.e., $e$ lies on the boundary of $\R(s)$).
If $\R(s) = P$, then $e = \partial P$.
Otherwise, $e$ is not a cycle.
Let $u$ and $v$ be the endpoints of $e$, and let $t_u$ and $t_v$ be the Voronoi terminal events of $u$ and $v$, respectively.
Let $\seg{s s'}$ be the last-edge extension of $\pi(v^*, s)$.

We distinguish four cases.

\begin{description}
\item[Case~1: $e \neq \partial P$ and $v^* \in e$.]
In this case, we show that at the events $t_u$ and $t_v$, two vertices $u'$ and $v'$ are added to $\W_t$ that trace $e$ from $u$ and $v$ toward $v^*$ from opposite directions (counterclockwise and clockwise around $P$), respectively.
These traces terminate at the final event $t_{v^*}$.

\item[Case~2: $e \neq \partial P$, $v^* \notin e$, and $s' \in e$.]
In this case, we show that at the site event $t_s$, two vertices $u'$ and $v'$ are added to $\W_t$ that trace $e$ from $s'$ toward $u$ and $v$, respectively.
These traces terminate at the events $t_u$ and $t_v$.

\item[Case~3: $e \neq \partial P$, $v^* \notin e$, and $s' \notin e$.]
In this case, we show that at one of the events $t_u$ or $t_v$ (without loss of generality, say $t_u$), a vertex $u'$ is added to $\W_t$ that traces $e$ from $u$ toward $v$.
This trace terminates at the event $t_v$.

A special situation occurs when $t_u = t_v$, in which case both are pocket Voronoi terminal events and the trace of $e$ is instantaneous.

\item[Case~4: $e = \partial P$.]
In this case, we show that at the site event $t_s$, two vertices $u'$ and $v'$ are added to $\W_t$ that trace $e$ from $s'$ toward $v^*$ in opposite directions (counterclockwise and clockwise), respectively.
Both traces terminate at the final event $t_{v^*}$.
This case can be viewed as a special instance of Case~1.
\end{description}

The detailed proof for the $\partial P$-portion case is given in Section~\ref{sec:boundary}.

The analysis for Voronoi edges is similar.
Let $e$ be a Voronoi edge in $\vd(S)$, and let $s_1$ and $s_2$ be the two sites associated with $e$.
Let $u$ and $v$ be the endpoints of $e$.
Let $\seg{s_1 s_1'}$ and $\seg{s_2 s_2'}$ be the last-edge extensions of $\pi(v^*, s_1)$ and $\pi(v^*, s_2)$, respectively.

We will show that at most one of $\seg{s_1 s_1'}$ and $\seg{s_2 s_2'}$ intersects $e$.
Accordingly, we distinguish two cases.

\begin{description}
\item[Case~1: Exactly one of $\seg{s_1 s_1'}$ and $\seg{s_2 s_2'}$ intersects $e$.]
Without loss of generality, assume that $\seg{s_1 s_1'}$ intersects $e$, and let $z$ be the intersection point.
We show that during the site event $t_{s_1}$, two vertices $u'$ and $v'$ are added to $\W_t$ that trace $e$ from $z$ toward $u$ and $v$, respectively.
These traces terminate at the events $t_u$ and $t_v$.

\item[Case~2: Neither $\seg{s_1 s_1'}$ nor $\seg{s_2 s_2'}$ intersects $e$.]
We show that during one of the events $t_u$ or $t_v$ (without loss of generality, say $t_u$), a vertex $u'$ is added to $\W_t$ that traces $e$ from $u$ toward $v$.
This trace terminates at the event $t_v$.

A special situation occurs when $t_u = t_v$, in which case both are pocket Voronoi vertex/terminal events and the trace is instantaneous.
\end{description}

For convenience, we introduce the following definition.

\begin{definition}
For a site $s \in S$ and a point $p \in P$, define
\[
\delta_s(p) = d_s(p) - d(v^*, p).
\]
\end{definition}

With this definition, we obtain the following observation and lemma, both of which will be used frequently.
Lemma~\ref{lem:crossing} follows from the triangle inequality.

\begin{observation} \label{obs:delta_time}
If $p \in \R(s)$, then at time $t = \delta_s(p)$, we have $p \in \W_t$.
\end{observation}

\begin{proof}
Since $p \in \R(s)$, we have $d_s(p) = d_S(p)$.
At time $t = \delta_s(p)$,
\[
d_{v^*}(p) = d(v^*, p) + \delta_s(p) = d_s(p) = d_S(p),
\]
and thus $p \in \W_t$.
\end{proof}

\begin{lemma} \label{lem:crossing}
For any two points $p_1, p_2 \in P$ and a site $s \in S$ such that $\pi(v^*, p_1)$ and $\pi(s, p_2)$ intersect, we have
\[
\delta_s(p_1) \le \delta_s(p_2).
\]
\end{lemma}

\begin{proof}
It suffices to show that
\[
d(s, p_1) - d(v^*, p_1) \le d(s, p_2) - d(v^*, p_2),
\]
since the lemma follows by adding $w(s)$ to both sides.

Let $q$ be a point in $\pi(v^*, p_1) \cap \pi(s, p_2)$. Then
\begin{align*}
d(s, p_1) - d(v^*, p_1)
&= d(s, p_1) - d(v^*, q) - d(q, p_1) && \text{since } q \in \pi(v^*, p_1) \\
&\le d(s, q) + d(q, p_1) - d(v^*, q) - d(q, p_1) && \text{by triangle inequality} \\
&= d(s, q) - d(v^*, q) \\
&= d(s, q) + d(q, p_2) - d(v^*, q) - d(q, p_2) \\
&\le d(s, q) + d(q, p_2) - d(v^*, p_2) && \text{by triangle inequality} \\
&= d(s, p_2) - d(v^*, p_2) && \text{since } q \in \pi(s, p_2).
\end{align*}
\end{proof}

\subsubsection{Analysis of the $\partial P$-portion case}
\label{sec:boundary}

\begin{figure}[t]
    \centering
    \begin{subfigure}[b]{.49\textwidth}
        \centering
        \includegraphics[scale=.9]{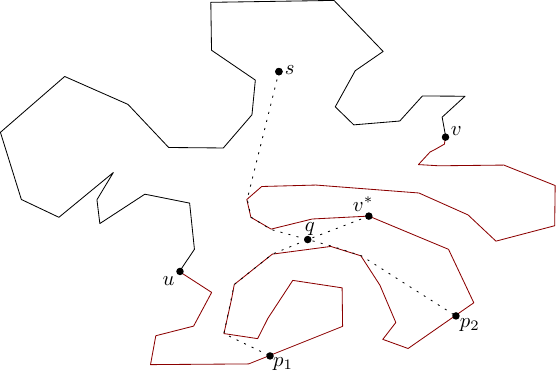}
        \subcaption{Case~1}
        \label{fig:boundary_monotone_inside}
    \end{subfigure}
    \hfill
    \begin{subfigure}[b]{.49\textwidth}
        \centering
        \includegraphics[scale=.9]{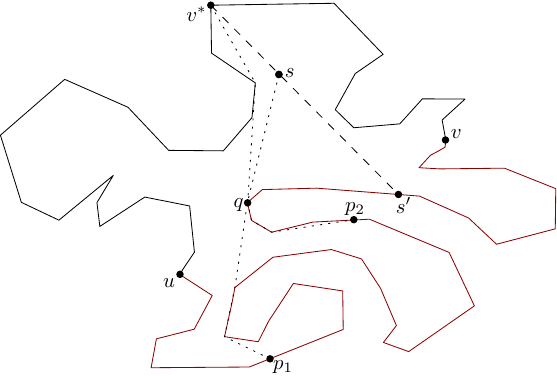}
        \subcaption{Case~2}
        \label{fig:bound_mon_no_cross}
    \end{subfigure}

    \begin{subfigure}[b]{.49\textwidth}
        \centering
        \includegraphics[scale=.9]{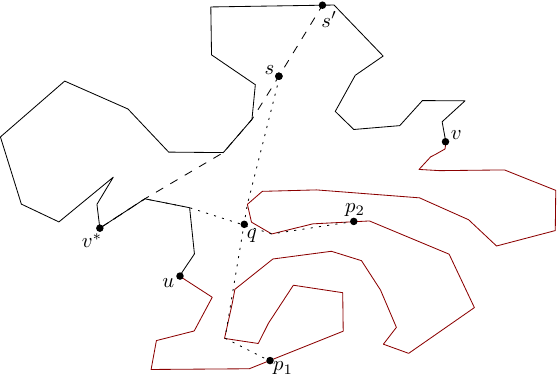}
        \subcaption{Case~3}
        \label{fig:bound_mon_cross}
    \end{subfigure}
    \hfill
    \begin{subfigure}[b]{.49\textwidth}
        \centering
        \includegraphics[scale=.9]{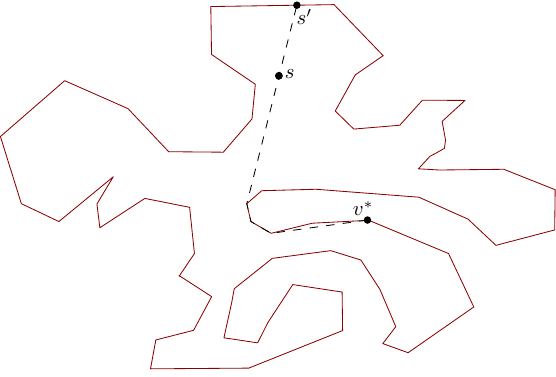}
        \subcaption{Case~4}
        \label{fig:bound_mon_one_region}
    \end{subfigure}
    \caption{Illustration of the $\partial P$-portion case.}
    \label{fig:bound_mon}
\end{figure}

We now analyze the case where $e$ is a $\partial P$-portion.
We follow the notation introduced in Section~\ref{sec:proofoverview}, namely $s$, $u$, $v$, $t_s$, $t_u$, and $t_v$.
We give detailed proofs for the four cases described in Section~\ref{sec:correctness}.
See Figure~\ref{fig:bound_mon} for illustrations.

When $e \neq \partial P$ (i.e., Cases~1–3), we orient $e$ from $u$ to $v$, and assume without loss of generality that this corresponds to the counterclockwise direction along $\partial P$.
We define an order $<_e$ on $e$ such that for any $p, q \in e$, $p <_e q$ if $p$ comes before $q$ in the traversal of $e$ from $u$ to $v$.

\paragraph{Case~1.}
In Case~1, $e \neq \partial P$ and $v^* \in e$ (see Figure~\ref{fig:boundary_monotone_inside}).
By definition, $u <_e v^* <_e v$.

Consider two points $p_1, p_2 \in e$ such that $u \le_e p_1 <_e p_2 \le_e v^*$.
Let $R$ be the region bounded by $\pi(s, u)$, $\pi(s, v)$, and $e$.
Since $\pi(v^*, p_1)$ separates $R$ into two subregions, one containing $s$ and the other containing $p_2$, it must intersect $\pi(s, p_2)$.
By Lemma~\ref{lem:crossing}, we obtain $\delta_s(p_1) \le \delta_s(p_2)$.

By Observation~\ref{obs:delta_time}, a point $p \in e$ lies on $\W_t$ exactly when $t = \delta_s(p)$.
Thus, $\delta_s(p_1) \le \delta_s(p_2)$ implies that $\W_t$ traces $e$ from $u$ to $v^*$ monotonically.
In particular, a vertex $u'$ is added at time $t_u = \delta_s(u)$ and traces $e$ until time $t_{v^*} = \delta_s(v^*)$.

We can also see that if $\delta_s(p_1) = \delta_s(p_2)$, then at $t = \delta_s(p_1)$,
$p_1, p_2 \in B(s, v^*)$.  This is only possible during the pocket event $t_{sv}$ for some vertex $v$ of $P$ such that both $p_1$ and $p_2$ are in $pocket(B(s, v^*))$.
Note also that $\delta_s(p)$ varies continuously as a point $p$ traces $e$ from $u$ to $v^*$.

From the above discussions, we can conclude that $u'$ traces $e$ from $u$ to $v^*$ continuously,
except at pocket events, when it may jump forward along $e$ in the same direction as it was tracing.

By symmetry, at $t_v$, a vertex $v'$ is added that traces $e$ from $v$ toward $v^*$ in the opposite direction.

\paragraph{Case~2.}
In Case~2, $e \neq \partial P$, $v^* \notin e$, and $s' \in e$ (see Figure~\ref{fig:bound_mon_no_cross}).

Consider $p_1, p_2 \in e$ with $u \le_e p_1 <_e p_2 \le_e s'$.
The curves $\pi(s, p_1)$ and $\seg{s s'}$ partition $P$ into two regions, one containing $v^*$ and the other containing $p_2$.
Thus, $\pi(v^*, p_2)$ must intersect either $\pi(s, p_1)$ or $\seg{s s'}$.

We claim it intersects $\pi(s, p_1)$.
If $\pi(v^*, p_2)$ intersects $\seg{s s'}$, then since $\pi(v^*, s') = \pi(v^*, s) \cup \seg{s s'}$ and shortest paths from the same source do not cross, $\pi(v^*, p_2)$ must share a prefix with $\pi(v^*, s')$, implying it passes through $s$.
Hence it intersects $\pi(s, p_1)$.

Since $\pi(v^*, p_2)$ intersects $\pi(s, p_1)$, by Lemma~\ref{lem:crossing}, we have $\delta_s(p_1) \ge \delta_s(p_2)$.

By a similar analysis as in Case~1, we can conclude that at the site event $t_s$,
a vertex $u'$ is added to $\W_t$ at $s'$, and then $u'$ traces $e$ from $s'$ to $u$ where it is removed by the Voronoi terminal event $t_u$. During the trace, $u'$ moves along $e$ toward $u$ continuously, except during pocket events, where it may jump forward along $e$ in the same direction as it was tracing.

Symmetrically, at $t=t_s$, a vertex $v'$ is added to $\W_t$ that traces $e$ from $s'$ to $v$ in a similar fashion.

\paragraph{Case~3.}
In Case~3, $e \neq \partial P$, $v^* \notin e$, and $s' \notin e$ (see Figure~\ref{fig:bound_mon_cross}).
In this case, $s'$ lies on $\partial P$ either counterclockwise from $v$ to $v^*$ or counterclockwise from $v^*$ to $u$.
Without loss of generality, we assume the former case.

Let $p_1, p_2 \in e$ with $u \le_e p_1 <_e p_2 \le_e v$.
The path $\pi(v^*, p_2)$ partitions $P$ into two regions.
Let $P_1$ and $P_2$ denote the regions to the right and left, respectively.

We claim $p_1 \in P_1$ and $s \in P_2$.
Since the order along $e$ matches $\lp$, we have $p_1 <_{\partial P} p_2 <_{\partial P} s'$.
Thus $s' \in P_2$, and since $\pi(v^*, s')$ cannot cross $\pi(v^*, p_2)$, it follows that $s \in P_2$.
Similarly, $p_1 \in P_1$.

Hence $\pi(s, p_1)$ intersects $\pi(v^*, p_2)$, implying $\delta_s(p_1) \ge \delta_s(p_2)$.

By a similar analysis as before, we can conclude that at the Voronoi terminal event $t_v$,
a vertex $v'$ is added to $\W_t$ at $v$, and then $v'$ traces $e$ from $v$ to $u$ where it is removed by the Voronoi terminal event $t_u$. During the trace, $v'$ moves along $e$ continuously, except at pocket events, where it may jump forward along $e$ in the same direction as it was tracing.

Note that a special case where $t_u = t_v$ is possible. In this case, $t_u$ and $t_v$ are pocket Voronoi terminal events,
and the trace of $e$ is instantaneous.

\paragraph{Case~4.}
In Case~4, $e = \partial P$ (see Figure~\ref{fig:bound_mon_one_region}).
At time $t_s$, two vertices are added at $s'$ and trace $e$ in opposite directions toward $v^*$.
Both traces terminate at $t_{v^*}$.

This case can be viewed as a special case of Case~1 by treating $s'$ as both endpoints of $e$.

\subsubsection{Analysis of Voronoi edges}

Before analyzing our problem, we first establish several properties of the bisector $B(s_1, s_2)$ for any two sites $s_1, s_2 \in S$.
Let $\seg{s_1 s_1'}$ and $\seg{s_2 s_2'}$ be the last-edge extensions of $\pi(v^*, s_1)$ and $\pi(v^*, s_2)$, respectively.

We begin with the following observation.

\begin{observation} \label{obs:one_site_event}
It is not possible for both $\seg{s_1 s_1'}$ and $\seg{s_2 s_2'}$ to intersect $B(s_1, s_2)$.
Moreover, each of $\seg{s_1 s_1'}$ and $\seg{s_2 s_2'}$ intersects $B(s_1, s_2)$ at most once.
\end{observation}

\begin{proof}
Assume for contradiction that both $\seg{s_1 s_1'}$ and $\seg{s_2 s_2'}$ intersect $B(s_1, s_2)$.
Since $\seg{s_1 s_1'}$ is a shortest path extension, it intersects $B(s_1, s_2)$ at most once by Observation~\ref{obs:star_shaped}; let $x$ denote this intersection.
Similarly, let $y$ be the unique intersection of $\seg{s_2 s_2'}$ with $B(s_1, s_2)$.

We claim that among all points $p \in B(s_1, s_2)$, the value $\delta_{s_1}(p)$ is minimized uniquely at $p = x$.
Indeed, for any $p \in B(s_1, s_2)$, by the triangle inequality,
\[
\delta_{s_1}(p)
= d(s_1, p) + w(s_1) - d(v^*, p)
\ge d(s_1, p) + w(s_1) - d(v^*, s_1) - d(s_1, p)
= w(s_1) - d(v^*, s_1).
\]
Equality holds if and only if $s_1 \in \pi(v^*, p)$, which occurs precisely when $p = x$.
Thus, $x$ uniquely minimizes $\delta_{s_1}(p)$ over $p \in B(s_1, s_2)$.

By the same argument, $y$ uniquely minimizes $\delta_{s_2}(p)$ over $p \in B(s_1, s_2)$.
However, for any $p \in B(s_1, s_2)$, we have $\delta_{s_1}(p) = \delta_{s_2}(p)$, and hence $x = y$.

Now observe that
\[
\pi(v^*, x) = \pi(v^*, s_1) \cup \seg{s_1 x}
\quad \text{and} \quad
\pi(v^*, y) = \pi(v^*, s_2) \cup \seg{s_2 y}.
\]
Since $x = y$, the two paths coincide, and thus their last edge contains $s_1$, $s_2$, and $x$.
This implies that either $s_1 \in \seg{s_2 y}$ or $s_2 \in \seg{s_1 x}$.
Without loss of generality, assume $s_1 \in \seg{s_2 y}$.
Then $s_1$, $s_2$, and the anchor of $s_2$ in $\pi(v^*, s_2)$ are collinear, contradicting our general position assumption that no three points in $S \cup V_P$ are collinear.
\end{proof}

Observation~\ref{obs:one_site_event} implies that there are two cases:
(1) exactly one of $\seg{s_1 s_1'}$ and $\seg{s_2 s_2'}$ intersects $B(s_1, s_2)$; and
(2) neither $\seg{s_1 s_1'}$ nor $\seg{s_2 s_2'}$ intersects $B(s_1, s_2)$.
The following is a key lemma in our analysis.

\begin{lemma}\label{lem:bisector_event}
\begin{enumerate}
\item
If $\seg{s_1 s_1'}$ intersects $B(s_1, s_2)$ at a point $x$, then $\delta_{s_1}(p)$ is non-decreasing as $p$ moves from $x$ to either endpoint of $B(s_1, s_2)$; similarly, if $\seg{s_2 s_2'}$ intersects $B(s_1, s_2)$ at a point $y$, then $\delta_{s_2}(p)$ is non-decreasing as $p$ moves from $y$ to either endpoint of $B(s_1, s_2)$.

\item
If neither $\seg{s_1 s_1'}$ nor $\seg{s_2 s_2'}$ intersects $B(s_1, s_2)$, then $\delta_{s_1}(p)$ is non-decreasing as $p$ moves along $B(s_1, s_2)$ from one endpoint to the other.
\end{enumerate}
\end{lemma}

\begin{proof}
Assuming that exactly one of $\seg{s_1 s_1'}$ and $\seg{s_2 s_2'}$ intersects $B(s_1, s_2)$, we prove the first statement of the lemma.
We only prove the case in which $\seg{s_1 s_1'}$ intersects $B(s_1, s_2)$ at a point $x$, since the other case can be proved analogously.

Let $u$ and $v$ be the two endpoints of the bisector $B(s_1, s_2)$. Note that both $u$ and $v$ lie on $\partial P$.
Consider any two points $p_1, p_2 \in B(s_1, s_2)$ that lie between $x$ and $v$ such that $p_2$ lies between $p_1$ and $v$, i.e., $x$, $p_1$, $p_2$, and $v$ appear in this order along $B(s_1, s_2)$.
Note that it is possible that $p_1 = x$, and also possible that $p_2 = v$.
See Figure~\ref{fig:bisector_monotone_site} for an illustration.

We make the following \emph{key claim}: $\pi(v^*, p_1)$ intersects $\pi(s_1, p_2) \cup \pi(s_2, p_2)$.
Before proving it, we first use it to prove the first statement of the lemma.

By Lemma~\ref{lem:crossing}, if $\pi(v^*, p_1)$ intersects $\pi(s_1, p_2)$, then $\delta_{s_1}(p_1) \le \delta_{s_1}(p_2)$.
Similarly, if $\pi(v^*, p_1)$ intersects $\pi(s_2, p_2)$, then $\delta_{s_2}(p_1) \le \delta_{s_2}(p_2)$.
Since $p_1, p_2 \in B(s_1, s_2)$, we have
\[
\delta_{s_1}(p_1) = \delta_{s_2}(p_1)
\quad\text{and}\quad
\delta_{s_1}(p_2) = \delta_{s_2}(p_2).
\]
Therefore, in either case, we obtain $\delta_{s_1}(p_1) \le \delta_{s_1}(p_2)$.
This implies that $\delta_{s_1}(p)$ is non-decreasing as $p$ moves from $x$ to $v$.
Symmetrically, $\delta_{s_1}(p)$ is non-decreasing as $p$ moves from $x$ to $u$.
This proves the first statement of the lemma.

We now prove the key claim.
First, if $p_1 = x$, then since $\pi(v^*, x) = \pi(v^*, s_1) \cup \overline{s_1 x}$, the path $\pi(v^*, x)$ contains $s_1$ and therefore intersects $\pi(s_1, p_2)$.
Hence, the key claim holds in this case.
In the remainder of the proof, we assume that $p_1 \neq x$.

\begin{figure}[t]
    \centering

    \begin{subfigure}[b]{\linewidth}
        \centering
        \includegraphics[scale=.9]{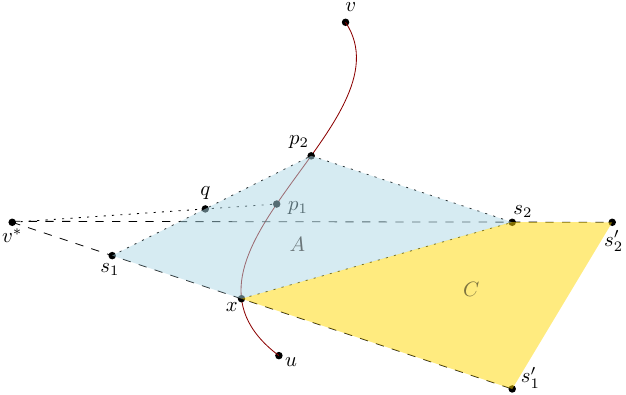}
        \subcaption{The two regions $A$ and $B$ do not overlap.}
        \label{fig:bisector_monotone_sitea}
    \end{subfigure}

    \begin{subfigure}[b]{\linewidth}
        \centering
        \includegraphics[scale=.9]{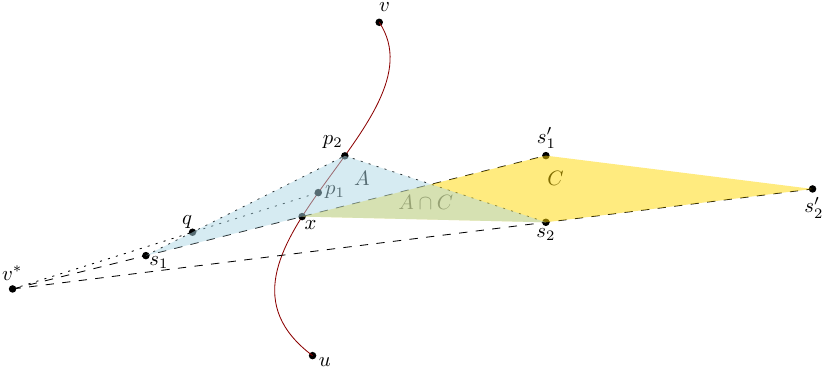}
        \subcaption{The two regions $A$ and $B$ overlap.}
        \label{fig:bisector_monotone_siteb}
    \end{subfigure}

    \caption{Illustrating the case in which $\seg{s_1 s_1'}$ intersects $B(s_1, s_2)$ (at a point $x$) while $\seg{s_2 s_2'}$ does not intersect $B(s_1, s_2)$.}
    \label{fig:bisector_monotone_site}
\end{figure}

Define $A$ to be the region bounded by $\pi(s_1, x)$, $\pi(s_1, p_2)$, $\pi(s_2, x)$, and $\pi(s_2, p_2)$.
Note that $\pi(s_1, x) = \seg{s_1 x}$.
The concatenation of $\pi(x, s_2)$, $\seg{x s_1'}$, and $\seg{s_2 s_2'}$ partitions $P$ into two regions; let $C$ denote the one that does not contain $v^*$ (see Figure~\ref{fig:bisector_monotone_site}, which illustrates two cases depending on whether $A$ and $C$ overlap).

Let $R$ denote the region bounded by $\pi(v^*, s_2') = \pi(v^*, s_2) \cup \seg{s_2 s_2'}$, $\pi(v^*, s_1') = \pi(v^*, s_1) \cup \seg{s_1 s_1'}$, and $C \cap \partial P$.
By definition, $\pi(s_2, x) \subseteq R$, and $\pi(s_2, x)$ partitions $R$ into two sides, one side being $C$ and the other containing $v^*$.
Also, for any point $p \in C$, $\pi(v^*, p)$ must cross $\pi(s_2, x)$.
On the other hand, for any point $p$ such that $\pi(v^*, p)$ crosses $\pi(s_2, x)$, one of the following three cases must occur:
(1) $p \in C$;
(2) $\pi(v^*, p)$ crosses $\seg{s_2 s_2'}$ (in this case, $s_2 \in \pi(v^*, p)$); and
(3) $\pi(v^*, p)$ crosses $\seg{x s_2'}$ (in this case, $x \in \pi(v^*, p)$).

We next show that $v^* \notin A$, $p_1 \in A$, and $p_1 \notin C$.
Indeed, $v^* \notin A$ because $v^*$ is a convex vertex of $P$, there are no vertices of $P$ in the interior of $A$, and any vertices of $P$ on the boundary of $A$ must be reflex vertices.
We get that $p_1 \in A$ because $A$ contains the portion of $B(s_1, s_2)$ between $x$ and $p_2$.
To see that $p_1 \notin C$, recall that the bisector $B(s_1, s_2)$ divides $P$ into two subpolygons, one $P_1$ containing $s_1$ and the other $P_2$ containing $s_2$.
As $\seg{s_2 s_2'}$ does not intersect $B(s_1, s_2)$, it must lie in $P_2$.
By Observation~\ref{obs:star_shaped}, $\pi(s_2, x)$ intersects $B(s_1, s_2)$ only at $x$.
Similarly, since $\seg{x s_1'} \subseteq \seg{s_1 s_1'}$, the segment $\seg{x s_1'}$ intersects $B(s_1, s_2)$ only at $x$.
All of the above together imply that the only intersection between $C$ and $B(s_1, s_2)$ is $x$.
Because $p_1 \in B(s_1, s_2)$ and $p_1 \neq x$, we obtain $p_1 \notin C$.

We now argue the key claim.
Since $v^* \notin A$ and $p_1 \in A$, the path $\pi(v^*, p_1)$ must cross the boundary of $A$, which consists of $\seg{s_1 x}$, $\pi(s_1, p_2)$, $\pi(s_2, x)$, and $\pi(s_2, p_2)$.

\begin{itemize}
    \item If $\pi(v^*, p_1)$ crosses either $\pi(s_1, p_2)$ or $\pi(s_2, p_2)$, then the key claim is proved.

    \item If $\pi(v^*, p_1)$ crosses $\seg{s_1 x}$, then because $\seg{s_1 x} \subseteq \pi(v^*, s_1')$ and shortest paths from the same source cannot cross, the paths $\pi(v^*, p_1)$ and $\pi(v^*, s_1')$ must share a common prefix subpath that contains $s_1$.
    Since $s_1 \in \pi(s_1, p_2)$, it follows that $\pi(v^*, p_1)$ intersects $\pi(s_1, p_2)$, and thus the key claim holds.

    \item If $\pi(v^*, p_1)$ crosses $\pi(s_2, x)$, then, as discussed above, one of the following three cases must occur:
    (1) $p_1 \in C$;
    (2) $s_2 \in \pi(v^*, p_1)$; and
    (3) $x \in \pi(v^*, p_1)$.
    Since we know that $p_1 \notin C$, the first case is impossible.
    In the second case, because $s_2 \in \pi(s_2, p_2)$, we obtain that $\pi(v^*, p_1)$ intersects $\pi(s_2, p_2)$, which proves the claim.
    In the third case, $\pi(v^*, p_1)$ intersects $\seg{s_1 x}$, and therefore the claim follows from the previous case.
\end{itemize}

In summary, the key claim holds in all cases.

\paragraph{Neither $\boldsymbol{\seg{s_1 s_1'}}$ nor $\boldsymbol{\seg{s_2 s_2'}}$ intersects $\boldsymbol{B(s_1, s_2)}$.}
We now prove the second statement of the lemma.
Assume that neither $\seg{s_1 s_1'}$ nor $\seg{s_2 s_2'}$ intersects $B(s_1, s_2)$.
See Figure~\ref{fig:bisector_monotone} for an illustration.

\begin{figure}
    \centering
    \begin{subfigure}[b]{\textwidth}
        \centering
        \includegraphics[scale=.8]{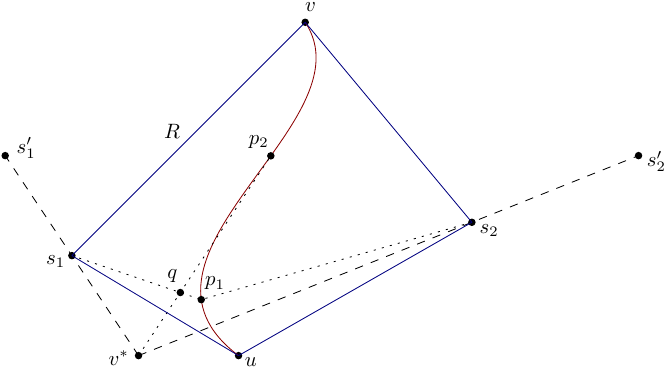}
        \label{fig:bisector_monotone_no_site}
    \end{subfigure}

    \caption{Illustrating the case in which neither $\seg{s_1 s_1'}$ nor $\seg{s_2 s_2'}$ intersects $B(s_1, s_2)$.}
    \label{fig:bisector_monotone}
\end{figure}

The bisector $B(s_1, s_2)$ divides $P$ into two subpolygons, one $P_1$ containing $s_1$ and the other $P_2$ containing $s_2$.
As neither $\seg{s_1 s_1'}$ nor $\seg{s_2 s_2'}$ intersects $B(s_1, s_2)$, we have $\seg{s_1 s_1'} \subseteq P_1$ and $\seg{s_2 s_2'} \subseteq P_2$.
Let $p$ be a point in $B(s_1, s_2)$.
By Observation~\ref{obs:star_shaped}, $\pi(s_1, p) \subseteq P_1$ and $\pi(s_2, p) \subseteq P_2$.
Hence, the union $\gamma_p$ of $\seg{s_1' s_1}$, $\pi(s_1, p)$, $\pi(p, s_2)$, and $\seg{s_2 s_2'}$ is a simple path in $P$ connecting two points on $\partial P$, and it partitions $P$ into two subpolygons.
If we orient $\gamma_p$ from $s_1'$ to $s_2'$, we let $P_l(p)$ be the subpolygon on the left side of $\gamma_p$ and $P_r(p)$ the subpolygon on the right side.
Clearly, $u$ and $v$ are on different sides of $\gamma_p$.
Without loss of generality, we assume that both $v^*$ and $u$ are in $P_r(p)$, while $v$ is in $P_l(p)$.

Since $u$ and $v$ are the two endpoints of $B(s_1, s_2)$, we have $u \in P_r(p)$ and $v \in P_l(p)$ for all points $p \in B(s_1, s_2)$.
We claim that $v^*$ is always in $P_r(p)$ for all points $p \in B(s_1, s_2)$.
Indeed, consider the region $R$ bounded by $\pi(s_1, v)$, $\pi(s_1, u)$, $\pi(s_2, u)$, and $\pi(s_2, v)$.
Every vertex of $P$ on the boundary of $R$ is reflex.
Since $v^*$ is a convex vertex of $P$, it cannot be a vertex of $R$.
Hence, $v^*$ is strictly outside $R$.
On the other hand, $R$ is essentially the union of $\pi(s_1, p) \cup \pi(s_2, p)$ over all points $p \in B(s_1, s_2)$.
Thus, as $p$ moves on $B(s_1, s_2)$ from $u$ to $v$, the union $\pi(s_1, p) \cup \pi(s_2, p)$ will not pass through $v^*$, and therefore $\gamma_p$ will not pass through $v^*$ either.
The claim follows.

Consider any two points $p_1, p_2 \in B(s_1, s_2)$ such that $p_2$ lies between $p_1$ and $v$, i.e., $u$, $p_1$, $p_2$, and $v$ appear in this order along $B(s_1, s_2)$.
By the above claim, $v^*$ lies in $P_r(p_1)$.
Since $p_2$ lies on $B(s_1, s_2)$ between $p_1$ and $v$, it must lie in $P_l(p_1)$.
Therefore, $v^*$ and $p_2$ lie on different sides of $\gamma_{p_1}$, and thus $\pi(v^*, p_2)$ must intersect $\gamma_{p_1}$.
We claim that $\pi(v^*, p_2)$ must intersect $\pi(s_1, p_1)$ or $\pi(s_2, p_1)$.
Indeed, by an argument similar to the above case, if $\pi(v^*, p_2)$ intersects $\seg{s_1 s_1'}$, then it must contain $s_1$ and thus intersects $\pi(s_1, p_1)$.
Similarly, if $\pi(v^*, p_2)$ intersects $\seg{s_2 s_2'}$, then it must contain $s_2$ and thus intersects $\pi(s_2, p_1)$.
Therefore, $\pi(v^*, p_2)$ must intersect $\pi(s_1, p_1)$ or $\pi(s_2, p_1)$.

By Lemma~\ref{lem:crossing}, if $\pi(v^*, p_2)$ intersects $\pi(s_1, p_1)$, then $\delta_{s_1}(p_1) \ge \delta_{s_1}(p_2)$.
Similarly, if $\pi(v^*, p_2)$ intersects $\pi(s_2, p_1)$, then $\delta_{s_2}(p_1) \ge \delta_{s_2}(p_2)$.
Since $p_1, p_2 \in B(s_1, s_2)$, we have
\[
\delta_{s_1}(p_1) = \delta_{s_2}(p_1)
\quad\text{and}\quad
\delta_{s_1}(p_2) = \delta_{s_2}(p_2).
\]
Therefore, in either case, $\delta_{s_1}(p_1) \ge \delta_{s_1}(p_2)$.
This implies that $\delta_{s_1}(p)$ is non-decreasing as $p$ moves from $v$ to $u$.
This proves the second statement of the lemma.
\end{proof}

\paragraph{Analysis of Voronoi edges.}
We are now in a position to analyze the Voronoi edge case for Lemma~\ref{lem:sweeping}.
Consider a Voronoi edge $e$ of $\vd(S)$.
We follow the notation introduced in Section~\ref{sec:proofoverview} for this case, namely $s_1$, $s_2$, $u$, $v$, $t_{s_1}$, $t_{s_2}$, $t_u$, $t_v$, $\seg{s_1 s_1'}$, and $\seg{s_2 s_2'}$.
We distinguish two cases depending on whether one of $\seg{s_1 s_1'}$ and $\seg{s_2 s_2'}$ intersects $e$.

\begin{itemize}
\item If one of $\seg{s_1 s_1'}$ and $\seg{s_2 s_2'}$ intersects $e$, we assume without loss of generality that $\seg{s_1 s_1'}$ intersects $e$, say at a point $x$.
Since $e \subseteq B(s_1, s_2)$, Lemma~\ref{lem:bisector_event}(1) implies that $\delta_{s_1}(p)$ is non-decreasing as $p$ moves from $x$ to either endpoint $u$ or $v$ of $e$.

By Observation~\ref{obs:delta_time}, and as discussed in Section~\ref{sec:boundary}, at the site event $t_s$, two vertices $u'$ and $v'$ are added to $\W_t$ at $x$ that trace $e$ from $x$ to $u$ and $v$, respectively, where the two traces are stopped by the events $t_u$ and $t_v$, respectively.
Each trace is continuous except at pocket events, where $u'$ (resp., $v'$) may jump discontinuously, but only in the direction it has already been tracing.

\item If neither $\seg{s_1 s_1'}$ nor $\seg{s_2 s_2'}$ intersects $e$, then by Lemma~\ref{lem:bisector_event} (regardless of whether we are in Case~1 or Case~2 there), for any site $s \in \{s_1, s_2\}$, $\delta_s(p)$ is non-decreasing as $p$ moves along $e$ from one endpoint to the other.
Without loss of generality, assume that $\delta_s(p)$ is non-decreasing as $p$ moves from $u$ to $v$.

By Observation~\ref{obs:delta_time}, at the event $t_u$, a vertex $u'$ is added to $\W_t$ at $u$ that traces $e$ from $u$ to $v$, where the trace is stopped by the event $t_v$.
The trace is continuous except at pocket events, where $u'$ may jump, but only in the direction it has already been tracing.
As before, in the special case where $t_u = t_v$, both events are pocket Voronoi vertex/terminal events, and the trace is instantaneous.
\end{itemize}

\section{Implicit point location}
\label{sec:point_location}

The above algorithm computes the Voronoi diagram $\vd(S)$ implicitly.
Given a query point $q$, we wish to use $\vd(S)$ to find its nearest neighbor in $S$, i.e., the site whose weighted distance to $q$ is minimized.
The traditional approach is to construct $\vd(S)$ explicitly and then build a point location data structure on it~\cite{ref:KirkpatrickOp83,ref:EdelsbrunnerOp86}.
However, this requires $\Omega(m)$ preprocessing time, since explicitly constructing $\vd(S)$ already requires $\Omega(m)$ space.
Therefore, we must support point location queries directly on the implicit representation of $\vd(S)$.

We show that after $O(n \log m)$ preprocessing time on the implicit $\vd(S)$, a nearest neighbor query for any point $q$ can be answered in $O(\log (n+m))$ time.

In fact, we prove a more general result.
Our approach applies to a decomposition of $P$ into \emph{geodesic triangles} and \emph{geodesic sectors}.
We first introduce these concepts.

\begin{figure}[t]
\centering
\includegraphics[height=2.5in]{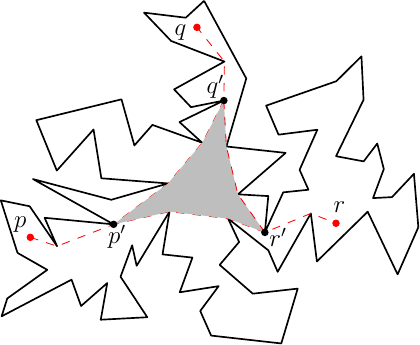}
\caption{Illustrating a geodesic triangle (the gray region), with apexes $p'$, $q'$, and $r'$.}
\label{fig:geotriangle}
\end{figure}

\paragraph{Geodesic triangles.}
Let $p, q, r$ be three points in $P$ such that none lies on the shortest path between the other two.
Removing the common subpaths shared by each pair among $\pi(p, q)$, $\pi(p, r)$, and $\pi(q, r)$ yields a region bounded by the remaining portions of these paths; this region is called a \emph{geodesic triangle}~\cite{ref:ChazelleRa94} (see Figure~\ref{fig:geotriangle}).
We denote it by $\triangle pqr$.

The junction vertex of each pair of paths is called an \emph{apex} of $\triangle pqr$.
Thus, $\triangle pqr$ has three apexes.
The portion of the boundary connecting each pair of adjacent apexes is called a \emph{side}, which is a subpath of one of the three shortest paths and forms a concave chain.
Using the GH data structure, a geodesic triangle can be represented implicitly by its three apexes (equivalently, by $(p, q, r)$, since the apexes can be computed in $O(\log m)$ time).

\begin{figure}[t]
\centering
\includegraphics[height=1.7in]{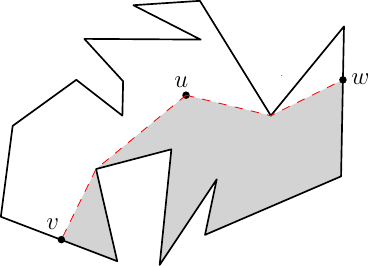}
\caption{Illustrating a geodesic sector (the gray region).}
\label{fig:polysector}
\end{figure}

\paragraph{Geodesic sectors.}
For any two points $v, w \in \partial P$, let $\partial_{[v,w]} P$ denote the portion of $\partial P$ from $v$ to $w$ in counterclockwise order.
Let $u$ be a point in the interior of $P$, and let $v, w \in \partial P$ such that $\pi(u, v)$ and $\pi(u, w)$ intersect only at $u$.
We define $\sector_u(v, w)$ to be the region bounded by $\pi(u, v)$, $\pi(u, w)$, and $\partial_{[v, w]} P$ (see Figure~\ref{fig:polysector}).
We call $\sector_u(v, w)$ a \emph{geodesic sector}, and $u, v, w$ are its \emph{apexes}.
For convenience, points of $\partial_{[v, w]} P \setminus \{v, w\}$ are also considered interior points of the sector boundary.

Using the GH data structure, a geodesic sector can be represented implicitly by its three apexes.

\paragraph{The implicit point location problem.}
Let $T$ be a set of $n$ pairwise interior-disjoint geodesic triangles and geodesic sectors in $P$.
Each region in $T$ is specified by its apexes using $O(1)$ space.
Note that $T$ need not form a full decomposition of $P$, i.e., some points of $P$ may lie in no region of $T$.

We consider the problem of constructing a data structure for $T$ to support \emph{point location queries}:
given a query point $q \in P$, determine which region of $T$ contains $q$, or report that $q$ is not contained in any region.
We prove the following result.

\begin{restatable}{theorem}{pointlocation} \label{thm:pointlocation}
Let $P$ be a simple polygon with $m$ vertices.
After $O(m)$ preprocessing of $P$, given a set $T$ of $n$ interior-disjoint geodesic triangles and geodesic sectors in $P$,
we can construct a data structure for $T$ in $O(n \log (n + m))$ time such that each point location query can be answered in $O(\log (n + m))$ time.
If $T$ is connected, the construction time improves to $O(n \log m)$.
\end{restatable}

\paragraph{Overview.}
We follow the notation introduced in Section~\ref{sec:pre}.
Let $V_T$ denote the set of vertices of the regions in $T$.
For ease of exposition, we assume that no three points in $V_T \cup V_P$ are collinear.
We also assume that no query point is collinear with any two points in $V_T \cup V_P$, which implies that no query point lies on the boundary of any region in $T$.
We perform the same global preprocessing for $P$ as described in Section~\ref{sec:global}, which takes $O(m)$ time.

\begin{figure}[t]
\begin{minipage}[t]{\linewidth}
\begin{center}
\includegraphics[totalheight=3.5in]{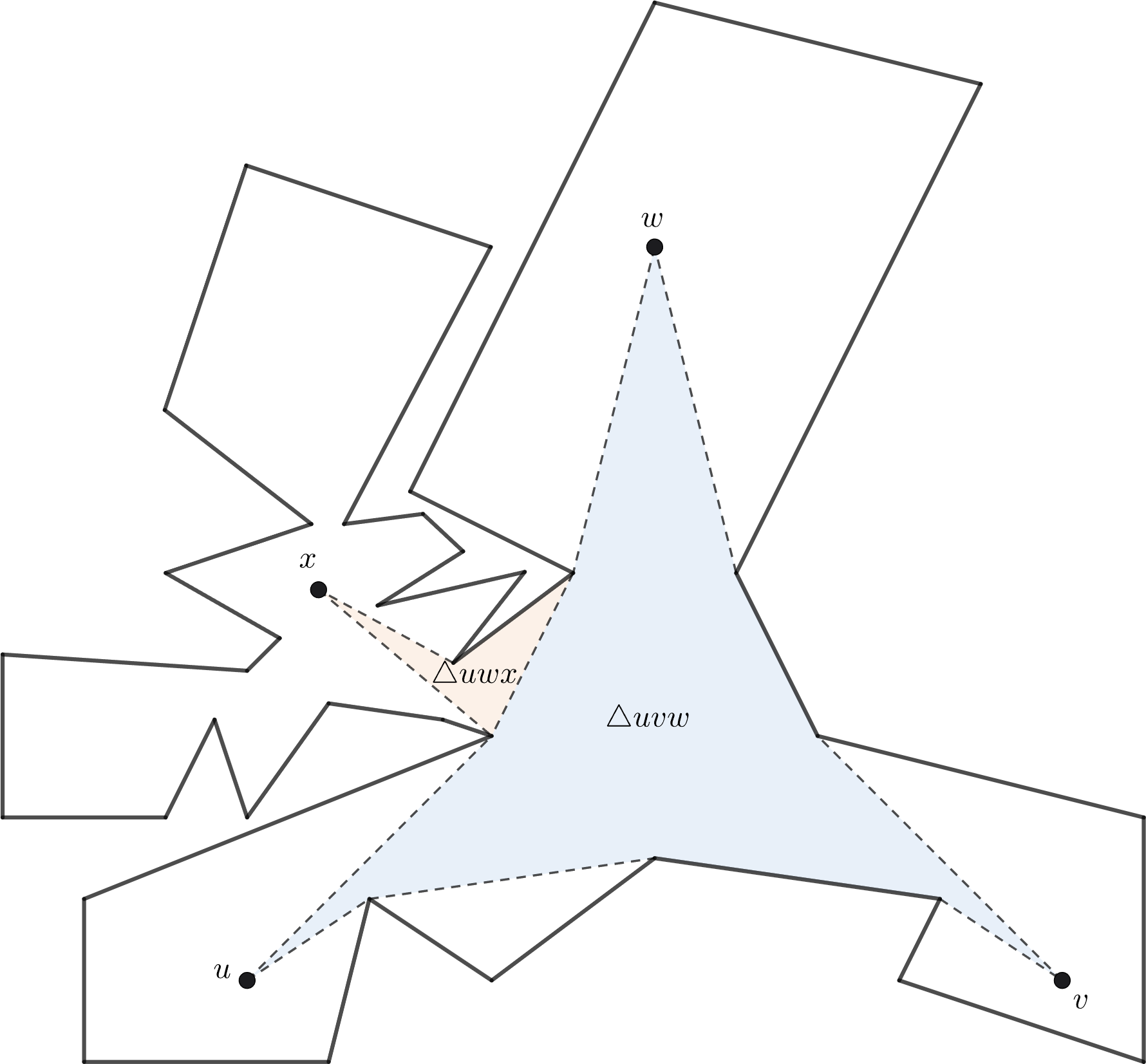}
\caption{A geodesic triangulation for four points $\{x,u,v,w\}$.}
\label{fig:gst}
\end{center}
\end{minipage}
\end{figure}

In Section~\ref{sec:pseudo_triangles}, we consider the case where $T$ is a set of interior-disjoint geodesic triangles.
An important special case is when $T$ forms a \emph{geodesic triangulation} of $V_T$, i.e., the triangles in $T$ form a partition of the geodesic convex hull of $V_T$ (see Figure~\ref{fig:gst}).
A region $R \subseteq P$ is \emph{geodesically convex} if $\pi(p,q) \subseteq R$ for every pair of points $p, q \in R$.
The \emph{geodesic convex hull} of a set of points in $P$ is the smallest geodesically convex region containing all the points; it is also known as the \emph{relative convex hull}~\cite{ref:AronovTh93}.

\begin{figure}[t]
\begin{minipage}[t]{\linewidth}
\begin{center}
\includegraphics[totalheight=3.5in]{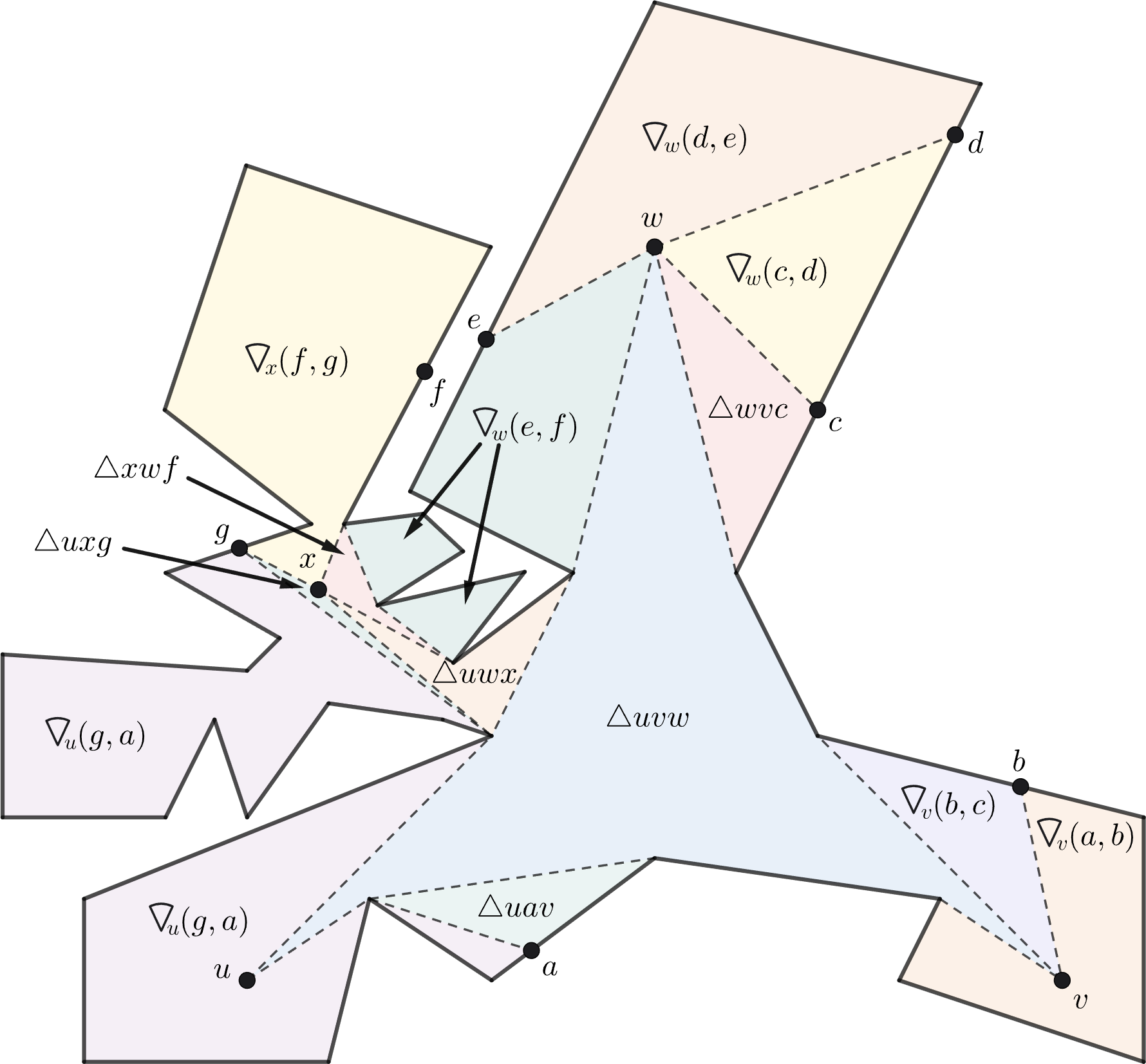}
\caption{A geodesic decomposition of $P$.}
\label{fig:egst}
\end{center}
\end{minipage}
\end{figure}

In Section~\ref{sec:geodesic_sectors}, we extend our technique to handle geodesic sectors.
An important special case is a \emph{geodesic decomposition} of $P$, where a collection of geodesic triangles and geodesic sectors forms a decomposition of $P$ (see Figure~\ref{fig:egst}).
In contrast, in a traditional \emph{geodesic triangulation} of $P$~\cite{ref:ChazelleRa94}, $P$ is partitioned into geodesic triangles with a total of $O(m)$ apexes, allowing an explicit point location data structure to be constructed in $O(m)$ time~\cite{ref:KirkpatrickOp83,ref:EdelsbrunnerOp86}.
In our setting, however, $n$ may be much smaller than $m$, and thus we must construct a point location data structure implicitly.

In Section~\ref{sec:dcel_improvement}, we improve the construction time when $T$ forms a connected subdivision.

Finally, in Section~\ref{sec:vd_pl}, we reduce point location in the implicit Voronoi diagram $\vd(S)$ to point location in a geodesic decomposition of $P$, thereby completing the proof of Theorem~\ref{thm:mainthm}.
In particular, we show that $\vd(S)$ can be further decomposed to obtain such a decomposition.

\subsection{Point locations among geodesic triangles} \label{sec:pseudo_triangles}

We consider the case where $T$ is a set of $n$ interior-disjoint geodesic triangles in $P$.
For each geodesic triangle $\triangle \in T$, we show how to ``approximate'' $\triangle$ using a constant number of line segments.

\begin{figure}
    \centering
    \includegraphics[width=.40\linewidth]{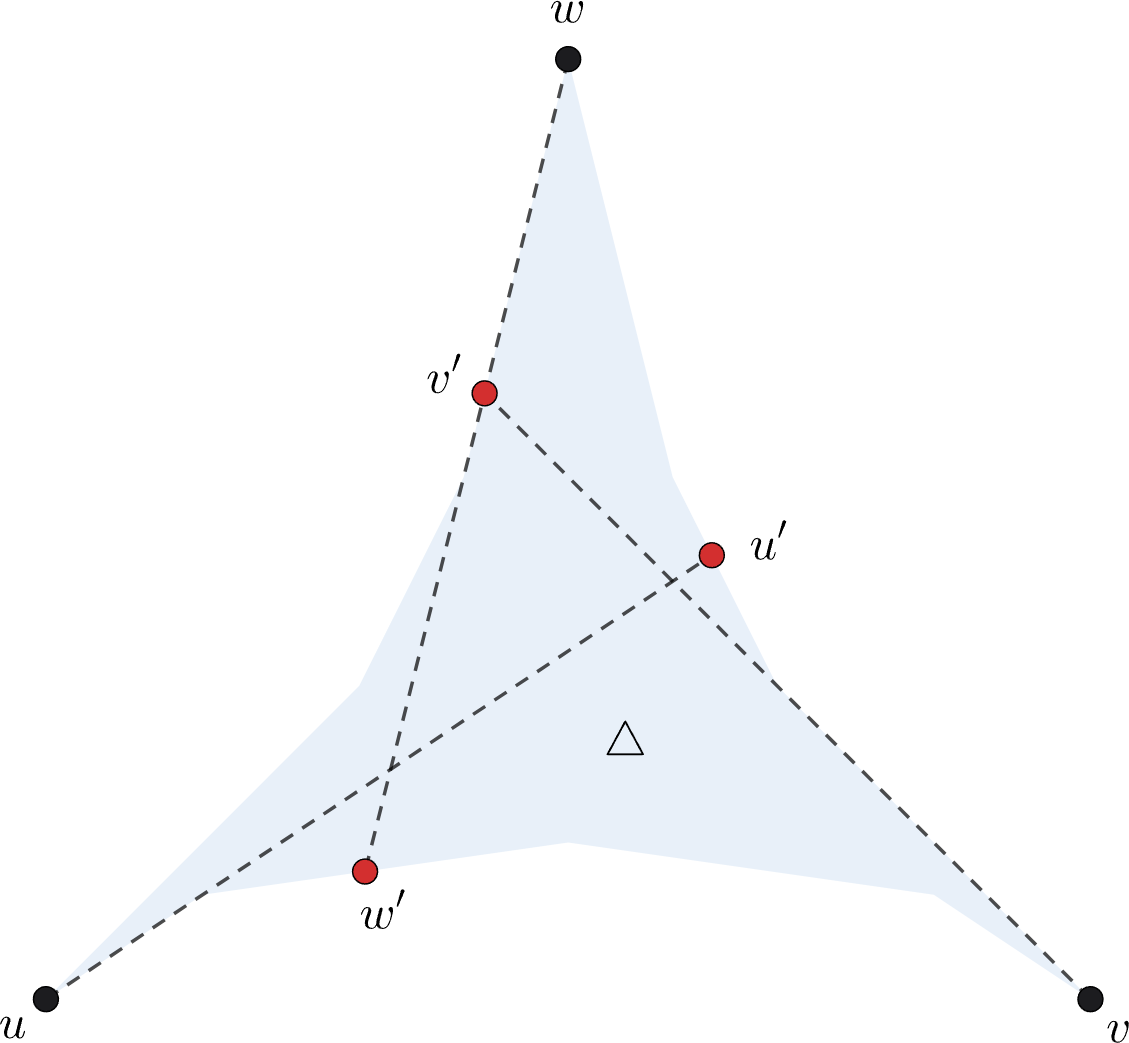}
    \caption{Illustration of $approx(\triangle)$.}
    \label{fig:tapprox}
\end{figure}

Figure~\ref{fig:tapprox} illustrates our construction.
Let $u$, $v$, and $w$ be the apexes of $\triangle$ in counterclockwise order.
Let $u'$ be the point on $\pi(v, w)$ obtained by extending a ray from $u$ out of the direction of the anchor of $u$ in $\pi(u, v)$
(note that, due to the convexity of $\pi(u, v)$ and $\pi(u, w)$, this ray intersects $\pi(v, w)$ exactly once).
Define $v'$ and $w'$ analogously.

Let $approx(\triangle)$ be the set of subsegments of $\seg{u u'}$, $\seg{v v'}$, and $\seg{w w'}$, partitioned by their pairwise intersections.
Clearly, $|approx(\triangle)| = O(1)$.

\begin{lemma}
Given the apexes $u$, $v$, and $w$ of $\triangle$, $approx(\triangle)$ can be computed in $O(\log m)$ time using the GH data structure.
\end{lemma}

\begin{proof}
Using the GH data structure, we can compute the anchor of $u$ in $\pi(u, v)$ in $O(\log m)$ time.
We can also obtain a binary search structure for the path $\pi(v, w)$, which supports searching in $O(\log m)$ time.
Using this structure, we compute $u'$ by binary search in $O(\log m)$ time.

The points $v'$ and $w'$ can be computed similarly.
Once $u'$, $v'$, and $w'$ are known, $approx(\triangle)$ can be constructed in $O(1)$ additional time.
\end{proof}

The following lemma is crucial for our point location data structure and explains the role of $approx(\triangle)$.

\begin{lemma}\label{lem:pointlocation10}
For any query point $q \in \triangle$, at least one vertical ray (either upward or downward) from $q$ intersects a segment of $approx(\triangle)$ before hitting the boundary of $\triangle$.
\end{lemma}

\begin{proof}
Let $q_1$ (resp., $q_2$) be the first point on $\partial \triangle$ hit by the upward (resp., downward) vertical ray from $q$.
By our general position assumption, $q$ lies in the interior of $\triangle$, and hence $q_1$ and $q_2$ lie on different edges of $\triangle$.
Moreover, since each side of $\triangle$ is a concave chain, $q_1$ and $q_2$ lie on different sides of $\triangle$.

Observe that $\seg{u u'}$ separates $\pi(w, u)$ from $\pi(u, v)$, i.e., the two paths are on different sides of $\seg{ u u'}$.
Similarly, $\seg{v v'}$ separates $\pi(u, v)$ from $\pi(v, w)$, and $\seg{w w'}$ separates $\pi(v, w)$ from $\pi(w, u)$.
Since $q_1$ and $q_2$ lie on different sides of $\triangle$, the segment $\seg{q_1 q_2}$ must intersect at least one of these segments.
Therefore, at least one of the vertical rays from $q$ intersects a segment in $approx(\triangle)$ before exiting $\triangle$.
\end{proof}

Since the geodesic triangles in $T$ are interior-disjoint and their approximation segments lie strictly in their interiors, it follows that at least one of the vertical rays from $q$ must intersect a segment of $approx(\triangle)$ before intersecting any segment from another triangle in $T$.

Define $approx(T) = \bigcup_{\triangle \in T} approx(\triangle)$. Since $|approx(\triangle)| = O(1)$ for each triangle $\triangle \in T$, we have $|approx(T)| = O(n)$, and $approx(T)$ can be constructed in $O(n \log m)$ time.

We build a vertical ray-shooting data structure on $approx(T)$ using the persistent data structure of Sarnak and Tarjan~\cite{ref:SarnakPl86}.
The construction time is $O(n \log n)$, the space is $O(n)$, and each query can be answered in $O(\log n)$ time.

Given a query point $q$, our goal is to find the geodesic triangle $\triangle_q$ of $T$ that contains $q$ or report that no such $\triangle_q$ exists (i.e., $q$ is not contained in any geodesic triangle of $T$). To this end, we perform two vertical ray-shooting queries (upward and downward) to find at most two segments of $approx(T)$ intersected first.
Let $\triangle_1$ and $\triangle_2$ be the corresponding triangles.
By Lemma~\ref{lem:pointlocation10}, if $q$ lies in some triangle $\triangle_q$, then $\triangle_q$ must be either $\triangle_1$ or $\triangle_2$.

\begin{lemma}\label{lem:pointlocationgeotriangle}
Given a geodesic triangle $\triangle$, we can determine whether $q \in \triangle$ in $O(\log m)$ time using the GH data structure.
\end{lemma}

\begin{proof}
Let $u$, $v$, and $w$ be the apexes of $\triangle$.
Notice that $\triangle$ is the geodesic convex hull of $u$, $v$, and $w$.
The apexes of the geodesic convex hull of a set $A$ of $k$ points in $P$ can be computed in $O(k \log (m + k))$ time using the GH data structure~\cite{ref:GuibasOp89} (an \emph{apex} refers to a point of $A$ that appears on the boundary of the hull).
To determine whether $q\in \triangle$, we compute the apexes of the geodesic convex hull of $\{u, v, w, q\}$ in $O(\log m)$ time.
    If $q$ is an apex of the hull, then $q$ is not in $\triangle$. Otherwise, $q$ is in $\triangle$.
\end{proof}

Using the above lemma, we can check whether $q$ lies in $\triangle_1$ or $\triangle_2$.
If neither contains $q$, then $q$ lies in no triangle of $T$.
The total query time is $O(\log m + \log n) = O(\log (m + n))$.

In summary, given $T$, we can construct the data structure in $O(n \log m + n \log n) = O(n \log (n + m))$ time using $O(n)$ space, and answer each query in $O(\log (n + m))$ time.

\subsection{Point location among geodesic sectors} \label{sec:geodesic_sectors}

We consider the case where $T$ is a set of $n$ interior-disjoint geodesic triangles and geodesic sectors.
We process each geodesic triangle as in Section~\ref{sec:pseudo_triangles}.
In the following, we focus on how to handle geodesic sectors.

\begin{figure}
    \centering
    \begin{subfigure}[b]{.49\textwidth}
        \centering
        \includegraphics[width=\linewidth]{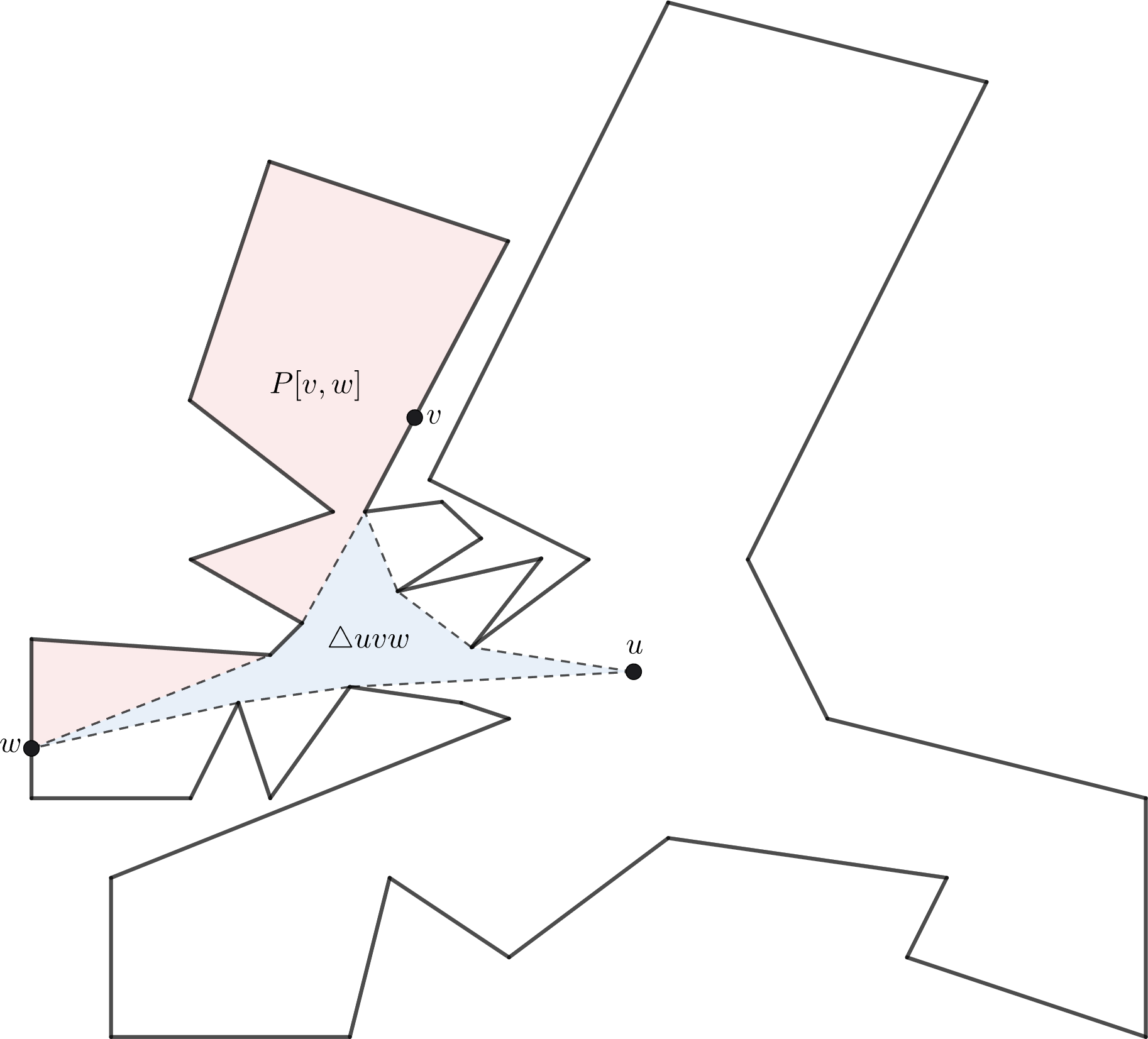}
        \subcaption{A geodesic minor sector $\sector_u(v,w)$ partitioned into a geodesic triangle $\triangle uvw$ and a polygonal segment $P[v,w]$.}
        \label{fig:et1}
    \end{subfigure}
    \hfill
    \begin{subfigure}[b]{.49\textwidth}
        \centering
        \includegraphics[width=\linewidth]{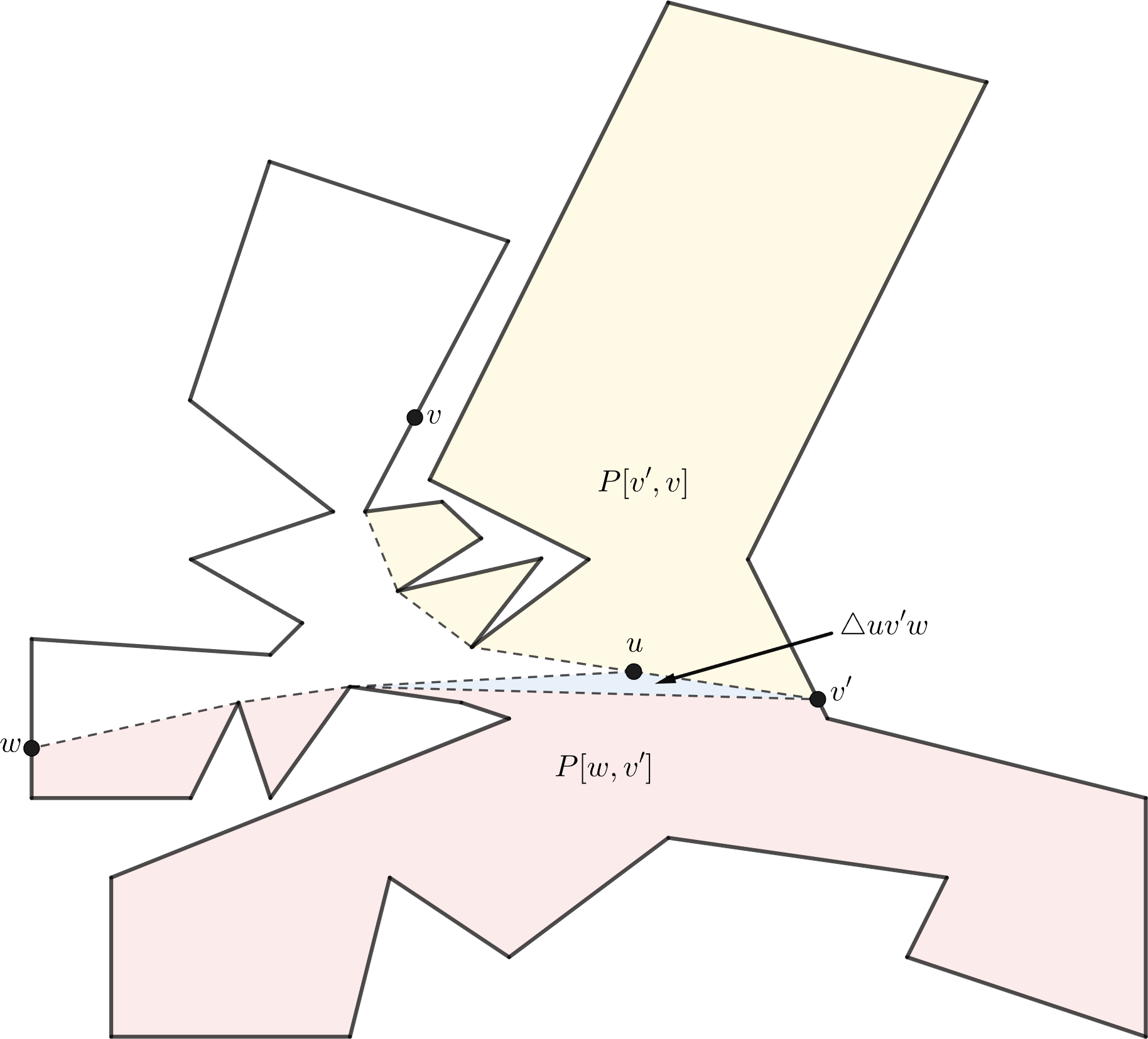}
        \subcaption{A geodesic major sector $\sector_u(w,v)$ partitioned into a geodesic triangle $\triangle uv'w$ and two polygonal segments $P[w, v']$ and $P[v', v]$.}
        \label{fig:et2}
    \end{subfigure}
    \caption{Illustration of geodesic sectors.}
    \label{fig:et}
\end{figure}

Consider a geodesic sector $\sector_u(v,w)$.
If $u \notin \pi(v, w)$, then $\pi(v, w)$ lies entirely in either $\sector_u(v, w)$ or $\sector_u(w, v)$, but not both.
If $\pi(v, w) \subseteq \sector_u(v, w)$, we call $\sector_u(v, w)$ a \emph{geodesic minor sector}; otherwise, it is a \emph{geodesic major sector} (in Figure~\ref{fig:et},
$\sector_u(v, w)$ is a geodesic minor sector and $\sector_u(w, v)$ is a geodesic major sector).
We define $P[v, w]$ to be the subpolygon of $P$ bounded by $\partial_{[v, w]} P$ and $\pi(v, w)$, and call it a \emph{polygonal segment}.
We also refer to $\partial_{[v, w]} P$ as a \emph{polygonal arc}.

\begin{itemize}
\item
For a minor sector $\sector_u(v, w)$ (see Figure~\ref{fig:et1}), the path $\pi(v, w)$ decomposes $\sector_u(v, w)$ into a geodesic triangle $\triangle u v w$ and a polygonal segment $P[v, w]$.

\item
For a major sector $\sector_u(w, v)$ (see Figure~\ref{fig:et2}), we further decompose it into a geodesic triangle and two polygonal segments as follows.

Let $\seg{u v'}$ be the last-edge extension of $\pi(v, u)$.
We claim that $v' \in \partial_{[w, v]} P$.
To see this, suppose for contradiction that $v' \in \partial_{[v, w]} P$.
Since $\pi(v, w)$ separates $\partial_{[v, w]} P$ from $u$ and $u \in \pi(v, v') = \pi(v, u) \cup \seg{u v'}$, the segment $\seg{u v'}$ would have to intersect $\pi(v, w)$.
However, shortest paths from the same source cannot intersect except along a common prefix, and since $u \notin \pi(v, w)$, this is impossible.
Thus $v' \in \partial_{[w, v]} P$.

It follows that $\seg{u v'}$ and $\pi(w, v')$ decompose $\sector_u(w, v)$ into a geodesic triangle $\triangle u v' w$ and two polygonal segments $P[w, v']$ and $P[v', v]$.

Since $v'$ can be computed in $O(\log m)$ time using the GH and ray-shooting data structures, this decomposition takes $O(\log m)$ time.

\item
If $u \in \pi(v, w)$, then $\sector_u(v, w)$ itself is a polygonal segment $P[v, w]$.
\end{itemize}

In summary, each geodesic sector can be decomposed into at most one geodesic triangle and at most two polygonal segments in $O(\log m)$ time.
Applying this to all sectors in $T$ yields $O(n)$ interior-disjoint geodesic triangles and polygonal segments.
We handle geodesic triangles as in Section~\ref{sec:pseudo_triangles}; we now focus on polygonal segments.

Let $T'$ denote the set of all polygonal segments.
Since the sectors in $T$ are interior-disjoint, their polygonal arcs are also interior-disjoint by our definition.
We begin with the following lemma.

\begin{lemma}\label{lem:pointlocation20}
For any point $q$ in the interior of a polygonal segment $P[v, w]$, there exists at least one vertical ray (upward or downward) from $q$ such that the point on the boundary of $P[v, w]$ hit first by the ray is on $\partial_{[v, w]} P$.
\end{lemma}
\begin{proof}
Suppose, for contradiction, that the lemma statement is not true. Then, both the vertical upward ray and the vertical downward ray from $q$ hit $\pi(v, w)$ first.
Let $a$ and $b$ be these intersection points.
Since $\seg{ab} \subseteq P$, it follows that $\seg{ab} \subseteq \pi(v, w)$, implying that $q$ lies on $\pi(v, w)$, contradicting that $q$ is in the interior of $P[v, w]$.
\end{proof}

Using this lemma, we locate the polygonal segment containing a query point $q$ as follows.
We shoot vertical rays upward and downward from $q$ and compute their first intersections with $\partial P$.
For each intersection point, we determine whether it lies on a polygonal arc of $T'$ (the details are given below).
This yields at most two candidate segments.
For each candidate segment $R$, we test whether $q \in R$.
If so, we return $R$; otherwise, $q$ lies in no segment.
The correctness follows from Lemma~\ref{lem:pointlocation20}.
We use the following lemma to check whether $q$ lies in a polygonal segment.

\begin{lemma}\label{lem:pointlocationpolysegment}
Given a polygonal segment $P[v,w]$, we can determine whether $q \in P[v,w]$ in $O(\log m)$ time using the GH data structure.
\end{lemma}

\begin{proof}
We first check whether $q \in \pi(v, w)$ in $O(\log m)$ time.
If so, then $q \in P[v, w]$.
Otherwise, we compute the junction vertex $v'$ of $\pi(v, w)$ and $\pi(v, q)$ and determine the incident edges at $v'$.
All these can be done in $O(\log m)$ time using the GH data structure.
From these, we determine whether $\pi(v, q)$ exits $v'$ to the right or left of $\pi(v, w)$.
If it exits to the right, then $q \in P[v, w]$; otherwise, $q \in P[w, v]$.
\end{proof}

It remains to solve the following subproblem: given a point $p \in \partial P$, find the polygonal arc of $T'$ that contains $p$ (or report that no arc contains $p$).

To this end, recall that the second step of our global preprocessing for $P$ computes a list of the edges of $P$ sorted counterclockwise starting from a vertex $v^*$ of $P$ (i.e., under the order $\lp$ defined in Section~\ref{sec:struct}), and assigns each edge an index according to this order.
The third step of the preprocessing constructs a polygon-boundary point location data structure.

For each endpoint of a polygonal arc in $T'$, we use the polygon-boundary point location data structure to determine the edge of $P$ that contains it.
This allows us to sort all endpoints of polygonal arcs in $T'$ under the order $\lp$.
We compute and store this sorted list as part of our point location data structure for $T$.
The total time for this sorting is $O(n \log n + n \log m)$.

Given a query point $p \in \partial P$, we first determine the edge $e_p$ of $P$ that contains $p$, which takes $O(\log m)$ time using the polygon-boundary point location data structure.
Then, using the index of $e_p$, we perform a binary search on the sorted list to find the polygonal arc of $T'$ that contains $p$ in $O(\log n)$ time, or report that no such arc exists.

In summary, for a set $T$ of $n$ interior-disjoint geodesic triangles and geodesic sectors, we can construct a data structure in $O(n \log m + n \log n) = O(n \log (n + m))$ time using $O(n)$ additional space, such that each point location query can be answered in $O(\log (n + m))$ time.

\subsection{Improvement for a connected $\boldsymbol{T}$}
\label{sec:dcel_improvement}
Let $T$ be a set of $n$ interior-disjoint geodesic triangles and geodesic sectors.
If $T$ is connected, then we can further improve the above construction time to $O(n \log m)$. Since $n \log(n + m) = O(n \log m)$ for $n < m$, it suffices to consider the case in which $n \ge m$.

Let $A_T$ be the set of the apexes of the geodesic triangles and geodesic sectors in $T$. Note that $|A_T|=O(n)$.
For each geodesic triangle or geodesic sector in $T$, we explicitly compute all its edges. This can be done in $O(\log m)$ time plus the number of its edges, e.g., by using the GH data structure and traversing $\partial P$.
This results in a planar subdivision whose vertices are in $A_T \cup V_P$,
so its size is bounded by $O(n + m) = O(n)$ (because all edges are interior-disjoint).
Therefore, the total number of edges in $T$ is $O(n)$, and computing all these edges explicitly as above takes $O(n \log m)$ time.

Next, we triangulate each face in this connected subdivision in linear time~\cite{ref:ChazelleTr91} and then
construct a point location data structure for this final connected triangulation in $O(n)$ time that supports queries in $O(\log n)$ time~\cite{ref:KirkpatrickOp83,ref:EdelsbrunnerOp86}.

In summary, if $T$ is connected, regardless of whether $n < m$ or $n \geq m$,
we can construct a data structure for $T$ in $O(n \log m)$ time and $O(n)$ additional space, and each point location query can be answered in $O(\log (n + m))$ time.

Combining the result in Section~\ref{sec:geodesic_sectors}, Theorem~\ref{thm:pointlocation} is proved.

\subsection{Point locations in the implicit Voronoi diagram $\boldsymbol{\vd(S)}$} \label{sec:vd_pl}

Given the implicit Voronoi diagram $\vd(S)$, we wish to construct a data structure that, for any query point $q \in P$, finds the site $s \in S$ whose Voronoi region $\R(s)$ contains $q$ (equivalently, the nearest neighbor of $q$ in $S$).
Using the fact that Voronoi regions in $\vd(S)$ are star-shaped, we reduce the problem to point location in a geodesic decomposition as follows.

\begin{figure}
    \centering
    \includegraphics[width=.4\linewidth]{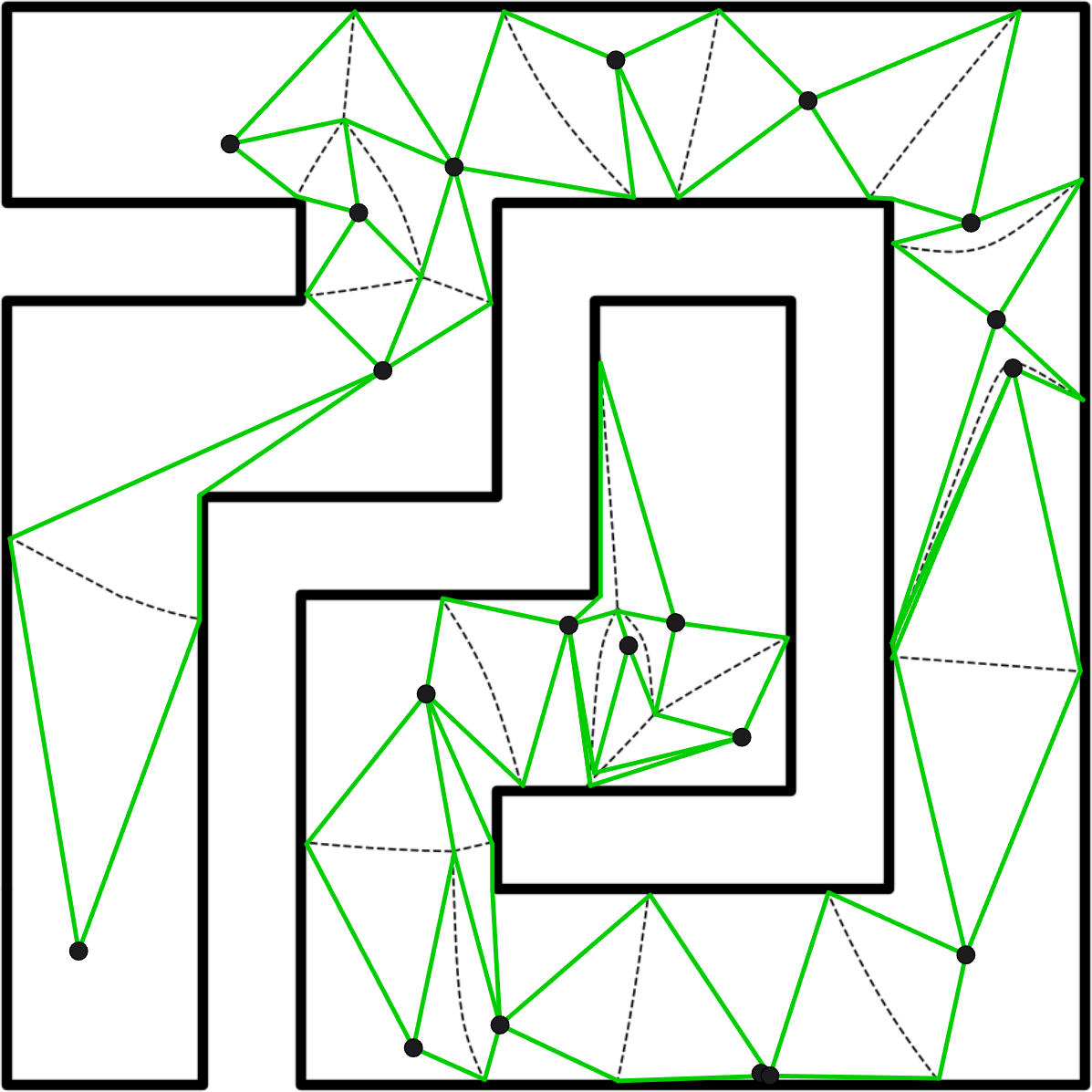}
    \caption{Illustrating a spoked additively weighted geodesic Voronoi diagram. The green segments are spokes.}
    \label{fig:spoked_geodesic_vd}
\end{figure}

Consider a Voronoi region $\R(s)$ for a site $s \in S$ with $\R(s) \neq \emptyset$.
For each Voronoi vertex or Voronoi terminal $v$ of $\R(s)$, we add the path $\pi(s, v)$ to $\vd(S)$.
We do this implicitly by storing the pair $(s, v)$ to represent the new edge.
We call $\pi(s, v)$ a \emph{spoke} of $\R(s)$.
Applying this construction to all sites $s \in S$ with $\R(s) \neq \emptyset$, we obtain a refined decomposition of $P$, which we call the \emph{spoked additively weighted geodesic Voronoi diagram}, or simply the \emph{spoked diagram}.
We denote it by $\vd^*(S)$ (see Figure~\ref{fig:spoked_geodesic_vd}).

Recall that there are two types of edges in $\vd(S)$: Voronoi edges and $\partial P$-portions.
Each $\partial P$-portion becomes a $\partial P$-side in a geodesic sector in $\vd^*(S)$.

Now let $e$ be a Voronoi edge of $\vd(S)$ with endpoints $u$ and $v$, and let $s_1$ and $s_2$ be the two sites associated with $e$.
Consider the region $R_e$ bounded by the four paths $\pi(s_1,u)$, $\pi(u,s_2)$, $\pi(s_2,v)$, and $\pi(v,s_1)$.
Each of these paths is an edge in $\vd^*(S)$.
Note that the Voronoi edge $e$ is contained in $R_e$.
We have the following lemma.

\begin{lemma}\label{lem:decomgeoquad}
At least one of $\pi(s_1, s_2)$ and $\pi(u, v)$ is contained in $R_e$.
Moreover, we can determine which of these two paths lies in $R_e$ in $O(\log m)$ time using the GH data structure.
\end{lemma}

\begin{proof}
We distinguish two cases depending on whether $R_e$ is geodesically convex.

If $R_e$ is geodesically convex, then both $\pi(s_1, s_2)$ and $\pi(u, v)$ are contained in $R_e$, and the lemma trivially holds.

Otherwise, one of the points $s_1$, $s_2$, $u$, and $v$ must lie in the interior of the geodesic convex hull $Q$ of $\{s_1, s_2, u, v\}$.
Without loss of generality, assume that $v$ lies in the interior of $Q$.
Then $s_1$, $s_2$, and $u$ must lie on the boundary of $Q$.

Observe that $\pi(v, s_1) \cup \pi(v, s_2)$ partitions $Q$ into two regions, one of which is $R_e$.
Since $\pi(u, v)$ is contained in $Q$ and shortest paths cannot cross except along a shared prefix, $\pi(v, u)$ cannot cross either $\pi(v, s_1)$ or $\pi(v, s_2)$.
Because both endpoints $u$ and $v$ lie in $R_e$, it follows that the entire path $\pi(u, v)$ must be contained in $R_e$.

Next, we explain how to determine which path lies in $R_e$.
The apexes of the geodesic convex hull of a set $A$ of $k$ points in $P$ can be computed in $O(k \log (m + k))$ time using the GH data structure~\cite{ref:GuibasOp89}.
In our case, $k = 4$, so we can compute the apexes of $Q$ in $O(\log m)$ time.

If both $s_1$ and $s_2$ are apexes of $Q$, then according to the above discussion, the only possible case in which $R_e$ is not geodesically convex is when either $u$ or $v$ is in the interior of $Q$, in which case $\pi(u, v)$ is contained in $R_e$. Hence, in this case, we can assert that $\pi(u, v)$ is contained in $R_e$; otherwise, according to the above analysis, $\pi(s_1, s_2)$ must be contained in $R_e$. Therefore, we can determine which path lies in $R_e$ in $O(\log m)$ time.
\end{proof}

Based on the above lemma, we further process $R_e$ as follows.
We first determine whether $\pi(s_1, s_2)$ or $\pi(u, v)$ lies in $R_e$ by applying the lemma.
We then add the corresponding path (implicitly) to $\vd^*(S)$ as a new edge, and remove the original Voronoi edge $e$ from $\vd^*(S)$.
This new edge partitions $R_e$ into two geodesic triangles, and we associate each triangle with the two sites $s_1$ and $s_2$.

We perform this operation for every Voronoi edge of $\vd(S)$, which takes a total of $O(n \log m)$ time.
As a result, we obtain a geodesic decomposition $T$ of $P$ consisting of $O(n)$ geodesic triangles and geodesic sectors.

We then construct a point location data structure for $T$ in $O(n \log m)$ time using Theorem~\ref{thm:pointlocation}.
Given a query point $q$, we use this data structure to locate the region of $T$ that contains $q$, which takes $O(\log (n+m))$ time.

If $q$ lies in a geodesic sector $\sector$, then $\sector$ is contained in the Voronoi region of its associated site, and we can return this site as the nearest neighbor of $q$.

If $q$ lies in a geodesic triangle $\triangle$, then there are exactly two candidate sites, namely the two sites $s_1$ and $s_2$ associated with $\triangle$.
We determine which of the two sites is closer to $q$ by computing their weighted distances to $q$, which can be done in $O(\log m)$ time using the GH data structure.

Therefore, the nearest neighbor of $q$ in $S$ can be found in $O(\log (n+m))$ time.

This completes the proof of Theorem~\ref{thm:mainthm}.

\section{Applications}
\label{sec:applications}

In this section, we present several applications of our results, in particular Theorem~\ref{thm:mainthm}.

\subsection{Dynamic nearest neighbor queries}
\label{sec:dynamic}

An immediate application is the design of data structures for dynamic nearest neighbor queries.
Let $S$ be a set of weighted point sites in $P$.
We wish to design a \emph{fully dynamic data structure} that maintains $S$ under the following operations:
(1) \textbf{Insertions}: insert a weighted point into $S$;
(2) \textbf{Deletions}: delete a point from $S$; and
(3) \textbf{Nearest neighbor queries}: given a query point $q$, find its nearest neighbor in $S$ with respect to weighted geodesic distances.
If deletions are not allowed, we refer to this as the \emph{insertion-only} case.
If insertions are not allowed, we refer to this as the \emph{deletion-only} case.

Using standard static-to-dynamic transformations~\cite{ref:BentleyDe80,ref:OvermarsTh83}, we obtain the following corollaries by directly applying Theorem~\ref{thm:mainthm}.

\begin{corollary} \label{cor:insert_only}
Let $P$ be a simple polygon with $m$ vertices.
After $O(m)$ preprocessing of $P$, we can maintain an insertion-only additively weighted geodesic nearest neighbor data structure for a set $S$ of $n$ weighted point sites in $P$ that supports amortized $O(\log^2 n + \log n \log^2 m)$ insertion time and $O(\log n \log (n + m))$ query time.
\end{corollary}

\begin{proof}
This follows by applying the logarithmic method of Bentley and Saxe~\cite{ref:BentleyDe80} to Theorem~\ref{thm:mainthm}.
\end{proof}

\begin{corollary} \label{cor:fully_dynamic}
Let $P$ be a simple polygon with $m$ vertices.
After $O(m)$ preprocessing of $P$, given a set $S$ of $n$ weighted point sites in $P$, we can construct a fully dynamic additively weighted geodesic nearest neighbor data structure for $S$ in $O(n \log n + n \log^2 m)$ time that supports $O(\log n + \log^2 m)$ amortized insertion time, $O(\sqrt{n} (\log n + \log^2 m))$ amortized deletion time, and $O(\sqrt{n} \log (n + m))$ query time.
\end{corollary}

\begin{proof}
By applying the equal block method of Overmars~\cite{ref:OvermarsTh83} to Theorem~\ref{thm:mainthm},
we get $O(\sqrt{n} (\log n + \log^2 m))$ update time and $O(\sqrt{n} \log (n + m))$ query time.
We briefly describe the main idea and then show how to reduce the insertion time to $O(\log n + \log^2 m)$.

The main idea is to partition $S$ into $O(\sqrt{n})$ groups, each of size $O(\sqrt{n})$.
Initially, we construct the data structure of Theorem~\ref{thm:mainthm} for each group.
For each deletion, we rebuild the data structure for the group containing the deleted point.
For each insertion, we insert the new point into the smallest group and rebuild that group's data structure.
For each query, we query all groups and return the best result.
Occasionally, we also need to rebalance the group size.
This results in updates taking amortized $O(\sqrt{n} (\log n + \log^2 m))$ time
and queries taking $O(\sqrt{n} \log (n + m))$ time.

We can improve the insertion time to amortized $O(\log n + \log^2 m)$ as follows.
We partition $S$ into two sets $S_1$ and $S_2$,
where $|S_1| \leq \sqrt{n}$ and $S_2 = S \setminus S_1$.
We use the equal block method for $S_2$,
and we maintain $S_1$ arbitrarily in a linked list.
When we insert a new site into $S$,
we insert it into $S_1$ in $O(1)$ time.
If $|S_1| > \sqrt{n}$ after the insertion,
then we build the data structure of Theorem~\ref{thm:mainthm} for $S_1$,
and add this as a new block in $S_2$,
after which we have $S_1 = \emptyset$ and $S_2 = S$.
This takes $O(|S_1| \cdot (\log n + \log^2 m))$ time,
so the amortized time per insertion is $O(\log n + \log^2 m)$.
Using an incremental global rebuilding method~\cite{ref:OvermarsTh83}, we could even make this insertion time worst-case.

To delete a site, if it is in $S_1$, then we simply remove it from the linked list for $S_1$.
If it is in $S_2$, then we delete it as in the equal block method.
Therefore, each deletion takes $O(\sqrt{n} (\log n + \log^2 m))$ amortized time.

Given a query point $q \in P$,
we find its nearest neighbor in $S_1$ and its nearest neighbor in $S_2$,
and then return the one closer to $q$.
We can find the nearest neighbor to $q$ in $S_2$ in $O(\sqrt{n} \log (n + m))$ time.
To find the nearest neighbor to $q$ in $S_1$,
we compute $d_s(q)$ for every site $s \in S_1$
in $O(|S_1| \log m) = O(\sqrt{n} \log m)$ time using the GH data structure,
and then return the one nearest to $q$.
Therefore, each query can be answered in $O(\sqrt{n} \log (n + m))$ time.
\end{proof}



In some applications (e.g., our algorithm in Section~\ref{sec:weighteddiskgraph}), a deletion-only data structure is required.
We do not currently have a better solution than the fully dynamic structure in Corollary~\ref{cor:fully_dynamic}.
However, for reference purposes, we restate the corresponding result below.

\begin{corollary} \label{cor:delete_only}
Let $P$ be a simple polygon with $m$ vertices.
After $O(m)$-time preprocessing of $P$, given a set $S$ of $n$ weighted point sites in $P$, we can construct a deletion-only additively weighted geodesic nearest neighbor data structure for $S$ in $O(n \log n + n \log^2 m)$ time that supports $O(\sqrt{n} (\log n + \log^2 m))$ deletion time and $O(\sqrt{n} \log (n + m))$ query time.
\end{corollary}

\subsection{Shortest paths in unweighted geodesic disk graphs}
\label{sec:unweighteddiskgraph}

For a point $p \in P$ and a radius $r_p$, a \emph{geodesic disk} centered at $p$ with radius $r_p$ consists of all points in $P$ whose geodesic distance from $p$ is at most $r_p$.

Let $S$ be a set of $n$ point sites in $P$, where each point $p \in S$ is associated with a radius $r_p$ defining a geodesic disk centered at $p$.
The \emph{geodesic disk graph} $G(S)$ induced by $S$ has vertex set $S$ and contains an edge between two vertices whenever their geodesic disks intersect.
Specifically, two vertices $u, v \in S$ are adjacent in $G(S)$ if $d(u, v) \le r_u + r_v$.

Given a source vertex $s \in S$, we consider the single-source shortest path (SSSP) problem, which asks to compute shortest paths from $s$ to all other vertices in $G(S)$.
We focus on the \emph{unweighted case}, where the length of a path in $G(S)$ is defined as the number of edges in the path.

For the Euclidean case (i.e., when $P$ is the entire plane), the problem can be solved in $O(n \log n)$ time~\cite{ref:BrewerAn25}, which matches an $\Omega(n \log n)$ lower bound~\cite{ref:CabelloSh15}.
For the geodesic unit-disk case, where all radii $r_p$ are equal, an algorithm with running time $O(m + n \log n \log^2 m)$ was given in~\cite{ref:BrewerSh26}.

We present an algorithm for the general geodesic disk case with running time $O(m + n \log n + n \log^2 m)$.
Theorem~\ref{thm:mainthm} plays a crucial role in this algorithm.
With this result, our algorithm can be viewed either as an extension of the Euclidean algorithm in~\cite{ref:BrewerAn25,ref:deBergAn25} to simple polygons, or as an extension of the geodesic unit-disk algorithm in~\cite{ref:BrewerSh26} to general geodesic disks.
In particular, our running time matches the Euclidean bound when $m = O(1)$, and improves upon the result of~\cite{ref:BrewerSh26}.
Our overall framework follows those in~\cite{ref:BrewerSh26,ref:BrewerAn25}; for completeness, we present the main ideas below.

\subsubsection{Notation}
\label{sec:notation}

We follow the notation from Section~\ref{sec:pre}.
In addition, for $i = 0, 1, 2, \ldots$, let $S_i$ denote the set of vertices of $G(S)$ whose shortest-path distance from $s$ in $G(S)$ is exactly $i$.
Clearly, $S_0 = \{s\}$.
Define $S_{\le i} = \bigcup_{j \le i} S_j$.
Our goal is to compute all sets $S_i$.

For each vertex $v \in S$, we assign a weight $w(v) = -r_v$ and define the \emph{weighted distance} $d_v(p) = w(v) + d(v, p)$ for any point $p \in P$, as before.

For ease of exposition, we assume a general position condition that no point in $V_P \cup S$ is equidistant to two other points in $V_P \cup S$.
This implies that the nearest neighbor of any point $p \in S$ in any subset $S' \subseteq S$ is unique.

We use $\DT(S)$ to denote the dual graph of the Voronoi diagram $\vd(S)$; that is, the vertex set of $\DT(S)$ is $S$, and two sites $u, v \in S$ are adjacent in $\DT(S)$ if their Voronoi regions share a Voronoi edge.
For any site $p \in S$, let $\N(p)$ denote the set of vertices in $\DT(S)$ adjacent to $p$.
For a subset $S' \subseteq S$, we slightly abuse notation and define $\N(S') = \bigcup_{p \in S'} \N(p)$.

We first assume that every site in $S$ has a nonempty Voronoi region in $\vd(S)$.
We will later explain how to modify the algorithm when this assumption does not hold.
\subsubsection{Algorithm description and runtime}

We first construct $\vd(S)$ and thus obtain $\DT(S)$.
The algorithm then proceeds in iterations, with $S_0 = \{s\}$ initially.
In the $i$-th iteration for $i \ge 1$, we compute $S_i$ as follows, assuming that $S_{i-1}$ is already available.

Let $Q = \N(S_{i-1}) \setminus S_{\le i-1}$, i.e., the set of sites in $S$ that are adjacent to those in $S_{i-1}$ in $\DT(S)$ but are not yet in $S_{\le i-1}$.
While $Q \neq \emptyset$, we remove a site $p$ from $Q$ and determine whether $d_{S_{i-1}}(p) \le r_p$.
If so, we add $p$ to $S_i$, and update
$Q \gets Q \cup (\N(p) \setminus S_{\le i})$,
i.e., add to $Q$ all neighbors of $p$ that are not already in $Q \cup S_{\le i}$.
The $i$-th iteration terminates once $Q$ becomes empty.
If $S_i = \emptyset$, the algorithm stops; otherwise, we proceed to the next iteration.
See Algorithm~\ref{algo:simple} for the pseudocode.

\begin{algorithm}
    \caption{An unweighted SSSP algorithm} \label{algo:simple}
    \KwIn{A simple polygon $P$, sites $S \subseteq P$, source site $s \in S$, and radius $r_p$ for each site $p \in S$.}
    \KwOut{$S_i$ for $i = 0, 1, \ldots$}
    $i \gets 0$\;
    $S_0 \gets \{s\}$\;
    Build the implicit $\vd(S)$ by Theorem~\ref{thm:mainthm}\;
    Build $\DT(S)$ from the implicit $\vd(S)$\label{line:euc_dt}\;
    \While{$S_i \neq \emptyset$}{
        $i \gets i + 1$\;
        $S_i \gets \emptyset$\;
        $Q \gets \N(S_{i - 1}) \setminus S_{\le i - 1}$ \label{line:euc_Q1}\;
        \While{$Q \neq \emptyset$ \label{line:euc_Q_while}}{
            $p \gets pop(Q)$\;
            \If{$d_{S_{i - 1}}(p) \le r_p$ \label{line:euc_check}}{
                $S_i \gets S_i \cup \{p\}$\; \label{line:add}
                $Q \gets Q \cup (\N(p) \setminus S_{\le i})$ \label{line:euc_Q2}\;
            }
        }
    }
    \Return{$S_i$ for $i = 0, 1, \ldots$\;}
\end{algorithm}

Compared to the algorithms in~\cite{ref:BrewerAn25,ref:BrewerSh26}, the only difference lies in Line~\ref{line:euc_check}.
The overall time analysis is similar, except that we need to efficiently implement this step.

To evaluate Line~\ref{line:euc_check}, we apply Theorem~\ref{thm:mainthm} and construct a nearest neighbor data structure for $S_{i-1}$.
This takes $O(|S_{i-1}|\cdot (\log n + \log^2 m))$ time.
Using this data structure, we can compute $d_{S_{i-1}}(p)$ by finding the nearest neighbor of $p$ in $S_{i-1}$ in $O(\log (n + m))$ time.
Since the sum of $|S_{i-1}|$ over all iterations is at most $n$, the total time spent constructing nearest neighbor data structures is $O(n(\log n + \log^2 m))$.
The total time spent on Line~\ref{line:euc_check} is $O(n \log (n + m))$.

Following the analysis in~\cite{ref:BrewerAn25,ref:BrewerSh26} for the remaining parts of the algorithm (note that $\DT(S)$ can be obtained from the implicit $\vd(S)$ in $O(n)$ time), the overall running time is $O(n \log n + n \log^2 m)$, excluding the initial $O(m)$ preprocessing time for $P$ in Theorem~\ref{thm:mainthm}.

\subsubsection{Algorithm correctness}

We now prove that the algorithm is correct.
The analysis is very similar to those in~\cite{ref:BrewerAn25,ref:BrewerSh26}.
In particular, with the following Lemma~\ref{lem:path}, the remainder of the proof is identical to that in~\cite[Section~3.3]{ref:BrewerAn25}.

We first state the following observation, which is a direct extension of the Euclidean case in~\cite{ref:BrewerAn25} and will be used in the proof of Lemma~\ref{lem:path}.

\begin{observation}[{\cite{ref:BrewerAn25}}]\label{obser:weightdis}
For any two sites $u, v \in S$, they are connected by an edge in $\G$ if and only if $d_v(u) \leq r_u$ or $d_u(v) \leq r_v$.
\end{observation}

\begin{lemma} \label{lem:path}
For any $i > 0$ and any site $v \in S_i$, there exists a site $u \in S_{i-1}$ and a path from $u$ to $v$ in $\DT(S)$ such that all internal vertices of the path lie in $S_i$.
\end{lemma}

\begin{proof}
The proof is similar to~\cite[Lemma~3]{ref:BrewerAn25} and~\cite[Lemma~36]{ref:BrewerSh26}.
We briefly describe it below.

Let $u$ be the nearest neighbor of $v$ in $S_{i-1}$.
By our general position assumption, $u$ is unique.
Consider the shortest path $\pi(u,v)$.
Let $u = w_0, w_1, \ldots, w_k = v$ be the sites whose Voronoi regions in $\vd(S)$ intersect $\pi(u, v)$ in order from $u$ to $v$.
Note that repetitions are possible, i.e., $w_j = w_{j'}$ for $j \neq j'$.

For ease of discussion, we assume that $\pi(u,v)$ does not pass through any Voronoi vertex of $\vd(S)$ (otherwise, we can perturb $v$ slightly and apply the same argument).
Then, for each $0 \le j \le k - 1$, the path $\pi(u, v)$ crosses the Voronoi edge between $w_j$ and $w_{j + 1}$, implying that $w_j$ and $w_{j + 1}$ are adjacent in $\DT(S)$.
Hence, $\omega = w_0, w_1, \ldots, w_k$ is a path from $u$ to $v$ in $\DT(S)$.

It remains to show that all internal vertices $w_j$ ($1 \le j \le k-1$) of $\omega$ lie in $S_i$.
We proceed in two steps.
First, we show that $G(S)$ contains an edge $(u, w_j)$, which implies $w_j \in S_{i-2} \cup S_{i-1} \cup S_i$ since $u \in S_{i-1}$.
Second, we show that $w_j \notin S_{i-2} \cup S_{i-1}$, which implies $w_j \in S_i$.

\paragraph{Proving that $\boldsymbol{G(S)}$ has an edge $\boldsymbol{(u,w_j)}$.}
We first show that $G(S)$ has an edge $(u,v)$.
Since $v \in S_i$, there exists a vertex $u' \in S_{i-1}$ adjacent to $v$ in $\G$.
By Observation~\ref{obser:weightdis}, $d_{u'}(v) \le r_v$.
Since $u$ is the nearest neighbor of $v$ in $S_{i-1}$, we have $d_u(v) \le d_{u'}(v) \le r_v$.
Therefore, by Observation~\ref{obser:weightdis}, $\G$ has an edge $(u,v)$.

Now consider any $w_j$ with $1 \le j \le k-1$.
Let $q$ be a point in $\pi(u,v) \cap \R(w_j)$.
We derive
\[
d_{w_j}(u) = d(w_j, u) - r_{w_j}
\overset{(1)}{\le} d(w_j, q) + d(q, u) - r_{w_j}
\overset{(2)}{\le} d(v, q) + d(q, u) - r_v
\overset{(3)}{=} d(v, u) - r_v
= d_v(u)
\overset{(4)}{\le} r_u,
\]
where
(1) follows from the triangle inequality,
(2) follows from $q \in \R(w_j)$,
(3) follows from $q \in \pi(u,v)$, and
(4) follows from Observation~\ref{obser:weightdis} and the fact that $(u,v)$ is an edge in $\G$.
Thus, by Observation~\ref{obser:weightdis}, $G(S)$ has an edge $(u,w_j)$.

\paragraph{Proving that $\boldsymbol{w_j \notin S_{i-2} \cup S_{i-1}}$.}
Let $q \in \pi(u, v) \cap \R(w_j)$ be a point that does not lie on any Voronoi edge of $\R(w_j)$.
Then we derive
   \[d_{w_j}(v) \os{1} \le d(w_j, q) + d(q, v) - r_{w_j} \os{2} < d(u, q) + d(q, v) - r_u \os{3} = d(u, v) - r_u = d_u(v) \os{4} \le r_v,\]
where
(1) follows from the triangle inequality,
(2) follows from $q \in \R(w_j)$ and the fact that $q$ is not on a Voronoi edge,
(3) follows from $q \in \pi(u, v)$, and
(4) follows from Observation~\ref{obser:weightdis} and the fact that $(u, v)$ is an edge in $\G$.

Thus $d_{w_j}(v) < d_u(v)$, which contradicts the fact that $u$ is the nearest neighbor of $v$ in $S_{i-1}$ if $w_j \in S_{i-1}$.
Hence $w_j \notin S_{i-1}$.

Furthermore, since $(w_j,v)$ is an edge in $G(S)$ and $v \in S_i$, $w_j$ cannot lie in $S_{i-2}$; otherwise $v$ would lie in $S_{\le i-1}$, a contradiction.

Therefore, $w_j \notin S_{i-2} \cup S_{i-1}$, and thus $w_j \in S_i$.
\end{proof}

\paragraph{Handling empty Voronoi regions.}
The above discussion assumes that every site in $S$ has a non-empty Voronoi region in $\vd(S)$.
We now explain how to handle the general case.
Suppose that $v$ is a site in $S$ whose Voronoi region in $\vd(S)$ is empty.
Let $u$ be the site in $S$ whose Voronoi region contains $v$.
We modify $\DT(S)$ by adding an edge between $u$ and $v$, i.e., we add $v$ to $\N(u)$ and $u$ to $\N(v)$.
We perform this modification for all such sites $v$.
After this preprocessing, we apply exactly the same algorithm as before.
Note that the algorithm also works even if the source site $s$ itself has an empty Voronoi region in $\vd(S)$.
The overall running time remains asymptotically unchanged.

\medskip

\begin{theorem}
Let $P$ be a simple polygon with $m$ vertices.
After $O(m)$ preprocessing of $P$, given a set $S$ of $n$ points in $P$, each associated with a radius, and a source point $s \in S$, we can compute shortest paths from $s$ to all points in $S$ in the geodesic disk graph of $S$ in $O(n \log n + n \log^2 m)$ time.
\end{theorem}

\subsection{Shortest paths in weighted geodesic disk graphs}
\label{sec:weighteddiskgraph}

In this section, we study the weighted version of the SSSP problem introduced in Section~\ref{sec:unweighteddiskgraph}.
Here, each edge $(u,v)$ of $G(S)$ has weight equal to $d(u,v)$, and the length of a path in $G(S)$ is defined as the total weight of its edges.
We follow the notation from Section~\ref{sec:unweighteddiskgraph}.

The current best algorithm for the Euclidean case of this weighted problem runs in $O(n \log^4 n)$ time~\cite{ref:AnSi25}.
For the geodesic unit-disk case, an $O(m + n \log^2 n \log^2 m)$-time algorithm was given in~\cite{ref:BrewerSh26}.

We solve our weighted SSSP problem by extending the algorithm in~\cite{ref:BrewerSh26}.
Our algorithm runs in $O(m + n^{3/2} (\log n + \log^2 m))$ time.
In particular, it uses both the insertion-only data structure in Corollary~\ref{cor:insert_only} and the deletion-only data structure in Corollary~\ref{cor:delete_only}.
Note that the algorithm in~\cite{ref:AnSi25} relies on quadtrees, which appear inherently tied to the Euclidean plane and do not extend naturally to simple polygons.
Our technique introduces a new algorithmic framework that is conceptually simple and may be of independent interest for other shortest path problems in disk graphs and, more broadly, geometric intersection graphs.

In the following, we describe the algorithm and prove its correctness in Section~\ref{sec:description}, and discuss its implementation and running time in Section~\ref{sec:imple}.

\subsubsection{Algorithm description and correctness}
\label{sec:description}

For each point $p \in S$, let $\delta_G(p)$ denote the shortest path length from $s$ to $p$ in $G(S)$.
We focus on computing $\delta_G(p)$ for all $p \in S$; the algorithm can be extended in a standard way to output a shortest path tree.

For two points $p, q \in S$, we say that they are \emph{adjacent} in $G(S)$ if they share an edge in the graph, i.e., if $d(p,q) \le r_p + r_q$.

The algorithm maintains a tentative distance value $dis(p)$ for each point $p$, such that $dis(p) = \delta_G(p)$ when the algorithm terminates.
Initially, we set $dis(s) = 0$ and $dis(p) = \infty$ for all $p \in S \setminus \{s\}$.

We maintain two subsets $A$ and $B$ of $S$, where $A$ consists of points whose shortest path distances have been finalized (i.e., $dis(p) = \delta_G(p)$ for all $p \in A$), and $B = S \setminus A$.
Furthermore, we partition $A$ into two subsets $A_1$ and $A_2$ such that $A_1$ has the following property: no point in $A_1$ is adjacent to any point in $B$ in $G(S)$.

The algorithm maintains an invariant that is stated in the following observation for reference.

\begin{observation}\label{obser:invariant}
For every point $p \in A$, we have $dis(p) = \delta_G(p)$.
Furthermore, if $p \in A_1$, then every point adjacent to $p$ in $G(S)$ also belongs to $A$.
\end{observation}

Initially, we set $A_1 = \emptyset$, $A_2 = \{s\}$, and $B = S \setminus \{s\}$.
The algorithm proceeds iteratively (see Algorithm~\ref{algo:spdisk} for the pseudocode).

In each iteration, we select a point $a \in A_2$ that minimizes $dis(a) + r_a$.
We then perform the following three main steps:

\begin{enumerate}
\item Compute the set
\[
B_a = \{p \in B : (a,p) \text{ is an edge in } G(S)\},
\]
i.e., the points in $B$ that are adjacent to $a$ in $G(S)$.

\item For each point $b \in B_a$, among all points of $A$ adjacent to $b$ in $G(S)$, find the point $q_b$ minimizing $dis(q_b) + d(q_b, b)$, and set
\[
dis(b) = dis(q_b) + d(q_b, b).
\]

\item Move all points of $B_a$ from $B$ to $A_2$, and move $a$ from $A_2$ to $A_1$.
\end{enumerate}

This completes one iteration.
Lemma~\ref{lem:invariant} (to be given below) shows that for every $b \in B_a$, the computed value $dis(b)$ equals $\delta_G(b)$.
Together with step (3), this ensures that the invariant in Observation~\ref{obser:invariant} is preserved, thereby establishing correctness.

The algorithm terminates when $A_2$ becomes empty.
At that point, if $B \neq \emptyset$, then the vertices in $B$ are not reachable from $s$ in $G(S)$ (by Observation~\ref{obser:invariant}), and thus $\delta_G(p) = \infty$ for all $p \in B$.

\begin{algorithm}[h]
	\caption{A weighted SSSP algorithm}
	\label{algo:spdisk}
	\SetAlgoNoLine
	\KwIn{A simple polygon $P$, sites $S \subseteq P$, source site $s \in S$, and radius $r_p$ for each site $p \in S$.}
	\KwOut{$dis(p)$ for all $p \in S$}
        \BlankLine
        $A_1 \leftarrow \emptyset$, $A_2 \leftarrow \{s\}$, $B \leftarrow S \setminus \{s\}$, $dis(s) \leftarrow 0$\;
        \While{$A_2 \neq \emptyset$}
        {
          $a \leftarrow \argmin_{p \in A_2} (dis(p) + r_p)$\;
          $B_a \leftarrow \{p \in B : (a,p) \text{ is an edge in } G(S)\}$\;
          \For{\em each $b \in B_a$}
          {
            $q_b \leftarrow \argmin_{p \in A,\ (p,b)\text{ is an edge in } G(S)} (dis(p) + d(p,b))$\;
            $dis(b) \leftarrow dis(q_b) + d(q_b, b)$\;
          }
          move all points of $B_a$ from $B$ to $A_2$\;
          move $a$ from $A_2$ to $A_1$\;
        }
\end{algorithm}

\begin{lemma}\label{lem:invariant}
For each point $b \in B_a$, we have $\delta_G(b) = dis(q_b) + d(q_b, b)$.
\end{lemma}

\begin{proof}
Consider a point $b \in B_a$. Our goal is to prove that $\delta_G(b) = dis(q_b) + d(q_b, b)$.

Assume for contradiction that $\delta_G(b) \neq dis(q_b) + d(q_b, b)$.
By the definition of $q_b$, the point $q_b$ is adjacent to $b$ in $G(S)$, and thus $\delta_G(b) \le dis(q_b) + d(q_b, b)$.
Since equality does not hold, we obtain
\begin{equation}\label{equ:contradict}
    \delta_G(b) < dis(q_b) + d(q_b, b).
\end{equation}

Let $\Pi_G(s,b)$ be a shortest path from $s$ to $b$ in $G(S)$, and among all such paths, let $\Pi_G(s,b)$ be the one with the smallest number of edges.
Let $p$ be the first vertex on $\Pi_G(s,b)$ that is not in $A$.
Since $b \notin A$, such a vertex $p$ must exist.
Let $q$ be the predecessor of $p$ on $\Pi_G(s,b)$ (see Figure~\ref{fig:path}).
By definition, $q$ is adjacent to $p$ in $G(S)$ and $q \in A$.
We distinguish two cases depending on whether $p = b$.

\paragraph{Case $\boldsymbol{p = b}$.}
In this case, $\delta_G(b) = \delta_G(q) + d(q,b)$.
By the definition of $q_b$, we have
\[
dis(q_b) + d(q_b, b) \le dis(q) + d(q, b).
\]
Since $q \in A$, by Observation~\ref{obser:invariant}, we have $dis(q) = \delta_G(q)$.
Thus,
\[
\delta_G(b) = \delta_G(q) + d(q,b) = dis(q) + d(q,b) \ge dis(q_b) + d(q_b,b),
\]
which contradicts \eqref{equ:contradict}.

\paragraph{Case $\boldsymbol{p \neq b}$.}
In this case, we further distinguish whether $q \in A_1$ or $q \in A_2$.

If $q \in A_1$, then by Observation~\ref{obser:invariant}, every vertex adjacent to $q$ in $G(S)$ must already be in $A$.
Since $q$ is adjacent to $p$, this implies $p \in A$, contradicting the fact that $p\not\in A$.
Hence, this case is impossible.

\begin{figure}[t]
\begin{minipage}[t]{\linewidth}
\begin{center}
\includegraphics[totalheight=0.9in]{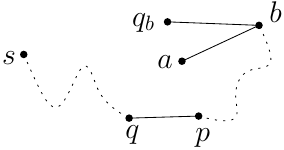}
\caption{The three solid segments represent edges in $G(S)$. The two dotted curves represent two subpaths of $\Pi_G(s,b)$.}
\label{fig:path}
\end{center}
\end{minipage}
\end{figure}

We now consider the case $q \in A_2$.
By the definition of $a$, since $q \in A_2$, we have
\begin{equation}\label{equ:invariant20}
    dis(a) + r_a \le dis(q) + r_q.
\end{equation}

Recall that $a$ is adjacent to $b$ in $G(S)$, and hence $d(a,b) - r_a - r_b \le 0$.
On the other hand, since $p \neq b$ and $q$ is the predecessor of $p$ on $\Pi_G(s,b)$, the vertex $q$ cannot be adjacent to $b$ in $G(S)$; otherwise, the path $\Pi_G(s,b)$ would use the edge $(q,b)$ directly.
Thus, $d(q,b) - r_q - r_b > 0$.
Combining these two inequalities yields
\begin{equation}\label{equ:invariant30}
    d(a,b) - r_a < d(q,b) - r_q.
\end{equation}

Combining \eqref{equ:invariant20} and \eqref{equ:invariant30}, we obtain
\[
dis(a) + d(a,b) < dis(q) + d(q,b).
\]
By the definition of $q_b$, we have
\[
dis(q_b) + d(q_b,b) \le dis(a) + d(a,b),
\]
and hence
\[
dis(q_b) + d(q_b,b) < dis(q) + d(q,b).
\]
Since both $q_b$ and $q$ are in $A$, by Observation~\ref{obser:invariant}, we have $dis(q_b) = \delta_G(q_b)$ and $dis(q) = \delta_G(q)$.
Therefore,
\begin{equation}\label{equ:invariant40}
    \delta_G(q_b) + d(q_b,b) < \delta_G(q) + d(q,b).
\end{equation}

On the other hand, the length of $\Pi_G(s,b)$ is $\delta_G(b)$, which can be written as the sum of the length of the subpath from $s$ to $q$, namely $\delta_G(q)$, and the length of the subpath from $q$ to $b$.
By the triangle inequality, the latter is at least $d(q,b)$, and thus
\begin{equation}\label{equ:invariant50}
    \delta_G(b) \ge \delta_G(q) + d(q,b).
\end{equation}

Combining \eqref{equ:invariant40} and \eqref{equ:invariant50}, we obtain
\[
\delta_G(b) > \delta_G(q_b) + d(q_b,b),
\]
which contradicts \eqref{equ:contradict}.

Therefore, $\delta_G(b) = dis(q_b) + d(q_b,b)$ holds for all $b \in B_a$.
\end{proof}

\subsubsection{Algorithm implementation and time analysis}
\label{sec:imple}

We now elaborate on the efficient implementation of the algorithm, focusing in particular on the three main steps in each iteration.
We maintain the set $A_2$ using a min-heap, where the key of each point $p \in A_2$ is $dis(p) + r_p$.
Thus, the point $a$ can be selected by a delete-min operation in $O(\log n)$ time, and each insertion into $A_2$ also takes $O(\log n)$ time.

In the following, we focus on the first two main steps.
We will show that the first main step can be implemented in $O((|B_a| + 1)\cdot (\tau_d + \tau_q^d))$ time, where $\tau_d$ is the amortized deletion time and $\tau_q^d$ is the query time for a deletion-only data structure of Corollary~\ref{cor:delete_only}.
Similarly, the second main step can be implemented in $O(|B_a|\cdot (\tau_i + \tau_q^i)\cdot \log n)$ time, where $\tau_i$ is the amortized insertion time and $\tau_q^i$ is the query time for an insertion-only data structure of Corollary~\ref{cor:insert_only}.
Since each point enters some set $B_a$ at most once throughout the algorithm, the total running time over all iterations is
\[
O\bigl(n \cdot (\tau_d + \tau_q^d + (\tau_i + \tau_q^i)\cdot \log n)\bigr).
\]

\paragraph{The first main step.}
The goal of the first main step is to compute $B_a$.
By definition, a point $p \in B$ is adjacent to $a$ in $G(S)$ if
$d(p,a) - r_p - r_a \le 0$, which is equivalent to
\[
d(p,a) - r_p \le r_a.
\]

We reduce this problem to a sequence of dynamic nearest neighbor queries under weighted distance.
For each point $p \in B$, we assign a weight $w(p) = -r_p$, so that the weighted distance between a query point $q \in P$ and $p$ is $w(p) + d(p,q)$, as defined previously.

To compute $B_a$, we repeatedly find the nearest neighbor $p \in B$ of $a$, add $p$ to $B_a$, and delete $p$ from $B$, as long as the condition $d(p,a) + w(p) \leq r_a$ holds.
We maintain $B$ in a deletion-only data structure from Corollary~\ref{cor:delete_only}, which supports deletions throughout the entire algorithm.

In this way, computing $B_a$ takes $O((|B_a| + 1)\cdot (\tau_d + \tau_q^d))$ amortized time, excluding the initial $O(n \log n + n \log^2 m)$ time required to build the deletion-only data structure for $B$ in Corollary~\ref{cor:delete_only}.

\paragraph{The second main step.}
The goal of the second main step is to compute $q_b$ for all points $b \in B_a$.

We maintain a balanced binary search tree $T_A$, whose leaves (from left to right) store the points $p \in A$ in increasing order of $dis(p) + r_p$.
For each internal node $v \in T_A$, let $A_v$ denote the subset of points stored at the leaves of the subtree rooted at $v$.
We store two implicit Voronoi diagrams at each node $v$:
\begin{itemize}
    \item $\vd_1(v)$, the additively weighted geodesic Voronoi diagram of $A_v$ with weights $-r_p$;
    \item $\vd_2(v)$, the additively weighted geodesic Voronoi diagram of $A_v$ with weights $dis(p)$.
\end{itemize}
Both diagrams can be constructed using Theorem~\ref{thm:mainthm}.

For each point $b \in B_a$, we compute $q_b$ in two procedures.

\begin{enumerate}
\item
In the first procedure, we find the leftmost leaf $u_b$ of $T_A$ such that $u_b$ is adjacent to $b$ in $G(S)$, i.e.,
\[
d(u_b, b) - r_{u_b} - r_b \le 0.
\]
For convenience, we use $u_b$ to denote both the leaf and the point stored at that leaf.
Once $u_b$ is found, $q_b$ must lie among the leaves to the right of $u_b$ (including $u_b$ itself).

We find $u_b$ by traversing $T_A$ from the root as follows.
At each node $u$, let $v$ be its left child.
Using $\vd_1(v)$, we find the nearest neighbor $q \in A_v$ of $b$, and check whether $d(q,b) - r_q \le r_b$ using the GH data structure.
If so, we recurse into $v$; otherwise, we recurse into the right child of $u$.
This process continues until a leaf is reached, which is returned as $u_b$.

Since each nearest neighbor query takes $O(\log (n+m))$ time by Theorem~\ref{thm:mainthm}, and the height of $T_A$ is $O(\log n)$, the total time for computing $u_b$ is $O(\log n \cdot \log (n+m))$.

\item
In the second procedure, we identify a set $V_b$ of $O(\log n)$ nodes of $T_A$ whose leaf ranges form a partition of all leaves to the right of (and including) $u_b$.
That is, $\bigcup_{v \in V_b} A_v$ is exactly the set of candidate points for $q_b$.

For each node $v \in V_b$, we use $\vd_2(v)$ to find the nearest neighbor $p_v \in A_v$ of $b$.
Among all such $p_v$, let $q^*$ be the one minimizing $dis(p_v) + d(p_v, b)$.
The next lemma shows that $q^* = q_b$, and thus we return $q^*$.

The running time of this procedure is $O(\log n \cdot \log (n+m))$, using Theorem~\ref{thm:mainthm} and the GH data structure.
\end{enumerate}

\begin{lemma}
The point $q^*$ is equal to $q_b$, i.e.,
\[
q^* = \argmin_{p \in A,\ (p,b)\text{ is an edge in } G(S)} \bigl(dis(p) + d(p,b)\bigr).
\]
\end{lemma}

\begin{proof}
By the definition of the leaf $u_b$, the point $q_b$ must lie in $A_v$ for some node $v \in V_b$.
By the definition of $q^*$, to prove that $q^* = q_b$, it suffices to show that $q^*$ is adjacent to $b$ in $G(S)$, i.e.,
$d(q^*, b) - r_{q^*} - r_b \le 0$, or equivalently,
\[
d(q^*, b) - r_{q^*} \le r_b.
\]

First, since the leaves of $T_A$ are sorted in increasing order of $dis(p) + r_p$ from left to right, and $q^*$ is not to the left of $u_b$, we have
\begin{equation*}
    dis(u_b) + r_{u_b} \le dis(q^*) + r_{q^*}.
\end{equation*}

By the definition of $q^*$, we also have
\begin{equation*}
    dis(u_b) + d(u_b, b) \ge dis(q^*) + d(q^*, b).
\end{equation*}

Subtracting the first inequality from the second gives
\begin{equation}\label{equ:qb}
    d(u_b, b) - r_{u_b} \ge d(q^*, b) - r_{q^*}.
\end{equation}

Recall that $u_b$ is adjacent to $b$ in $G(S)$, and hence
$d(u_b, b) - r_{u_b} - r_b \le 0$, or equivalently,
\[
d(u_b, b) - r_{u_b} \le r_b.
\]
Combining this with \eqref{equ:qb}, we obtain
\[
d(q^*, b) - r_{q^*} \le r_b,
\]
which shows that $q^*$ is adjacent to $b$ in $G(S)$.
The lemma follows.
\end{proof}

The above discussion shows that each $q_b$ can be computed in $O(\log n \cdot \log (m+n))$ time using the tree $T_A$.
Since each point $b \in B_a$ is subsequently moved to $A_2$ and then to $A$, we must insert $b$ into $T_A$.
To support such insertions, we replace the static Voronoi diagrams $\vd_1(v)$ and $\vd_2(v)$ at each node $v \in T_A$ with insertion-only Voronoi diagrams as in Corollary~\ref{cor:insert_only}.
Thus, if $\tau_q^i$ denotes the query time of the insertion-only data structure, computing $q_b$ takes $O(\tau_q^i \log n)$ time.

After computing $dis(b) = dis(q_b) + d(q_b, b)$, we insert $b$ as a new leaf into $T_A$, using $dis(b) + r_b$ as its key.
For each ancestor $v$ of the new leaf, we insert $b$ into both $\vd_1(v)$ and $\vd_2(v)$.
Therefore, if $\tau_i$ is the amortized insertion time from Corollary~\ref{cor:insert_only}, the total cost of inserting $b$ into $T_A$ is $O(\tau_i \log n)$.\footnote{
An insertion may trigger a rebalancing operation at some node $v \in T_A$.
Since a node $v$ is rebalanced only after $\Theta(|A_v|)$ new points have been inserted into $A_v$ since its last rebalancing, we can charge the rebalancing cost to these insertions.
Hence, the amortized insertion time remains $O(\tau_i \log n)$.
}

Therefore, the second main step of the algorithm can be implemented in $O(|B_a| \cdot (\tau_i + \tau_q^i)\cdot \log n)$ amortized time.

\paragraph{Putting it all together.}
In summary, the total time over all iterations of the algorithm is
\[
O\bigl(n \cdot (\tau_d + \tau_q^d + (\tau_i + \tau_q^i)\cdot \log n)\bigr),
\]
excluding the time for initially constructing a deletion-only data structure for $B=S\setminus\{s\}$.

We conclude with the following theorem.

\begin{theorem}
Let $P$ be a simple polygon.
Given a set $S$ of $n$ points in $P$, each associated with a radius, and a source point $s \in S$, we can compute shortest paths from $s$ to all points of $S$ in the geodesic disk graph of $S$ in
\[
O\bigl(D(n)+ n \cdot (\tau_d + \tau_q^d + (\tau_i + \tau_q^i)\cdot \log n)\bigr)
\]
time, where $\tau_d$ and $\tau_q^d$ are the amortized deletion and query times for a deletion-only additively weighted geodesic nearest neighbor data structure, $D(n)$ is the time for initially constructing the deletion-only data structure for a set of $n$ points, and $\tau_i$ and $\tau_q^i$ are the corresponding insertion and query times for an insertion-only such data structure.
\end{theorem}

By Corollary~\ref{cor:delete_only}, we have
\[
D(n) = O(n \log n + n \log^2 m), \quad
\tau_d = O(\sqrt{n} (\log n + \log^2 m)), \quad
\tau_q^d = O(\sqrt{n} \log (n + m)),
\]
and by Corollary~\ref{cor:insert_only}, we have
\[
\tau_i = O(\log^2 n + \log n \log^2 m), \quad
\tau_q^i = O(\log n \log (n + m)).
\]
Substituting these bounds yields the following result.

\begin{corollary}\label{cor:weightedshortestpath}
Let $P$ be a simple polygon with $m$ vertices.
After $O(m)$ preprocessing of $P$, given a set $S$ of $n$ points in $P$, each associated with a radius, and a source point $s \in S$, we can compute shortest paths from $s$ to all points of $S$ in the geodesic disk graph of $S$ in $O(n^{3/2}(\log n + \log^2 m))$ time.
\end{corollary}

The running time is dominated by the bounds in Corollary~\ref{cor:delete_only}.
As discussed in Section~\ref{sec:dynamic}, it may be possible to reduce the update and query times in Corollary~\ref{cor:fully_dynamic} (and hence Corollary~\ref{cor:delete_only}) to $O(\log^{O(1)}(n+m))$.
If this can be achieved, then the running time in Corollary~\ref{cor:weightedshortestpath} would improve to $O(n \cdot \log^{O(1)}(n+m))$.

\section{Implicit farthest geodesic Voronoi diagrams}
\label{sec:farthest}

In this section, we prove Theorem~\ref{thm:farthest}.
Unless otherwise stated, we follow the same notation as before.
A key difference in this section is that the sites in $S$ are unweighted.

In what follows, we first introduce some additional notation and concepts in Section~\ref{sec:fvdpre}. We present our algorithm for constructing the implicit $\vd_f(S)$ in Section~\ref{sec:fvdconstruction} and discuss point location in $\vd_f(S)$ in Section~\ref{sec:fvdpointlocation}. Theorem~\ref{thm:farthest} immediately leads to dynamic data structures for geodesic farthest neighbor searching, which are discussed in Section~\ref{sec:fvddynamic}.

\subsection{Notation and concepts}
\label{sec:fvdpre}

For a site $s \in S$, let $\R_f(s)$ denote its \emph{farthest Voronoi region}, i.e.,
\[
\R_f(s) = \{p \in P : d(s,p) \ge \max_{s' \in S} d(s',p)\}.
\]
The collection of these regions forms the \emph{farthest geodesic Voronoi diagram} of $S$, denoted by $\vd_f(S)$.

\begin{figure}
    \centering
    \includegraphics[width=.4\linewidth]{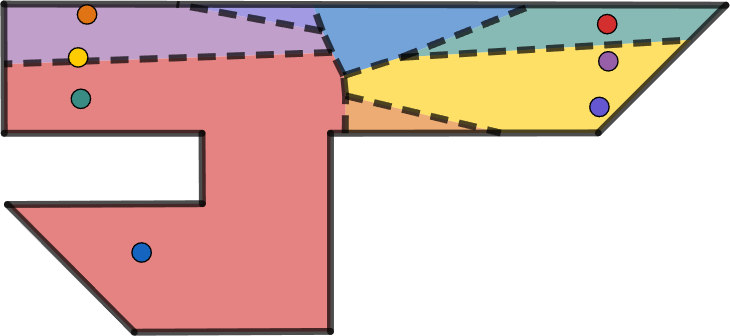}
    \caption{Illustration of a farthest geodesic Voronoi diagram. The circled points are sites.}
    \label{fig:fgvd}
\end{figure}

As before, the common boundary between two regions is called a \emph{Voronoi edge}.
The intersection of three Voronoi edges is a \emph{Voronoi vertex}, and the intersection of a Voronoi edge with $\partial P$ is a \emph{Voronoi terminal}.
A maximal connected portion of $\partial P$ on the boundary of $\R_f(s)$ is called a \emph{$\partial P$-portion}.
See Figure~\ref{fig:fgvd} for an illustration.

If $v$ is a Voronoi vertex or lies on a Voronoi edge, and $s$ is a farthest site of $v$, we say that $s$ is \emph{associated with} $v$.

We make similar general position assumptions as in Section~\ref{sec:high_level}. Under these assumptions, each Voronoi vertex of $\vd_f(S)$ is associated with three sites and each Voronoi terminal is associated with two sites. In particular, the bisector of any two sites of $S$ does not contain any vertex of $P$.

\paragraph{Implicit representation of $\boldsymbol{\vd_f(S)}$.}
We represent $\vd_f(S)$ implicitly in the same way as for the nearest diagram.
Specifically, we explicitly store Voronoi vertices and terminals together with their associated sites.
For each Voronoi edge or $\partial P$-portion, we store only its incident vertices and associated sites.
The size of $\vd_f(S)$ is $O(n)$~\cite{ref:AronovTh93,ref:OhVo20}.

\paragraph{Geodesic center of $\boldsymbol{S}$.}
A point in $P$ is a \emph{geodesic center} of $P$ if it minimizes the maximum geodesic distance to all points of $P$. The geodesic center is unique~\cite{ref:PollackCo89} and it can be computed in $O(m)$ time~\cite{ref:AhnA16}.
A point in $P$ is a \emph{geodesic center} of $S$ if it minimizes the maximum geodesic distance to all sites of $S$; we denote it by $c^*$. It is known that $c^*$ coincides with the geodesic center of the geodesic hull of $S$. Therefore, $c^*$ can be computed in $O(m + n \log n)$ time by first constructing the geodesic hull in $O(m + n \log n)$ time and then applying the linear-time algorithm in~\cite{ref:AhnA16}.
\medskip

In the following two subsections, we assume that we have already performed the same $O(m)$-time global preprocessing for $P$ as discussed in Section~\ref{sec:global}.

\subsection{Algorithm for constructing the implicit $\boldsymbol{\vd_f(S)}$}
\label{sec:fvdconstruction}

To compute the implicit $\vd_f(S)$, we follow the algorithm of Oh and Ahn~\cite{ref:OhVo20}. The bottleneck of their algorithm is to compute the geodesic center $c^*$ of $S$, which takes $O(n \log n \log^2 m)$ time. The rest of their algorithm for computing the implicit $\vd_f(S)$ runs in $O(n \log n + n \log^2 m)$ time in total. In the following, we show that $c^*$ can be computed in $O(n \log n + n \log^2 m)$ time, and this leads to an algorithm to compute the implicit $\vd_f(S)$ in a total of $O(n \log n + n \log^2 m)$ time. Note that although computing $c^*$ can also be done in $O(m + n \log n)$ time as discussed above, here we need an algorithm whose running time depends only logarithmically on $m$.

If a line segment inside $P$ connects two points on the boundary of $P$ and it is not an edge of $P$, then we call it a {\em diagonal}. A diagonal partitions $P$ into two {\em sides}, each being a subpolygon of $P$.

The following is a key lemma to the success of our approach.
Our algorithm for the lemma uses the tentative prune-and-search approach of Kirkpatrick and Snoeyink~\cite{ref:KirkpatrickTe95}. Since the proof is lengthy and somewhat technical, we defer it to Section~\ref{sec:diagonalintersection}.

\begin{lemma}\label{lem:diagonalintersection}
Given a diagonal $d$ of $P$ and two sites $s_1, s_2 \in S$ in the same side of $d$, we can compute the intersection between $d$ and the bisector $B(s_1, s_2)$, or report that no intersection exists, in $O(\log m)$ time.
\end{lemma}

\begin{figure}[t]
\begin{minipage}[t]{\linewidth}
\begin{center}
\includegraphics[totalheight=1.2in]{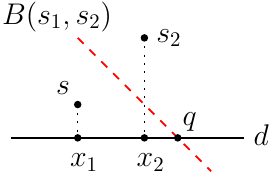}
\caption{Illustrating a counterexample: the dashed red line is the bisector $B(s_1, s_2)$ of $s_1$ and $s_2$, and $x_1$ and $x_2$ are the points on $d$ closest to $s_1$ and $s_2$, respectively. In this example, the intersection $q = B(s_1, s_2) \cap d$ does not lie between $x_1$ and $x_2$. }
\label{fig:counterexample}
\end{center}
\end{minipage}
\end{figure}

Note that $B(s_1, s_2)$ intersects $d$ in at most one point~\cite{ref:AronovOn89,ref:AgarwalIm18}.
In fact, the result of Lemma~\ref{lem:diagonalintersection} was already claimed by Oh and Ahn~\cite[Lemma 3.8]{ref:OhVo20}. However, their algorithm has a critical error. Their algorithm is based on the observation that the intersection $B(s_1, s_2) \cap d$ must lie between $x_1$ and $x_2$, where $x_1$ and $x_2$ are the points on $d$ closest to $s_1$ and $s_2$, respectively. Unfortunately, this observation is not correct. We can construct a counterexample even in the Euclidean case; see Figure~\ref{fig:counterexample}.

Lemma~\ref{lem:diagonalintersection} further leads to the following lemma.

\begin{lemma}\label{lem:diagonal}
Given a diagonal $d$ of $P$, we can determine which side of $d$ contains $c^*$ in $O(n \log m)$ time.
\end{lemma}
\begin{proof}
The result of the lemma was claimed by Oh and Ahn~\cite[Lemma 3.8]{ref:OhVo20}. Their algorithm uses their incorrect result of Lemma~\ref{lem:diagonalintersection} as a subroutine. The rest of the algorithm is correct. Now by plugging our new algorithm for Lemma~\ref{lem:diagonalintersection} into the framework of \cite[Lemma~3.8]{ref:OhVo20}, we can determine which side of $d$ contains $c^*$ in $O(n \log m)$ time.
\end{proof}

\paragraph{Balanced polygon decomposition (BPD).}
We use the {\em balanced polygon decomposition} (BPD) for a simple polygon $P$~\cite{ref:GuibasLi87,ref:GuibasOp89,ref:ChazelleVi89}.
Chazelle~\cite{ref:ChazelleCu93} proved that $P$ contains a diagonal $d$ connecting two vertices such that it partitions $P$ into two subpolygons, each with at most $2m/3$ vertices. Recursively decomposing the subpolygons until each becomes a triangle yields the BPD.
The BPD can be represented by a tree $T_P$, called the {\em BPD-tree}. Each node of $T_P$ corresponds to a subpolygon of $P$: the root corresponds to $P$ itself, and each leaf corresponds to a triangle.
The triangles at the leaves of $T_P$ form a triangulation of $P$. The entire decomposition and $T_P$ can be constructed in $O(m)$ time~\cite{ref:GuibasOp89}. In the following, we assume that $T_P$ is already constructed in the global preprocessing for $P$. In fact, the GH data structure also relies on $T_P$; so $T_P$ is actually part of the GH data structure.
\medskip

As discussed in \cite{ref:OhVo20}, using $T_P$ and Lemma~\ref{lem:diagonal}, we can find in $O(n \log^2 m)$ time a leaf of $T_P$ whose triangle $Q$ contains $c^*$. The remaining task is to compute $c^*$ in $Q$.

\subsubsection{Computing $\boldsymbol{c^*}$ in the triangle $\boldsymbol{Q}$}

\begin{figure}[t]
\begin{minipage}[t]{\linewidth}
\begin{center}
\includegraphics[totalheight=1.6in]{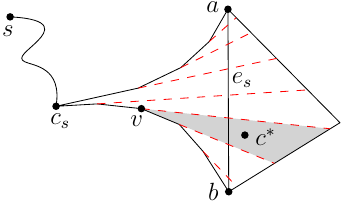}
\caption{The gray region is $R_s^*$. For any point $q\in R_s^*$, $\pi(s,q)=\pi(s,v)\cup \overline{vq}$, meaning that $d(s, q) = d(s, v) + |\seg{v q}|$.}
\label{fig:extensions}
\end{center}
\end{minipage}
\end{figure}

Consider a site $s\in S\setminus Q$. Let $e_s$ be the edge of $Q$ that separates $Q$ from $s$ (i.e., $Q$ and $s$ lie in two different subpolygons of $P$ divided by $e_s$); let $a$ and $b$ be the two endpoints of $e_s$. Let $c_s$ be the junction vertex of $\pi(s,a)$ and $\pi(s,b)$ (see Figure~\ref{fig:extensions}).

For each edge of $\pi(c_s,a)\cup \pi(c_s,b)$, we extend it toward $e_s$, and the portion of the extension inside $Q$ is a line segment that partitions $Q$ into two regions; let $L_s$ denote the set of these segments in $Q$ obtained from all edges of $\pi(c_s,a)\cup \pi(c_s,b)$.
The segments of $L_s$ partition $Q$ into a set $R_s$ of convex regions, each of which has at most five vertices. We wish to determine the region $R_s^*$ of $R_s$ that contains $c^*$. The reason we are interested in $R_s^*$ is that it has the following property: the combinatorial structure of the shortest path $\pi(s,q)$ for all points $q\in R_s^*$ is identical, and the analytic form of the corresponding function $d(s,q)$ is also identical for all points $q\in R_s^*$ (more specifically, $d(s,q)=d(s,v)+|vq|$ for a vertex $v$ of $R_s^*$; see Figure~\ref{fig:extensions}).

For each segment $g\in L_s$, it partitions $Q$ into two regions. To determine which side of $g$ contains $c^*$, we extend $g$ along both directions to the boundary of $P$ to obtain a diagonal $g'$ of $P$ with $g\subseteq g'$. Then, we apply Lemma~\ref{lem:diagonal} on $g'$ to determine which side contains $c^*$. Note that $g'$ can be obtained in $O(\log m)$ time by ray-shooting queries.

If we perform this for every segment $g \in L_s$, we can determine $R_s^*$, but the running time would be $O(k n \log m)$, where $k = |L_s|$ and $k = \Omega(m)$ in the worst case. We improve the algorithm by performing binary search on $L_s$. Since the segments of $L_s$ do not intersect, once we determine which side of a segment $g \in L_s$ contains $c^*$, we can discard half of the remaining segments. Thus, in $O(\log m)$ iterations, we can determine $R_s^*$.

To support this binary search, we need a sorted order of the segments in $L_s$. The GH data structure provides a structure supporting binary search on the edges of $\pi(c_s, a)$ and $\pi(c_s, b)$. The order of edges along these paths is consistent with the order of their extensions in $Q$. Hence, we can perform binary search on $L_s$. For simplicity, we assume that we have a data structure $D_s$ that supports binary search on $L_s$, which can be obtained in $O(\log m)$ time using the GH data structure.

With $D_s$, we can compute $R_s^*$ in $O(\log m)$ iterations. Each iteration takes $O(n \log m)$ time due to Lemma~\ref{lem:diagonal}, so the total time is $O(n \log^2 m)$.
Applying this independently to each site $s \in S$ would take $\Omega(n^2)$ time. In what follows, we improve this using a cutting-based approach, yielding a total running time of $O(n \log^2 m)$.

\paragraph{Computing $\boldsymbol{R_s^*}$ for all $\boldsymbol{s\in S}$.}
For simplicity, for each site $s\in S$ with $s\in Q$, we let $R_s^*=Q$.

For each point $s\in S\setminus Q$, to compute $R_s^*$, we first determine the edge $e_s$ of $Q$.  This can be done in $O(\log m)$ time.
Indeed, for each vertex $v$ of $Q$, using the GH data structure, we can compute the edge $e_v$ of $\pi(s,v)$ incident to $v$.
If $e_v$ intersects the edge of $Q$ opposite to $v$, then that edge is $e_s$.

After obtaining $e_s$, we can compute the junction vertex $c_s$ and construct the data structure $D_s$ in $O(\log m)$ time using the GH data structure.

During the execution of the algorithm, some segments of $L_s$ will be pruned (meaning that we already know which side contains $c^*$ for each pruned segment).
Let $k_s$ denote the number of remaining (unpruned) segments of $L_s$ at the beginning of each iteration. Initially, $k_s=|L_s|$.

Using the data structure $D_s$, we select the median segment among the remaining segments of $L_s$ and assign it a {\em weight} equal to $k_s$.
Let $L$ denote the set of these median segments for all $s\in S\setminus Q$.
Then, we have $|L|\leq n$.

Let $L'$ denote the set of supporting lines of the segments in $L$.
Each line in $L'$ inherits the same weight as its corresponding segment.
Let $w(L')$ denote the total weight of $L'$.

We compute a cutting for the weighted set of lines $L'$.
A {\em cutting} is a collection of interior-disjoint cells (each of which is a possibly unbounded triangle) whose union covers the entire plane.
The {\em size} of the cutting is the number of cells.
For a parameter $r$, a cutting $\Xi$ is a {\em $(1/r)$-cutting} for $L'$ if the total weight of all lines in $L'_\sigma$ is at most $w(L')/r$ for each cell $\sigma\in \Xi$, where $L'_\sigma$ is the set of lines intersecting the interior of $\sigma$.

We compute a $(1/r)$-cutting $\Xi$ of size $O(r^2)$ for $L'$ for a sufficiently large constant $r>0$.
This can be done in $O(n)$ time~\cite{ref:MatousekCu91,ref:ChazelleCu93}, since $|L'|\leq n$ and $r$ is constant.
For simplicity, we only consider the portion of $\Xi$ inside $Q$.

Next, we determine the cell $\sigma^*$ of $\Xi$ that contains $c^*$.
Since $\Xi$ has size $O(r^2)$ and $r$ is a constant, this can be done in $O(n\log m)$ time by applying Lemma~\ref{lem:diagonal} to each edge of $\Xi$ (specifically, each edge is extended to a diagonal of $P$ via ray shooting).

For each segment $g$ whose supporting line is not in $L'_{\sigma^*}$, we immediately know which side of $g$ contains $c^*$, and thus we can prune half of the segments in $L_s$, where $s$ is the site such that $g\in L_s$.
Since the total weight of the lines in $L'_{\sigma^*}$ is at most $w(L')/r$, the total weight of the segments that are pruned in this iteration is at least $\frac{r-1}{2r}\cdot w(L')$.
In other words, each iteration removes at least a constant fraction of the remaining segments in $\bigcup_{s\in S\setminus Q}L_s$.

Therefore, after $O(\log m)$ iterations, we can determine $R_s^*$ for all $s\in S$.
Since each iteration takes $O(n\log m)$ time, the total time is $O(n\log^2 m)$.

\paragraph{Computing $\boldsymbol{c^*}$ -- a final step.}
Next, we compute the common intersection $R^*$ of $R_s^*$ for all $s\in S$.
Since each $R_s^*$ is convex and has at most five edges, computing $R^*$ takes $O(n \log n)$ time.
Note that $R^*$ is convex and contains $c^*$.
Since for each site $s \in S$ the analytic form of the function $d(s, q)$ is determined and is identical for all $q \in R^*$, we can compute $c^*$ in an additional $O(n)$ time~\cite{ref:PollackCo89}.

In summary, computing $c^*$ takes $O(n \log n + n \log^2 m)$ time in total. We thus obtain the following lemma.

\begin{theorem}\label{thm:geocenter}
Let $P$ be a simple polygon with $m$ vertices.
After $O(m)$ preprocessing of $P$, given any set $S$ of $n$ points in $P$, we can compute the geodesic center of $S$ in $O(n \log n + n \log^2 m)$ time.
\end{theorem}

Consequently, as discussed before, computing the implicit $\vd_f(S)$ can be accomplished in $O(n \log n + n \log^2 m)$ time.

As in the nearest diagram case, once we have the implicit  $\vd_f(S)$, we can also construct it explicitly in an additional $O(m + n \log n)$ time using the GH data structure.

\subsubsection{Proving Lemma~\ref{lem:diagonalintersection} -- a tentative prune-and-search algorithm}
\label{sec:diagonalintersection}

In this section, we prove Lemma~\ref{lem:diagonalintersection}. We believe the result is interesting in its own right and will find many other applications (e.g., as remarked in the proof of Lemma~\ref{lem:bpp}, it can be used to solve the last step of the algorithm there in $O(\log m)$ time). We restate the lemma in the following more general and self-contained form.

\begin{lemma}\label{lem:gendiagonalintersection}
Let $P$ be a simple polygon with $m$ vertices. After $O(m)$-time preprocessing of $P$, given any diagonal $d$ of $P$ and two points $s, t \in P$ that lie on the same side of $d$, we can determine in $O(\log m)$ time whether the bisector $B(s, t)$ intersects $d$, and if so, compute the intersection point.
\end{lemma}

We construct the GH data structure as the preprocessing for $P$. This takes $O(m)$ time.

Without loss of generality, we assume that $d = \overline{ab}$ is an edge of $P$, $d$ is vertical, $a$ is above $b$, and $P$ lies locally to the left of $d$. Let $s$ and $t$ be two points in $P$. To simplify the notation, let $B = B(s, t)$. Our goal is to determine whether $B \cap d = \emptyset$, and if not, compute $B \cap d$.

We first determine whether $B \cap d = \emptyset$. Note that $B \cap d$ consists of at most one point~\cite{ref:AgarwalIm18,ref:AronovOn89}, and $B \cap d = \emptyset$ if and only if either $d(s, a) < d(t, a)$ and $d(s, b) < d(t, b)$ or $d(s, a) > d(t, a)$ and $d(s, b) > d(t, b)$. Therefore, whether $B \cap d = \emptyset$ can be determined in $O(\log m)$ time using the GH data structure. In the following, we assume that $B \cap d \neq \emptyset$ and let $z = B \cap d$. Our goal is to compute $z$ in $O(\log m)$ time.

For convenience, we extend $B$ beyond $z$ outside $P$ through $d$ (alternatively, we can redefine $P$ as an unbounded polygon by removing $d$, and then $B$ is simply the bisector of $s$ and $t$ in this unbounded polygon). Since $d$ is vertical, the portion of $B$ beyond $d$ (i.e., outside $P$) is $x$-monotone~\cite{ref:AgarwalIm18}. Our goal is to compute the intersection between $d$ and this unbounded bisector $B$.

\begin{figure}[t]
\begin{minipage}[t]{0.49\linewidth}
\begin{center}
\includegraphics[totalheight=1.7in]{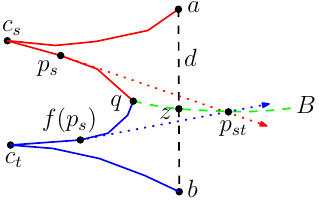}
\caption{Illustrating the definition of $f(p_s)$. The green curve is the bisector $B$.}
\label{fig:bisectorintersect10}
\end{center}
\end{minipage}
\hspace{0.02in}
\begin{minipage}[t]{0.49\linewidth}
\begin{center}
\includegraphics[totalheight=1.7in]{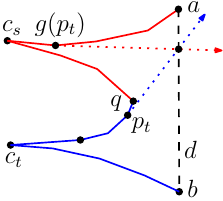}
\caption{Illustrating the definition of $g(p_t)$. }
\label{fig:bisectorintersect20}
\end{center}
\end{minipage}
\end{figure}

Let $q$ be the middle point of $\pi(s, t)$, which can be computed in $O(\log m)$ time using the GH data structure (see Figure~\ref{fig:bisectorintersect10}). Clearly, $q \in B$. Let $c_s$ be the junction vertex of $\pi(s, a)$ and $\pi(s, b)$. Let $c_t$ be the junction vertex of $\pi(t, a)$ and $\pi(t, b)$. Both vertices can be computed in $O(\log m)$ time using the GH data structure. Note that since $B$ intersects $d$, it must be the case that $c_s \in \pi(s, q)$ and $c_t \in \pi(t, q)$.

To compute $z$, we use a similar approach as in \cite[Lemma 3.10]{ref:WangA23}, which makes use of the tentative prune-and-search technique of Kirkpatrick and Snoeyink~\cite{ref:KirkpatrickTe95}. For completeness, we present some details below. In fact, the algorithm of \cite[Lemma 3.10]{ref:WangA23} misses one case, and we fix the issue here. To avoid a lengthy background explanation, we follow the notation in \cite{ref:KirkpatrickTe95} without definition. We will rely on \cite[Theorem 3.6]{ref:KirkpatrickTe95}. To do so, we need to define a decreasing function $f$ and an increasing function $g$.

\paragraph{Defining two functions $\boldsymbol{f}$ and $\boldsymbol{g}$.}
Let $\gamma_s = \pi(c_s, a)\cup \pi(c_s, q)$. For each point $p \in \gamma_s$, we let $\rho_p$ be the ray from $p$, tangent to $\gamma_s$, and toward $d$; we call $\rho_p$ the \emph{tangent ray} of $\gamma_s$ at $p$. Similarly, let $\gamma_t = \pi(c_t, q) \cup \pi(c_t, b)$ and define tangent rays for points on $\gamma_t$.\footnote{An alternative way is to define $\gamma_s = \pi(c_s, a) \cup \pi(c_s, b)$ and $\gamma_t = \pi(c_t, a)\cup \pi(c_t, b)$, and then follow the similar approach afterwards. In this way, the point $q$ is not needed.}

We parameterize $\gamma_s$ over $[0, 1]$ from $a$ to $q$, so that each parameter value corresponds to a point on $\gamma_s$. Similarly, we parameterize $\gamma_t$ over $[0, 1]$ from $q$ to $b$. For each point $p_s \in \gamma_s$, we define $f(p_s)$ to be the parameter of the point $p_t \in \gamma_t$ such that the two tangent rays $\rho_{p_s}$ and $\rho_{p_t}$ intersect at a point on the bisector $B$ (see Figure~\ref{fig:bisectorintersect10}). Note that the tangent ray at a vertex of $\gamma_s$ (resp., $\gamma_t$) is not unique, but this issue can be handled as in \cite{ref:KirkpatrickTe95} so that we can assume that as $p$ changes along $\gamma_s$ (resp., $\gamma_t$) from one end to the other, the slope of the tangent ray at $p$ changes continuously. It is possible that $\rho_{p_s}$ does not intersect $B$, in which case we let $f(p_s)$ be the parameter of the point $b'$, where $b'$ is the first point on $\gamma_t$ from $b$ toward $q$ whose tangent ray intersects $B$ (note that this is only a definition and we do not need to compute $b'$; our alternative approach at the end of this section actually provides an algorithm to compute $b'$). It is not difficult to see that $f$ is monotone decreasing.

For each point $p_t \in \gamma_t$, we define $g(p_t)$ to be the parameter of the point $p_s \in \gamma_s$ such that $\rho_{p_t}$ and $\rho_{p_s}$ intersect at a point on $d$.\footnote{It is possible that $\rho_{p_t}$ does not intersect $d$. To avoid this situation, we could perform a preprocessing step to remove the portion of $\gamma_t$ that contains points $p_t$ whose tangent ray $\rho_{p_t}$ does not intersect $d$. Such a preprocessing step can be done in $O(\log m)$ time. We can do the same for $\gamma_s$.} Note that $g$ is monotone increasing. Observe that the fixed point of the composition $g \circ f$ corresponds to $z$.

To compute $z$, we can apply the algorithm of \cite[Theorem 3.6]{ref:KirkpatrickTe95}. To this end, we need to show that given any two points $p_s \in \gamma_s$ and $p_t \in \gamma_t$, we can determine whether $f(p_s) \ge p_t$ in $O(1)$ time and also determine whether $g(p_t) \ge p_s$ in $O(1)$ time.

\paragraph{Resolving comparisons for $\boldsymbol{g}$.}
To determine whether $g(p_t) \ge p_s$, we first compute the intersections $q_s = \rho_{p_s} \cap d$ and $q_t = \rho_{p_t} \cap d$. If $q_s$ is above or at $q_t$, then $g(p_t) \ge p_s$; otherwise, $g(p_t) < p_s$. Hence, whether $g(p_t) \ge p_s$ can be determined in $O(1)$ time.

\paragraph{Resolving comparisons for $\boldsymbol{f}$.}
To determine whether $f(p_s) \ge p_t$, we do the following. If $\rho_{p_s}$ and $\rho_{p_t}$ intersect, then let $p$ be the intersection. Observe that $f(p_s) \ge p_t$ if and only if $d(s, p) \le d(t, p)$. Note that $d(s, p) = d(s, p_s) + |p_s p|$ and $d(t, p) = d(t, p_t) + |p_t p|$. The GH data structure can represent $\pi(c_s, a)$ in a structure that supports binary search on its vertices, so that $d(s, p_s)$ and $d(t, p_t)$ are available when we apply the algorithm. Hence, in this case, we can determine whether $f(p_s) \ge p_t$ in $O(1)$ time.

If $\rho_{p_s}$ and $\rho_{p_t}$ do not intersect, then we consider the situation in which either $\rho_{p_s}$ intersects the backward extension of $\rho_{p_t}$ or $\rho_{p_t}$ intersects the backward extension of $\rho_{p_s}$. In the former case, $f(p_s) < p_t$, whereas in the latter case, $f(p_s) > p_t$. Hence, this case can also be resolved in constant time.

The remaining case is when the backward extension of $\rho_{p_s}$ intersects the backward extension of $\rho_{p_t}$. This case is missing from the algorithm of~\cite[Lemma 3.10]{ref:WangA23}. If we assume that this case never happens, then the algorithm of~\cite[Theorem 3.6]{ref:KirkpatrickTe95} can be directly applied, which actually only uses the prune-and-search technique and does not need the tentative technique. More specifically, in each iteration, half of the remaining portion of either $\gamma_s$ or $\gamma_t$ is pruned. If we use the notation in~\cite{ref:KirkpatrickTe95}, the algorithm is always in the \emph{normal mode} and never enters a \emph{tentative mode}.

If this case does happen, then we do not immediately know which portion of $\gamma_s$ or $\gamma_t$ should be pruned. In this case, we can do the following. We let the algorithm enter a tentative mode by tentatively pruning the portion of $\gamma_s$ between $p_s$ and $a$ and tentatively pruning the portion of $\gamma_t$ between $p_t$ and $b$. The key observation is that at least one of the tentative discards is correct. To see this, let $z_s$ be the point of $\gamma_s$ whose tangent ray $\rho_{z_s}$ contains $z$. Define $z_t\in \gamma_t$ similarly. We claim that the following two cases cannot both happen: (1) $z_s$ is between $p_s$ and $a$, and (2) $z_t$ is between $p_t$ and $b$. Indeed, assume to the contrary that both cases happen. Then, since $\rho_{p_s}$ does not intersect $\rho_{p_t}$, $\rho_{z_s}$ does not intersect $\rho_{z_t}$ either. But this yields a contradiction since $\rho_{z_s}$ and $\rho_{z_t}$ intersect at $z$.

The above observation implies that the tentative mode is {\em viable}~\cite{ref:KirkpatrickTe95}. Indeed, in the future algorithm, each iteration permanently or tentatively prunes half of the remaining portion of a list (i.e., $\gamma_s$ or $\gamma_t$), or certifies that all of the tentative discards on one list are correct, revokes the tentative discards on the other list, and returns to the normal mode. Therefore, the tentative prune-and-search algorithm framework in \cite{ref:KirkpatrickTe95} still applies, and the total number of iterations of the whole algorithm is $O(\log m)$.

We remark that the above algorithm could be simplified by merging the last two cases. More specifically, if $\rho_{p_s}$ and $\rho_{p_t}$ intersect, then we proceed in the same way as before, in which case a permanent discard is made. Otherwise, we make the tentative discards on both lists as above.

In summary, the intersection $z=B\cap d$ can be computed in $O(\log m)$ time.

\paragraph{\bf An alternative approach.}
We present an alternative approach that computes $z$ in $O(\log m)$ time. This algorithm, which uses the above algorithm as a subroutine, is even more interesting because it may be used as a preprocessing step for other related problems.

We orient $\gamma_s$ from $a$ to $q$ and orient $\gamma_t$ from $b$ to $q$.
Define $a'$ as the first point of $\gamma_s$ such that the tangent ray $\rho_{a'}$ intersects $B$.
Similarly, define $b'$ as the first point of $\gamma_t$ such that the tangent ray $\rho_{b'}$ intersects $B$.
Alternatively, $a'$ and $b'$ can be defined as follows.
Let $d'$ be a vertical line whose $x$-coordinate is $+\infty$, and let $z' = B \cap d'$.
Then $a'$ is the point of $\gamma_s$ whose tangent ray $\rho_{a'}$ contains $z'$, and $b'$ is the point of $\gamma_t$ whose tangent ray $\rho_{b'}$ contains $z'$.
In particular, $\rho_{a'}$ and $\rho_{b'}$ intersect at $z'$.

Define $\gamma_s'$ to be the portion of $\gamma_s$ between $a'$ and $q$, and $\gamma_t'$ the portion of $\gamma_t$ between $b'$ and $q$.
By definition, we have $z_s \in \gamma_s'$ and $z_t \in \gamma_t'$.
For this reason, we call $\gamma_s \setminus \gamma_s'$ the \emph{useless portion} of $\gamma_s$; similarly, $\gamma_t \setminus \gamma_t'$ is the \emph{useless portion} of $\gamma_t$.

The key observation is that for any two points $p_s \in \gamma_s'$ and $p_t \in \gamma_t'$, their tangent rays $\rho_{p_s}$ and $\rho_{p_t}$ must intersect.
This can be argued by contradiction in a manner similar to the previous analysis.
Assume to the contrary that they do not intersect.
Since $a'$ lies between $p_s$ and $a$ along $\gamma_s$, and $b'$ lies between $p_t$ and $b$ along $\gamma_t$, it follows that the rays $\rho_{a'}$ and $\rho_{b'}$ would also not intersect.
However, this contradicts the fact that $\rho_{a'}$ and $\rho_{b'}$ intersect at $z'$.

Assuming that $a'$ and $b'$ are already known, the above observation implies that if we apply the algorithm to $\gamma_s'$ and $\gamma_t'$, then in every iteration the tangent rays $\rho_{p_s}$ and $\rho_{p_t}$ always intersect.
Therefore, the algorithm never enters the tentative mode and computes $z$ in $O(\log m)$ time.

It remains to discuss how to compute $a'$ and $b'$. We can apply essentially the same algorithm as above (including the tentative mode), but with respect to the line $d'$ to compute $z'$ instead.
The running time remains $O(\log m)$.
Thus, this step can be viewed as a \emph{preprocessing phase} for computing $z$.
Its purpose is to eliminate the useless portions of $\gamma_s$ and $\gamma_t$, thereby simplifying the subsequent computation.

This preprocessing step can also be applied to other problems involving convex chains (since $\gamma_s$ and $\gamma_t$ can be viewed as convex chains).
For example, in the problem of computing the intersection of two bisectors of three convex chains (equivalently, computing a Voronoi vertex)~\cite[Lemma 3.11]{ref:WangA23}, the algorithm in \cite{ref:WangA23} misses the case where the backward extension of $\rho_{p_s}$ intersects the backward extension of $\rho_{p_t}$.
This issue can be resolved by first eliminating the useless portions of the three chains using the preprocessing described above, and then applying the algorithm in \cite[Lemma 3.11]{ref:WangA23}.

\subsection{Point location in the implicit $\boldsymbol{\vd_f(S)}$}
\label{sec:fvdpointlocation}

We now discuss how to answer point location queries in the implicit $\vd_f(S)$.
Namely, given a query point $q \in P$, we wish to find the farthest Voronoi region of $\vd_f(S)$ that contains $q$, or equivalently, determine the farthest neighbor of $q$ in $S$.

As in the nearest diagram case, we reduce the problem to point location in a geodesic decomposition of $P$ and then apply Theorem~\ref{thm:pointlocation}.
To this end, we transform $\vd_f(S)$ into a geodesic decomposition of $P$ in an additional $O(n \log m)$ time.
The details are given below.

The following observation is known and will be used later.

\begin{observation}[{\cite{ref:AronovTh93}}]\label{obser:fvdextension}
For any point $p \in P$, if $p \in \R_f(s)$ for some site $s \in S$, then $\seg{pp'} \subseteq \R_f(s)$, where $\seg{pp'}$ is the last-edge extension of $\pi(s,p)$.
\end{observation}

\begin{figure}[t]
    \centering
    \begin{subfigure}[b]{.49\textwidth}
        \centering
        \includegraphics[width=0.8\linewidth]{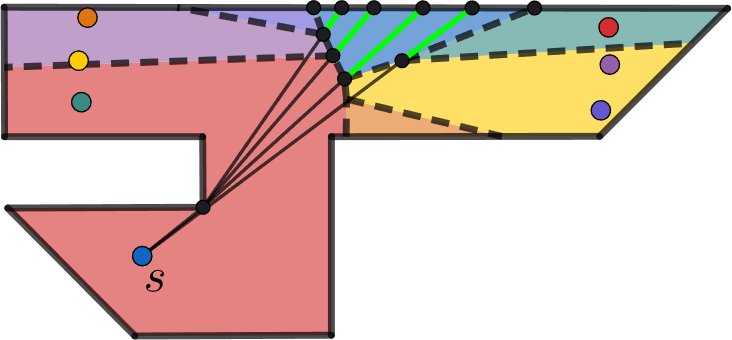}
        \subcaption{Spokes in one farthest Voronoi region $\R_f(s)$.}
        \label{fig:farthest_spoking}
    \end{subfigure}
    \hfill
    \begin{subfigure}[b]{.49\textwidth}
        \centering
        \includegraphics[width=0.8\linewidth]{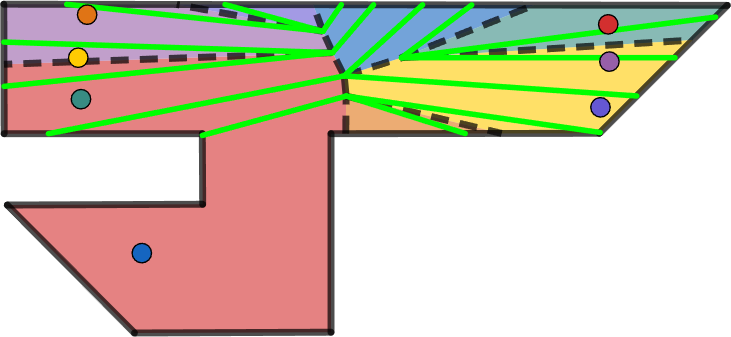}
        \subcaption{The entire spoked farthest Voronoi diagram.}
        \label{fig:farthest_spoked}
    \end{subfigure}
    \caption{Illustrating spokes added to $\vd_f(S)$. The green segments are spokes.}
    \label{fig:farthest_voronoi_diagram}
\end{figure}

Consider a Voronoi region $\R_f(s)$ for some $s \in S$.
Let $v$ be a Voronoi vertex incident to $\R_f(s)$.
Let $\seg{v v_s}$ be the last-edge extension of $\pi(s, v)$.
By Observation~\ref{obser:fvdextension}, $\seg{v v_s}$ lies entirely in $\R_f(s)$.
We add $\seg{v v_s}$ to the implicit $\vd_f(S)$ as a new edge and call it a \emph{spoke}.
We perform this for each Voronoi vertex $v$ of $\R_f(s)$ (see Figure~\ref{fig:farthest_spoking}).

Notice that these spokes are pairwise disjoint because they are all last-edge extensions of shortest paths from $s$, i.e., they belong to a shortest path tree rooted at $s$.
Hence, the spokes partition $\R_f(s)$ into a set of regions, which we call \emph{spoked regions}.
Each spoked region is bounded by at most two spokes, one $\partial P$-portion, and one Voronoi edge.
We associate each spoked region with $s$.
We perform this construction for all Voronoi regions $\R_f(s)$ of $\vd_f(S)$.
The resulting subdivision is called the \emph{spoked farthest geodesic Voronoi diagram}, or simply the \emph{spoked diagram} (see Figure~\ref{fig:farthest_spoked}).

Let $e$ be a Voronoi edge in $\vd_f(S)$.
By definition, $e$ is incident to two spoked regions in the spoked diagram;
let $R_e$ denote the union of these two regions.
Let $u$ and $v$ be the endpoints of $e$, and let $s$ and $t$ be the sites associated with $e$.
Let $R_e(s)$ (resp., $R_e(t)$) be the spoked region of $R_e$ associated with $s$ (resp., $t$).

\begin{figure}[t]
    \centering
    \includegraphics[totalheight=2.5in]{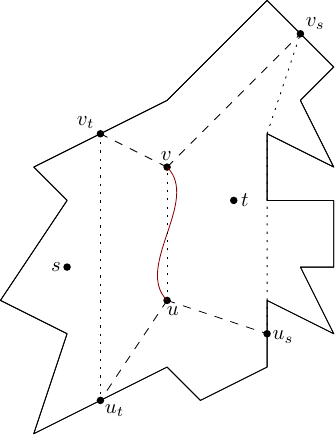}
    \caption{Illustrating the decomposition of $R_e$.}
    \label{fig:farthest_edge}
\end{figure}

We now perform a \emph{decomposition process} on $R_e$ to decompose it into at most four geodesic triangles and at most two geodesic sectors, as defined in Section~\ref{sec:point_location}.
We first consider the case where both $u$ and $v$ are Voronoi vertices (see Figure~\ref{fig:farthest_edge}).

We define $v_t$, $u_s$, and $u_t$ analogously to $v_s$.
Let $\partial_e(s)$ be the $\partial P$-portion of $R_e(s)$, and $\partial_e(t)$ the $\partial P$-portion of $R_e(t)$.
Then $R_e$ is bounded by $\seg{v v_t}$, $\seg{v v_s}$, $\seg{u u_t}$, $\seg{u u_s}$, $\partial_e(s)$, and $\partial_e(t)$.

We first remove $e$ from the spoked diagram and add $\pi(u, v)$.
Observe that $\pi(u, v)$ does not cross $\partial R_e$.
Indeed, it does not cross $\seg{v v_t}$ or $\seg{v v_s}$ since they are all shortest paths from $v$.
Similarly, it does not cross $\seg{u u_t}$ or $\seg{u u_s}$ since they are all shortest paths from $u$.
Moreover, it cannot cross $\partial_e(s)$ or $\partial_e(t)$, since $\pi(u, v)$ lies entirely within $P$.
Because $u_t$, $v_t$, $u_s$, and $v_s$ are all on $\partial P$, the above implies that $\pi(u,v)$ is inside $R_e$ and partitions $R_e$ into two regions. One of them, denoted by $R_e'(t)$, is bounded by $\seg{u_t u}$, $\pi(u, v)$, $\seg{v v_t}$, and $\partial_e(t)$. The other one, denoted by $R_e'(s)$, is bounded by $\seg{u_s u}$, $\pi(u, v)$, $\seg{v v_s}$, and $\partial_e(s)$.

We further decompose $R_e'(t)$ as follows. First, we use $\pi(u_t, v_t)$ to partition $R_e'(t)$ into two regions. One of them is bounded by $\pi(u_t, v_t)$ and $\partial_e(t)$; this region is a polygonal segment as defined in Section~\ref{sec:point_location}.
The other region is bounded by four shortest paths: $\seg{u_t u}$, $\pi(u, v)$, $\seg{v v_t}$, and $\pi(u_t, v_t)$. We further decompose this region into two geodesic triangles in the same way as we did in Section~\ref{sec:vd_pl} (specifically, Lemma~\ref{lem:decomgeoquad} and the discussion immediately following it).

Similarly, we further decompose $R_e'(s)$ into a polygonal segment and two geodesic triangles.

In summary, if both $u$ and $v$ are Voronoi vertices, then $R_e$ can be decomposed into at most four geodesic triangles and two polygonal segments. Note that polygonal segments are special cases of geodesic sectors.

If one of $u$ and $v$ is a Voronoi vertex and the other is a Voronoi terminal, then without loss of generality we assume that $v$ is a Voronoi vertex and $u$ is a Voronoi terminal. In this case, we decompose $R_e$ using $\pi(u, v)$, which results in two geodesic sectors $\sector_v(v_t, u)$ and $\sector_v(u, v_s)$.
If both $u$ and $v$ are Voronoi terminals, then $R_e = P$, and no further decomposition is needed (we may consider $P$ a special polygonal segment and thus a special geodesic sector).

Therefore, in all cases, we can decompose $R_e$ into at most four geodesic triangles and at most two geodesic sectors. This decomposition can be performed in $O(\log m)$ time using the GH data structure.

We perform the above decomposition process for each Voronoi edge $e$. This results in a decomposition of $P$ into a set of geodesic triangles and geodesic sectors. Hence, the resulting decomposition is a geodesic decomposition of $P$; we denote it by $T$. Note that we can obtain $T$ from the implicit $\vd_f(S)$ in $O(n \log m)$ time. Indeed, each spoke can be computed in $O(\log m)$ time using the GH and ray-shooting data structures. Since $\vd_f(S)$ has $O(n)$ Voronoi vertices, there are $O(n)$ spokes in total. Decomposing the region $R_e$ for each Voronoi edge $e$ takes $O(\log m)$ time. As $\vd_f(S)$ has $O(n)$ Voronoi edges, $T$ consists of $O(n)$ geodesic triangles and geodesic sectors, and obtaining $T$ takes $O(n \log m)$ time in total.

With $T$, we now build a point location data structure in $O(n \log m)$ time and $O(n)$ additional space using Theorem~\ref{thm:pointlocation}.
Given a query point $q$, we can use the data structure to identify the geodesic triangle or geodesic sector $\triangle_q$ of $T$ that contains $q$ in $O(\log (n + m))$ time. The region $\triangle_q$ is contained in some $R_e$ for a Voronoi edge $e$ of $\vd_f(S)$. Let $s$ and $t$ be the sites associated with $e$. Then $q$ lies in either $\R_f(s)$ or $\R_f(t)$. In other words, the farthest neighbor of $q$ in $S$ is either $s$ or $t$. We can finally determine the correct one in $O(\log m)$ time using the GH data structure.

In summary, given the implicit $\vd_f(S)$, we can build a point location data structure in $O(n \log m)$ time and $O(n)$ additional space that supports farthest neighbor queries in $O(\log (n + m))$ time.
This completes the proof of Theorem~\ref{thm:farthest}.

\subsection{Dynamic farthest neighbor queries}
\label{sec:fvddynamic}

Similar to the applications of Theorem~\ref{thm:mainthm} discussed in Section~\ref{sec:dynamic}, an immediate application of Theorem~\ref{thm:farthest} is to design dynamic data structures for geodesic farthest neighbor searching in simple polygons.
We have the following farthest counterparts of Corollaries~\ref{cor:insert_only} and \ref{cor:fully_dynamic}.

\begin{corollary} \label{cor:far_insert_only}
Let $P$ be a simple polygon with $m$ vertices.
After $O(m)$ preprocessing of $P$, we can maintain an insertion-only geodesic farthest neighbor data structure for a set $S$ of $n$ (unweighted) point sites in $P$ that supports amortized $O(\log^2 n + \log n \log^2 m)$ insertion time and $O(\log n \log (n + m))$ query time.
\end{corollary}

\begin{corollary} \label{cor:far_fully_dynamic}
Let $P$ be a simple polygon with $m$ vertices.
After $O(m)$ preprocessing of $P$, given a set $S$ of $n$ (unweighted) point sites in $P$, we can construct a fully dynamic geodesic farthest neighbor data structure for $S$ in $O(n \log n + n \log^2 m)$ time that supports $O(\log n + \log^2 m)$ amortized insertion time, $O(\sqrt{n} (\log n + \log^2 m))$ amortized deletion time, and $O(\sqrt{n} \log (n + m))$ query time.
\end{corollary}

\section{Conclusion}
\label{sec:conclude}

In this paper, we studied the problem of constructing additively weighted geodesic Voronoi diagrams in simple polygons in an implicit manner.
After $O(m)$ preprocessing of a simple polygon $P$, we present an algorithm that computes the diagram $\vd(S)$ for any set $S$ of $n$ weighted sites in $O(n \log n + n \log^2 m)$ time, while supporting nearest neighbor queries in $O(\log (n + m))$ time.
In contrast to explicit constructions, which incur an $\Omega(m)$ overhead, our approach avoids dependence on the combinatorial complexity of $P$ in the output and is particularly effective when multiple diagrams must be constructed in the same polygon.
We also obtained the same asymptotic bounds for the farthest version of the problem on unweighted points.

From a technical perspective, our work introduces several new algorithmic tools.
We developed a sweep-line framework for geodesic Voronoi diagrams that operates entirely on an implicit representation, together with a new method for point location in such diagrams.
We also designed a general point location data structure for subdivisions defined by geodesic triangles and geodesic sectors, which may be of independent interest beyond the specific problems studied here.

We showed that our techniques lead to improved results for several applications, including dynamic nearest and farthest neighbor searching and shortest path problems in geodesic disk graphs.
In particular, our algorithms either improve the best known bounds or provide the first efficient solutions.
More broadly, our results suggest that implicit representations can be a powerful tool for handling geometric structures in complex domains, especially when the ambient environment is large but the input size is relatively small.



\bibliographystyle{plainurl}
\bibliography{references}

\end{document}